\documentclass[11pt,letterpaper,usenames,dvipsnames]{article}
\usepackage{jheppub}  
\usepackage{tikz} 
\usetikzlibrary{arrows,snakes,shapes.arrows,decorations.markings}
     \tikzset{>=triangle 90}
     \tikzstyle{gr}=[draw,circle,green!50!black,fill=green!50!black,scale=.6]
     \tikzstyle{Bl}=[draw,circle,blue,scale=.6]
     \tikzstyle{R}=[draw,circle,fill=red,scale=.6]
     \tikzstyle{bl}=[draw,circle,fill=black,scale=.6]
     \tikzstyle{rc}=[circle,fill=red,scale=.6]
\usepackage{bm} 
\usepackage{array}
\usepackage{multirow} 
\usepackage{colortbl} 
\usepackage{stfloats}
\usepackage{xcolor}   
\usepackage{url}
\usepackage{pifont}       

\def\cmark{\text{\ding{51}}}
\def\del{{\partial}}
\def\bar{\overline}
\def\til{\widetilde}
\def\hat{\widehat}
 
\def\vev#1{{\langle{#1}\rangle}} 

\def\v{\vee}
\def\^{\wedge}
\def\tr{\mathop{\rm tr}}
\def\Im{\mathop{\rm Im}}

\def\U{\mathop{\rm u}}
\def\SU{\mathop{\rm su}}

\def\SO{\mathop{\rm so}}
\def\SL{\mathop{\rm SL}}
\def\GL{\mathop{\rm GL}} 
\def\Sp{\mathop{\rm sp}}

\def\C{\mathbb{C}}

\def\N{\mathbb{N}}

\def\Z{\mathbb{Z}} 
\def\ff{{\mathfrak f}}

\def\bm{{\bf m}}
\def\bR{{\bf R}}
\def\cA{{\mathcal A}}
\def\cB{{\mathcal B}}
\def\cBh{{\hat\cB}}
\def\cC{{\mathcal C}}

\def\cD{{\mathcal D}}
\def\cE{{\mathcal E}}

\def\cK{{\mathcal K}}
\def\cM{{\mathcal M}}
\def\cN{{\mathcal N}}

\def\cP{{\mathcal P}}
\def\cR{{\mathcal R}}
\def\cS{{\mathcal S}}
\def\cW{{\mathcal W}}

\def\tj{{\til\jmath}}

\def\tS{{\til S}}
\def\tQ{{\til Q}}

\def\a{{\alpha}}
\def\ad{{\dot\a}}
\def\ba{{\boldsymbol\a}}
\def\b{{\beta}}

\def\g{{\gamma}}

\def\G{{\Gamma}}
\def\d{{\delta}}

\def\D{{\Delta}}

\def\th{{\theta}}

\def\l{{\lambda}}
\def\L{{\Lambda}}
\def\m{{\mu}}
\def\n{{\nu}}

\def\s{{\sigma}}
\def\S{{\Sigma}}
\def\t{{\tau}}

\def\f{{\phi}}

\def\w{{\omega}}
\def\bw{{\boldsymbol\w}}
\def\bz{{\boldsymbol{z}}}
\def\ccw{\cellcolor{white}}

\def\rcy{\rowcolor{black!25!yellow!10}}

\def\ccr{\cellcolor{red!15}}
\def\rcr{\rowcolor{red!15}}

\title{Geometric constraints on the space of N=2 SCFTs\\ 
I: physical constraints on relevant deformations}

\author{Philip Argyres,}
\author{Matteo Lotito,}
\author{Yongchao L\"u,}
\author{and Mario Martone}

\affiliation{Physics Department, University of Cincinnati,\\
Cincinnati OH 45221-0011, USA}

\emailAdd{philip.argyres@gmail.com}
\emailAdd{lotitomo@mail.uc.edu}
\emailAdd{lychaoaa@gmail.com}
\emailAdd{martonmo@mail.uc.edu}

\abstract{
We initiate a systematic study of four dimensional  $\cN=2$ superconformal field theories (SCFTs) based on the analysis of their Coulomb branch geometries.  Because these SCFTs are not uniquely characterized by their scale-invariant Coulomb branch geometries we also need information on their deformations.  We construct all inequivalent such deformations preserving $\cN=2$ supersymmetry and additional physical consistency conditions in the rank 1 case.  These not only include all the ones previously predicted by S-duality, but also 16 additional deformations satisfying all the known $\cN=2$ low energy consistency conditions.  All but two of these additonal deformations have recently been identified with new rank 1 SCFTs; these identifications are briefly reviewed.

Some novel ingredients which are important for this study include: a discussion of RG-flows in the presence of a moduli space of vacua;  a classification of local $\cN=2$ supersymmetry-preserving deformations of unitary $\cN=2$ SCFTs; and an analysis of charge normalizations and the Dirac quantization condition on Coulomb branches.

This paper is the first in a series of three. The second paper \cite{Argyres:2015gha} gives the details of the explicit construction of the Coulomb branch geometries discussed here, while the third \cite{Argyres:2015ccharges} discusses the computation of central charges of the associated SCFTs.
}

\begin{document}
\maketitle
\flushbottom


\section{Introduction and summary}

Four-dimensional $\cN=2$ supersymmetric quantum field theories are notable for the richness of the techniques that have been brought to bear on solving for their supersymmetry-protected observables.  These techniques have revealed the existence of many new non-lagrangian field theories as strongly coupled fixed points of RG flows or as decoupled factors upon tuning marginal couplings in known theories.  Yet, despite the variety of techniques deployed, systematic results describing these fixed points are lacking.  One way of approaching this problem, which we will explore here, is to try to classify the possible geometries that can appear on the moduli space of vacua of $\cN=2$ field theories.   Finding a given such moduli space geometry does not imply that a corresponding $\cN=2$ QFT exists, but classifying all possible such geometries puts constraints on the possible QFTs that can exist.

We will focus on the Coulomb branch (CB) of the moduli space since it is not lifted under supersymmetric deformations \cite{Seiberg:1994rs,Seiberg:1994aj}.  The massless fields in the generic vacuum on the CB are free vector multiplets with ${\rm U}(1)^r$ gauge group.  We approach the problem of classifying CBs in two steps: 
\begin{itemize}
\item[1)] classify the possible physical scale-invariant CBs, 
\item[2)] classify their physical deformations by relevant parameters.  
\end{itemize}

\emph{Note added:}  In the two-and-a-half years since this paper first appeared as a preprint, most (all but two) of the new rank-1 CB geometries that we found have been identified with new SCFTs constructed by various means.  The rest of the introduction has been modified to reflect the current state of knowledge as of October 2017.

\paragraph{Scale invariant Coulomb branches.} 

The first step addresses the question of what are the geometries which can be interpreted as CBs of $\cN=2$ SCFTs.  An $\cN=2$ SUSY gauge theory with a rank $r$ gauge group has an $r$ complex-dimensional CB.  We extend the definition of ``rank" to non-lagrangian theories by defining the rank of a theory as the complex dimension of its CB.  For the rest of the paper we only consider the rank 1 case.

CBs of physically interesting theories are singular, and each singular point corresponds to a vacuum for which the theory contains additional massless states \cite{Seiberg:1994rs,Seiberg:1994aj} (for pedagogical reviews see \cite{AlvarezGaume:1996mv,Tachikawa:2013kta}).  Conformal invariance constrains the geometry to be scale invariant.  In the rank 1 case this restricts a CB to have a single singular point, since the distance between two distinct singularities would define a scale.  Furthermore, scale invariance restricts any component of a CB to be a flat cone with arbitrary deficit angle $\d$.  The singular point is the tip of the cone and represents the conformal vacuum, while all other points at finite distance from the tip are vacua where the conformal invariance is spontaneously broken.

In this paper we consider only rank 1 CB geometries whose complex structure is that of the complex plane, $\C$.  We will call such geometries ``planar".  In physical terms, this assumption means that the (reduced) chiral ring of Coulomb branch operators in the CFT is freely generated.  In particular, this implies that a good complex coordinate on the CB is the vev of a field in the CFT.  Without this assumption CBs with multiple components or more non-trivial topologies are allowed, and, for a given topology, infinitely many more geometries are allowed.  These more exotic CBs might be of physical interest in their own right, and are discussed in some detail in \cite{Argyres:2017tmj}.

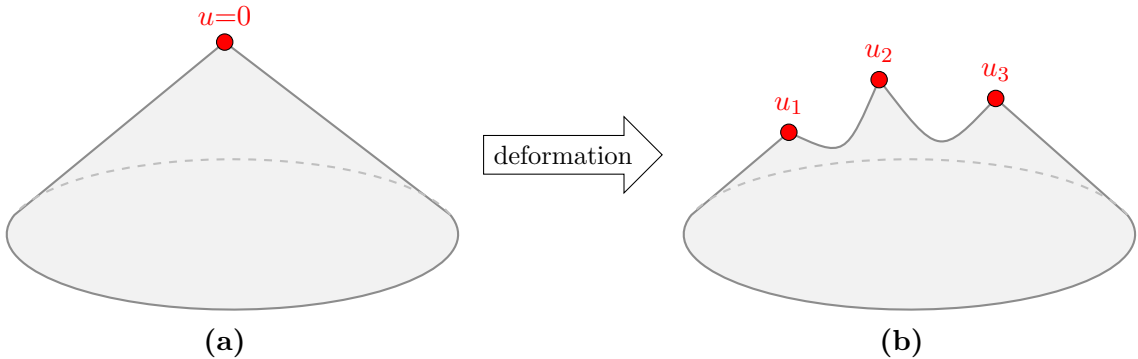
\begin{figure}[ht]
\centering
\begin{tikzpicture}
\clip (0,-.7) rectangle (15.5,4);
\node at (3,-0.5) {\bf (a)};
\filldraw[thick,draw=black!45,fill=black!05] (3,3.5) -- (6,1.2) arc (375:165:3cm and 1cm) -- cycle;
\node[R] (br) at (3,3.5) {};
\node[red] at (3,3.85) {$u{=}0$};
\draw[thick,dashed,black!25] (6,1.2) arc (15:165:3cm and 1cm);
\node[single arrow, draw, black] at (7.5,2) {{\small deformation}};
\node at (12,-0.5) {\bf (b)};
\filldraw[thick,draw=black!45,fill=black!05] (10.5,2.3) .. controls (11.25,1.95) .. (11.7,3.0) .. controls (12.5,1.95) .. (13.25,2.75) -- (15,1.2) arc (375:165:3cm and 1cm) -- cycle;
\node[R] (br1) at (10.5,2.3) {};
\node[red] at (10.5,2.65) {$u_1$};
\node[R] (br2) at (11.7,3) {};
\node[red] at (11.7,3.35) {$u_2$};
\node[R] (br3) at (13.25,2.75) {};
\node[red] at (13.25,3.10) {$u_3$};
\draw[thick,dashed,black!25] (15,1.2) arc (15:165:3cm and 1cm);
\end{tikzpicture}
\caption{(a) A planar, scale-invariant CB in the rank 1 case. The conformal vacuum is the tip of the cone, at $u=0$, while at points with $u\neq0$ conformality is spontaneously broken.  (b) A typical ``near" deformation of the scale-invariant geometry, in which the conical singularity splits into several conical or cusp-like singularities, $u=u_i$, without affecting the asymptotically far geometry.}
\label{fig:cone}
\end{figure}

The constraints from $\cN=2$ supersymmetry and electric-magnetic duality further restrict the geometry of the CB to be special K\"ahler (defined in more detail in appendix \ref{app:details}).  We choose a complex coordinate $u$ on the CB with $u=0$ at the tip of the cone; see figure \ref{fig:cone}a.  The special K\"ahler condition, the above planarity assumption, together with unitarity of the underlying CFT restrict the deficit angle $\d$ to a finite set of values, determined by Kodaira long ago \cite{KodairaI,KodairaII}, and reported here in table \ref{Table:Kodaira}.  

The seven conical geometries are denoted in the table by Kodaira's names for them: $II^*$, $III^*$, $IV^*$, $I_0^*$, $IV$, $III$, and $II$.  These seven geometries have been realized as CBs of physical $\cN=2$ SCFTs \cite{Seiberg:1994aj,Argyres:1995xn,Ganor:1996xd,Seiberg:1996bd,Minahan:1996fg,Minahan:1996cj}.  In the physics literature these geometries are often labeled  by the flavor symmetry of a corresponding field theory as $E_8$, $E_7$, $E_6$, $D_4$, $A_2$ (or $H_2$), $A_1$ (or $H_1$), and $A_0$ (or $H_0$), respectively.   But this naming scheme is confusing since many of these singularities are known to correspond to more than one SCFT with different flavor symmetries \cite{Seiberg:1994aj,Argyres:2007tq}, so we will stick to Kodaira's original names.

More importantly, the existence of inequivalent SCFTs with the same CB geometry calls for a more refined geometric analysis which can distinguish among the multiple theories.  The goal of this paper is to present this analysis.

The two infinite series at the bottom of table \ref{Table:Kodaira}, denoted by $I_n$ and $I^*_n$ in the Kodaira classification, have cusp-like geometries and are not scale invariant.  These can be interpreted as the CBs of $\U(1)$ and $\SU(2)$ IR-free gauge theories, respectively.  They play an important role in our analysis and will be discussed extensively below.

\paragraph{Physical deformations.}

An example of two different SCFTs with the same CB was already given in the original paper on the subject \cite{Seiberg:1994aj} where it was shown that an $\SU(2)$ gauge theory with either $N_f=4$ massless fundamental hypermultiplets or with one massless adjoint hypermultiplet (the $\cN=4$ theory) have the same CB geometry of type $I_0^*$.  These theories have different spectra of conformal primary fields, different global (flavor) symmetries, and different $\cN=2$ mass deformations.  Likewise, other scale-invariant CB geometries can correspond to multiple distinct SCFTs even though they have no direct lagrangian description.

Our strategy is to identify the possible $\cN=2$ supersymmetric deformations of a SCFT from their CB geometry, thereby probing some part of the spectrum of local operators in the theory.  This information helps distinguish between SCFTs.  Deformations (other than by exactly marginal operators) explicitly break the conformal invariance of the theory, and thus the scale invariance of the CB geometry.  As we will argue in detail in this paper, the typical deformation of a CB geometry by relevant opreators splits singularities, as illustrated in figure \ref{fig:cone}b.  A simple  example of this is turning on a mass, $m$, for a charged hypermultiplet in a $\U(1)$ gauge theory.  This mass term makes the hypermultiplet massive at the origin of the CB, thus modifying the singularity at $u=0$, and also causes the hypermultiplet to become massless at a point on the CB with $u\sim m$, thus creating a new singularity there.  The upshot of our analysis of section \ref{sec:CBdefs} is that this splitting of singularities is a general effect of turning on relevant operators in non-lagrangian theories. 

For a given Kodaira singularity we define the \emph{deformation pattern} of a deformation of its CB geometry to be the list of the Kodaira types of the singularities resulting from the splitting obtained by turning on a deformation.  There are many deformation patterns for which no special K\"ahler CB geometry exists.  Although we do not have a uniform proof, it turns out that in every case where a CB geometry exists for a given deformation pattern, it is unique.   We will therefore denote deformed CB geometries just by the topological data of their deformation patterns.  For example,
\begin{align}\label{pair}
IV^*\quad\to\quad\{I_1,I^*_1\}
\end{align}
is the deformation which splits an initial type $IV^*$ Kodaira singularity into two singularities of types $\{I_1,I^*_1\}$. 

Note that the existence of a deformed CB geometry is a necessary condition for the existence of a corresponding SCFT, but is not sufficient:  there may not exist a unitary SCFT which, upon deformation by its relevant local operators, gives rise to that CB.  Nor can we exclude the possibility that there could exist distinct SCFTs with the same deformed CB; for instance, they might differ only in their spectrum of irrelevant operators.  It is reasonable to hope that in certain cases conformal bootstrap techniques could be used to rule out the existence of potential SCFTs, or alternatively S-duality and string techniques could be used to give evidence for their existence. 

In light of this discussion, we can reformulate the question of classifying rank 1 $\cN=2$ SCFTs into classifying all the physically allowed deformation patterns, as in \eqref{pair}.  The problem we face is that the condition of low energy $\cN=2$ supersymmetry by itself puts weak constraints on possible deformations:  even for 1-dimensional Coulomb branches there seem to be too many possibilities.  This raises the main question which we will address in this paper:
\begin{quote}
What are the further physical constraints on the CB geometry that come from assuming the underlying theory is an $\cN=2$ SCFT, or a local $\cN=2$ deformation of one?
\end{quote}
We are able to formulate a coherent answer to this question by adopting the following 
\begin{quote}
{\bf Safely irrelevant conjecture:}\ \emph{$\cN=2$ conformal theories do not have dangerously irrelevant operators.}
\end{quote}
Motivating and providing evidence for this conjecture is one of the main subjects of this paper.  It gets support mainly from the algebraic structure of complex deformations of special K\"ahler geometries.  

This conjecture is fairly strong, as it severely limits the occurence of accidental flavor symmetries in $\cN=2$ theories.  As we will show, it implies that the only singularities on the Coulomb branch at generic values of the relevant deformation parameters (e.g., masses) correspond to ``undeformable" IR fixed-point theories which admit only a very restricted set of deformations by relevant $\cN=2$ operators.  From the representation theory of the $\cN=2$ superconformal algebra we can extract constraints on possible local $\cN=2$ deformations of unitary SCFTs which limit the list of such possible undeformable fixed-point theories to certain special IR-free gauge theories plus a short list of possible (hypothetical) ``frozen" SCFTs.  This gives a list of the possible kinds of singularities that can appear on the CB of an $\cN=2$ SCFT after turning on a generic relevant deformation and thus of deformation patterns of the kind appearing in \eqref{pair}.  We will call deformations of CB singularities which satisfy this conjecture ``safe deformations".   

We are forced to make two physical assumptions limiting the kinds of ``frozen" SCFTs which can appear at rank 1.  These assumptions are:
\begin{quote}
{\bf No rank 0 theories:}\ interacting $\cN=2$ SCFTs with no CB do not occur, and\\[2mm]
{\bf No discretely gauged flavor:}\ $\cN=2$ supersymmetric gaugings of discrete nonabelian flavor symmetries do not occur.\footnote{However, supersymmetric gaugings of certain discrete symmetries which include flavor \emph{outer} automorphisms are allowed, and are discussed in \cite{Argyres:2016yzz}.}
\end{quote}
These assumptions and the evidence for them are discussed in section \ref{froSing}.  With these assumptions, safe deformations of CB geometries are tightly constrained by the Dirac quantization condition in the low energy theory, and by consistency of the web of RG flows among the various fixed points.

\paragraph{Results of the analysis.}

In a companion paper \cite{Argyres:2015gha} we use these constraints to find all safe deformations of rank-1, planar, scale-invariant CBs.  The resulting 28 geometries are listed in table \ref{table:theories}.  Eleven of these have been previously constructed \cite{Seiberg:1994aj,Argyres:1995xn,Minahan:1996fg,Minahan:1996cj} or predicted by string theory constructions \cite{Ganor:1996xd,Seiberg:1996bd} or by S-duality arguments \cite{Argyres:2007cn,Argyres:2007tq}.  Of the seventeen ``new'' CB geometries that we construct --- shown as the shaded rows in table \ref{table:theories} --- fourteen have been identified with rank-1 SCFTs constructed in the last two years by the F-theory S-fold construction of $\cN=3$ theories \cite{Garcia-Etxebarria:2015wns}, by a twisted class $\cS$ construction \cite{Chacaltana:2016shw}, by RG flows between these theories \cite{Argyres:2016xua}, and by discrete gaugings of other rank-1 theories \cite{Argyres:2016yzz}.

\begin{table}
\small
\centering
$\begin{array}{|c|c|c|c|c|c|c|}
\hline
\multicolumn{7}{|c|}{\qquad \text{\bf Deformations of planar, rank 1, scale-invariant CBs satisfying}\qquad \,} \\
\multicolumn{7}{|c|}{\text{\bf low energy supersymmetry and Dirac quantization constraints}} \\
\hline\hline
\ \text{\#}\ \,  &
\quad \text{SCFT}\quad\,  &
\qquad \text{deformation}\qquad\, &
\text{max flavor} &
\text{curve} & \text{SCFT existence} & \text{class}\\
& \text{singularity} & \text{pattern} & \text{symmetry $F$}
& \text{ref.} & \text{ref.} & \cS ?\\
\hline
1. & &\{{I_1}^{10}\} & E_8  
& \text{\cite{Minahan:1996cj}}  & \text{\cite{Ganor:1996xd,Seiberg:1996bd,Argyres:2007tq}} & \cmark \\
2. & &\{{I_1}^6,I_4\} & \Sp(10) 
& \text{\cite{Argyres:2015gha}} & \text{\cite{Argyres:2007tq}} & \cmark \\
\rcr \ccr
3. & \ccw & \{{I_1}^2,{I_4}^2\} &\Sp(4) 
& \ccw \text{\cite{Argyres:2015gha}} &  & \ccw \\
\rcy \ccw
4. & \ccw &\{{I_1}^4,I^*_0\} &F_4 
& \ccw \text{\cite{Argyres:2015gha}} & \text{\cite{Argyres:2016yzz}} & \ccw \\
\rcy \ccw
5. & \ccw &\{{I_1}^3,I^*_1\} &\ccr \Sp(6)\ \text{or}\ \SO(7) 
& \ccw \text{\cite{Argyres:2015gha}} & \text{\cite{Chacaltana:2016shw}}_{F{=}\SU(4)} & \ccw \cmark \\
\rcy \ccr
6. & \ccw &\{I_3,I^*_{1\ Q{=}\sqrt3}\} &\SU(2)
& \ccw \text{\cite{Argyres:2015gha}} & \ccr & \ccw \\
\rcy \ccw
7. & \ccw &\{{I_1}^2,I^*_2\} &\Sp(4) 
& \ccw \text{\cite{Argyres:2015gha}} & \text{\cite{Argyres:2016yzz}} & \ccw \\
\rcy \ccr
8. & \ccw &\{I_1,I^*_3\} &\SU(2) 
& \ccw \text{\cite{Argyres:2015gha}} & \ccr & \ccw \\
\rcy \ccw
9. & \ccw &\{I_2,IV^*_{Q{=}\sqrt2}\} &\SU(2)
& \ccw \text{\cite{Argyres:2015gha}} & \text{\cite{Garcia-Etxebarria:2015wns}$_{F{=}\U(1)}$ and \cite{Argyres:2016yzz}}_{F{=}\SU(2)} & \ccw \\
\rcy \ccw
10. & \ccw &\{{I_1}^{2},IV^*\} &G_2 
& \ccw \text{\cite{Argyres:2015gha}} & \text{\cite{Argyres:2016xua}$_{F{=}\SU(3)}$ and \cite{Argyres:2016yzz}}_{F{=}G_2} & \ccw \\ 
\rcy \ccw 
11. &\ccw \multirow{-11}{*}{$II^*$}   
&\{I_1,III^*\} &\SU(2) 
& \ccw \text{\cite{Argyres:2015gha}} & \text{\cite{Argyres:2016yzz}} & \ccw \\
\hline 
12. & \ccw &\{{I_1}^9\} &E_7   
& \text{\cite{Minahan:1996cj}}  & \text{\cite{Ganor:1996xd,Seiberg:1996bd,Argyres:2007cn}} & \ccw \cmark\\ 
13. & \ccw &\{{I_1}^5,I_4\} &\ \Sp(6)\oplus\Sp(2)\ \,
& \ccw \text{\cite{Argyres:2015gha}} & \text{\cite{Argyres:2007tq}} & \ccw \cmark\\
\rcy \ccw
14. & \ccw &\{{I_1}^3,I^*_0\} &\SO(7) 
& \ccw \text{\cite{Argyres:2015gha}} & \text{\cite{Argyres:2016yzz}} & \ccw \\
\rcy \ccw
15. & \ccw &\{{I_1}^2,I^*_1\} &\ccr \SU(2)\oplus\SU(2) 
& \ccw \text{\cite{Argyres:2015gha}} & \text{\cite{Argyres:2016xua}}_{F{=}\SU(2)\oplus\U(1)} & \ccw \\
\rcy \ccw
16. & \ccw &\{I_2,I^*_{1\ Q{=}\sqrt2}\} &\SU(2) 
& \ccw \text{\cite{Argyres:2015gha}} & \text{\cite{Garcia-Etxebarria:2015wns}$_{F{=}\U(1)}$ and \cite{Argyres:2016yzz}}_{F{=}\SU(2)} & \ccw \\
\rcy \ccw
17. & \ccw &\{I_1,I^*_2\} &\SU(2) 
& \ccw \text{\cite{Argyres:2015gha}} & \text{\cite{Garcia-Etxebarria:2015wns,Argyres:2016yzz}}_{F{=}\U(1)} & \ccw \\
\rcy \ccw 
18. & \ccw \multirow{-7}{*}{$III^*$}  
&\{I_1,IV^*\} &\SU(2) 
& \ccw \text{\cite{Argyres:2015gha}} & \text{\cite{Garcia-Etxebarria:2015wns}$_{F{=}\U(1)}$ and \cite{Argyres:2016yzz}}_{F{=}\SU(2)} & \ccw \\
\hline
19. & &\{{I_1}^8\} &E_6  
& \text{\cite{Minahan:1996fg}}  & \text{\cite{Ganor:1996xd,Seiberg:1996bd,Argyres:2007cn}} & \cmark\\
20. & &\{{I_1}^4,I_4\} &\Sp(4)\oplus\U(1) 
& \ccw \text{\cite{Argyres:2015gha}} & \text{\cite{Argyres:2007tq}} & \cmark\\
\rcy \ccw
21. & \ccw &\{{I_1}^2,I^*_0\} &\SU(3) 
& \ccw \text{\cite{Argyres:2015gha}} & \text{\cite{Argyres:2016yzz}} & \ccw \\
\rcy \ccw 
22. & \ccw \multirow{-5}{*}{$IV^*$} 
&\{I_1,I^*_1\} &\U(1) 
& \ccw \text{\cite{Argyres:2015gha}} & \text{\cite{Garcia-Etxebarria:2015wns}} & \ccw \\
\hline
23. & &\{{I_1}^6\} &\SO(8)
& \text{\cite{Seiberg:1994aj}} & \SU(2)\ N_f=4 & \cmark \\
24. & &\{{I_2}^3\} &\Sp(2) 
& \text{\cite{Seiberg:1994aj}} & \SU(2)\ \cN=2^* & \\
\rcy \ccw
25. & \ccw \multirow{-3}{*}{$I^*_0$} 
&\{{I_1}^2,I_4\} &\Sp(2) 
& \ccw \text{\cite{Argyres:2015gha}} & \SU(2)\ \cN=2^* & \ccw \cmark\\
\hline
26. & IV 
&\{{I_1}^4\} &\SU(3)
& \ccw \text{\cite{Argyres:2015gha}} & \text{\cite{Argyres:1995xn}} & \cmark \\
\hline
27. & III
&\{{I_1}^3\} &\SU(2)
& \ccw \text{\cite{Argyres:2015gha}} & \text{\cite{Argyres:1995xn}} & \cmark \\
\hline
28. & II 
&\{{I_1}^2\} &-
& \ccw \text{\cite{Argyres:2015gha}} & \text{\cite{Argyres:1995xn}} & \cmark \\
\hline
\end{array}$
\caption{\label{table:theories} 
The 28 families of deformed planar rank-1 CB geometries consistent with the low energy Dirac quantization condition.  Column 2 lists the Kodaira type of the scale invariant CB geometry, column 3 the resulting singularity types under a generic relevant deformation, and column 4 the maximal flavor symmetry of the SCFT.  Note that by $\Sp(2r)$ we mean the algebra of rank $r$.  The meaning of the remaining columns and the row shadings are explained in the text.  The $Q=\sqrt q$ subscripts appearing on some of the singularities in the deformation column record the unit of charge quantization in these theories; the $I^*_n$, $III^*$, or $IV^*$ singularities appearing without such subscripts all have $Q=1$, while for $I_n$ singularities they are $Q=\sqrt n$.  Charge normalizations and charge quantization conditions are defined in appendix \ref{app:details} and explained in section \ref{froSing}. The values for the central charges for the entries in the table are known and can be found in table 1 of \cite{Argyres:2015ccharges}.}
\end{table}

In particular, table \ref{table:theories} lists the 28 possible inequivalent deformation patterns of scale-invariant rank-1 planar CBs, along with the maximal flavor symmetry the corresponding SCFT could posses, as determined from the explicit form of the deformed CB geometry.  The earliest papers which record the explicit CB geometries (in the form of Seiberg Witten curves and one-forms) are shown in the ``curve ref.'' column.  Note that in the case of many theories which have class $\cS$ realizations (as noted in the last column of the table), the CB geometries, at least for a subset of the mass deformations, could have in principle been obtained earlier, but were just not written explicitly.  

The ``SCFT existence'' column lists places where arguments for the existence of SCFTs corresponding to particular geometries were given.  Subscripts on the references in this column denote theories whose flavor symmetry is smaller than the maximal flavor symmetry allowed by the CB geometry.   Note that there are three entries --- numbers 3, 6, and 8, shaded red in the table --- for which there is no evidence of a corresponding SCFT.   Geometry number 3, corresponding to the generic $II^*\to\{{I_1}^2,{I_4}^2\}$ deformation, is shown in \cite{Argyres:2015gha} to be inconsistent under all RG flows (not just the generic one) with the maximal flavor symmetry assignment $\Sp(4)$.  In fact, the same argument also shows it to be inconsistent with any of its other allowed flavor symmetry assignments (namely, $\U(1)\oplus\U(1)$ or $\SU(2)\oplus\SU(2)$ as derived in \cite{Argyres:2016xua}), and so we conclude that there can be no SCFT corresponding to this geometry.  But we have no such argument in geometries number 6 and 8.

There are no obvious patterns to be discerned from the results summarized in table \ref{table:theories}.  Some CB geomtries seem to correspond to no SCFT, while others, such as geometries 9, 10, 16, and 18, correspond to more than one.  In many theories the maximal flavor symmetry allowed by the geometry is realized, but in some cases it is not.  In two of those cases, numbers 5 and 15, we show in \cite{Argyres:2015gha} that the maximal flavor symmetry assignment is, in fact, inconsistent when all RG flows for that geometry are considered, but in other cases there is no such inconsistency argument.  Also, there is one case where two distinct CB geometries seem to be associated to the same SCFT:  these are geometries numbered 24 and 25 in the table which, since they have an exactly marginal coupling and $\Sp(2)$ flavor symmetry are both identified as the lagrangian $\cN=2^*$ $\SU(2)$ gauge theory.  In \cite{Argyres:2015gha} it is shown that these two CB geometries give the same low energy predictions on the CB locally in the space of the marginal coupling parameter, but these spaces have distinct global structures.  In particular the $I_0^*\to\{{I_1}^2,I_4\}$ deformation has S-duality group $\G_0(2)\subset\SL(2,\Z)$ as predicted from GNO duality \cite{Goddard:1976qe}, while the $I_0^*\to\{{I_2}^3\}$ deformation has the full $\SL(2,\Z)$ as S-duality group.  The possible interpretation of this latter novel theory is further discussed in \cite{Argyres:2016yzz}.

One can also ask about the nature of the IR fixed points that these theories flow to in the IR on the CB.   By definition and the safely irrelevant conjecture, all these IR fixed points are undeformable or frozen SCFTs or IR-free QFTs.  In fact, in almost all cases where SCFTs corresponding to rank-1 CB goemetries are realized, the fixed points are all IR-free theories, such as the undeformable $I_n$ or $I^*_n$ theories discussed at length in section \ref{froSing} below, or discretely gauged versions of them described in \cite{Argyres:2016yzz}.  The exception are CB geometries number 10 and 18 which both flow to a new frozen $IV^*$ SCFT.  The conformal central charges of this new SCFT are deduced in \cite{Argyres:2016xua}.

It is interesting to note that the two consistent geometries --- numbers 6 and 8 in table \ref{table:theories} --- for which there is no evidence of an associated SCFT are ones which flow to $I^*_1$ ($Q=\sqrt3$) and $I^*_3$ ($Q=1$) frozen singularities.  These have no interpretation as IR free frozen theories or as discretely gauged versions of such, as shown in the analysis of section \ref{froSing} below and \cite{Argyres:2016yzz}.  (Section \ref{froSing} defines and explains the meaning of the unit of EM charge quantization $Q=\sqrt3$ or $Q=1$ appearing above.)  Since the low energy theory on the CB is free at these two singularities, it presumably follows that the only physical interpretation of these frozen IR fixed points are as weakly-gauged versions of special rank-0 SCFTs; again, this conclusion is discussed in more detail in section \ref{froSing}.  Thus, to the extent that the assumption that rank-0 SCFTs do not exist is correct, we expect that the number 6 and 8 CB geometries in table \ref{table:theories} cannot arise from any SCFT. 

Finally, it is important to re-emphasize that the classification shown in table \ref{table:theories} is subject to two artificial assumptions, made to simplify the analysis:
\begin{itemize}
\item[(1)] the CB has ``rank 1", i.e., it is 1 complex dimensional, and
\item[(2)] the CB is ``planar", i.e., the CB is isomorphic to $\C$ as a complex space.
\end{itemize}
The restriction to rank 1 is only for simplicity.  If the restriction to rank 1 Coulomb branches is lifted, some of the techniques we use in \cite{Argyres:2015gha} no longer work.  Furthermore our starting point in the rank 1 case --- the classification of the scale-invariant special K\"ahler geometries --- does not yet exist for ranks 2 and higher.  The planarity assumption amounts to the assumption that the CB chiral ring of the corresponding SCFT is freely generated.  If the planarity assumption is dropped, there are additional examples \cite{Argyres:2017tmj} of Coulomb branch geometries satisfying all known physical consistency conditions, though no corresponding SCFTs are known to exist.

\paragraph{SCFT data from CB geometries.}

A question which is closely related to the determination of physical consistency conditions on the CB geometry is: What part of the $\cN=2$ SCFT data (primary operator spectrum and OPE coefficients) can be deduced from knowing the CB geometry? The obvious part of the answer to this question is that the spectrum of dimensions of the Coulomb branch operators can be read off from the scale-invariant CB geometry, and the web of RG flows among different fixed point theories can be read off from their deformations.   

Less obviously, the flavor symmetry and the conformal and flavor central charges of the SCFT can be tightly constrained, though not completely determined, from the deformed CB geometry.  As we will review in this paper, the dependence of the CB geometry on the mass deformation parameters allows one to deduce the Weyl group of the maximal flavor symmetry.  This, however, fails to uniquely determine the flavor symmetry for two reasons.  First, the Weyl groups of $\SO(2r+1)$ and $\Sp(2r)$ factors coincide, and so cannot be differentiated by knowing the Weyl group.  Second, the flavor symmetry could be a smaller algebra of the same rank whose Weyl group is a normal subgroup of the maximal Weyl group (and the quotient group is a discrete global symmetry of the theory).  This was explained in \cite{Argyres:2016xua} where all allowed sub-maximal flavor groups consistent with a given maximal Weyl group were determined.  The 
central charges of gauge $\U(1)$-neutral BPS sectors span a flavor weight lattice which includes the root lattice, and there is the expectation that they span precisely the flavor root lattice; see the discussion in appendix \ref{app:details}.  The central charges are encoded in the CB geometry through the residues of the SW one-form.  Knowledge of the flavor root lattice and the Weyl group action on it are sufficient to determine the flavor algebra, thus eliminating the first ambiguity mentioned above.  The maximal flavor algebras listed in table \ref{table:theories} are determined in this way in \cite{Argyres:2015gha}.  Note, however, that there is still one case, geometry number 3, where the flavor lattice cannot be uniquely determined from the SW one-form, and an $\Sp(6)$/$\SO(7)$ ambiguity remains unresolved.

Following the method of Shapere and Tachikawa \cite{Shapere:2008zf}, we show in \cite{Argyres:2015ccharges} that the deformed CB geometry together with information about the action of the flavor symmetry on certain maximal mixed Coulomb-Higgs branches can be used in many cases to determine the conformal and flavor algebra central charges in terms of those of the IR fixed points on the CB.  This was used in \cite{Argyres:2016xua}, where the Higgs branch data was deduced from S-dualities or RG flows, to determine most of the central charges of the theories in table \ref{table:theories}.  But it is pointed out in \cite{Argyres:2016yzz} that the Shapere-Tachikawa calculus does not work for those theories obtained by discretely gauging a symmetry which acts non-trivially on the CB.

It is worth mentioning how our approach relates to the conformal bootstrap program for $\cN=2$ SCFTs \cite{Beem:2013sza,Beem:2014rza,Beem:2014zpa}.  The bootstrap uses analytic and numerical techniques to constrain the local operator algebras of CFTs and is largely complimentary to the one described here, where we assume in addition to superconformal invariance the existence of a Coulomb branch of vacua.  The geometry of the Coulomb branch gives us little direct information on the algebra of operators at the conformal point since the conformal invariance is spontaneously broken at any point on the Coulomb branch away from the conformal singularity.  Also, the amount of information about a SCFT encoded in its Coulomb branch geometry is far smaller than that of its full local operator algebra.  Nevertheless, some properties of the SCFT are easily accessible and tightly constrained by the Coulomb branch geometry, but seem difficult to constrain using bootstrap techniques.  For instance the scaling dimension, $\D(u)$, of the Coulomb branch operator is a free parameter in \cite{Beem:2014zpa}, while, by contrast, a geometrical analysis of rank-1 Coulomb branches constrains $\D(u)$ to be one of 7 possible values.  It is reasonable to hope that $\cN=2$ bootstrap results can be considerably strengthened by utilizing some of the SCFT properties obtained from studying the geometry of the Coulomb branch along the lines discussed here. Recent results outlined in \cite{Hellerman:2017sur} could shed light into developing a direct connection between the results derived following our studies of the Coulomb branch and the bootstrap ones.

\paragraph{Outline.}

The rest of the paper is organized as follows: in the next section we briefly review the geometric structure of $\cN=2$ SCFT Coulomb branches and review the classification of rank 1 planar scale invariant singularities.  In section \ref{sec:CBdefs} we describe the possible deformations of the Coulomb branch and clarify how these are related to deformations of the SCFT by relevant, marginal, and irrelevant operators.  

Section \ref{sec:sings} is the core of the paper, and describes the constraints of $\cN=2$ supersymmetry on deformations of Coulomb branches by relevant operators.  We present a simple geometrical picture of these deformations, and use it to argue for the safely irrelevant conjecture mentioned above.  We explain the implications of this conjecture for the (non)existence in $\cN=2$ theories of accidental flavor symmetries in the IR.  Finally, we discuss the way in which the flavor symmetry manifests itself in the dependence of the CB geometry on the mass deformation parameters, and point out that there can exist families of special K\"ahler geometries whose mass dependence is not compatible with the flavor symmetry algebra being a reductive Lie algebra.  Field theories with CBs with these geometries would then violate the Coleman-Mandula theorem.  Rejecting such deformed geometries as physical CBs is compatible with and lends further support to the safely irrelevant conjecture. 

In section \ref{sec:udefsing} we discuss the main practical implications of the safely irrelevant conjecture on the classification of rank 1 planar CB geometries.  As mentioned above, Dirac quantization of electric and magnetic charges in the low energy theory on the CB plays an important role.  There are further consistency tests of the simple singularities conjecture following from the pattern of RG flows linking the SCFTs in table \ref{table:theories} and IR-free theories.  The nature of these consistency tests and the intricate way in which they are satisfied will be explained heuristically in this section.   A detailed exposition requires the construction of the curves, and is given in \cite{Argyres:2015gha} for the case of maximal flavor symmetry, and for non-maximal flavor symmetries in \cite{Argyres:2016xua}.   We end in section \ref{sec:conclusion} with a brief discussion of the relation of our results to other results in the literature, and of various open questions raised here.  

While more technical, the appendices also contain original results.  Appendix \ref{app:details} discusses rank 1 special K\"ahler geometries.  Although known to experts, our discussion of charge normalizations has not appeared before in the literature.  In appendices \ref{app:repre} and \ref{app:defo} we review the representations of unitary $\cN=2$ SCFTs, and derive the forms of all possible local $\cN=2$ supersymmetric deformations of SCFTs.\footnote{A more systematic and comprehensive discussion of superconformal representations and supersymmetric deformations has since appeared in \cite{Cordova:2016xhm,Cordova:2016emh}.}


\section{Scale invariant Coulomb branch geometries}\label{outline}

Lagrangian $\cN=2$ gauge theories always possess a nontrivial moduli space of supersymmetric vacua where some of the complex scalar components, $\varphi$, of the vector multiplet have non-zero vevs.  The part of the moduli space where only these fields get vevs is called the Coulomb branch (CB).  There may also be parts of the moduli space where complex scalars in hypermultiplets (matter multiplets) develop vevs, known as Higgs or mixed branches.  We focus here primarily on CB geometries.

In lagrangian theories the CB is $r$ (complex)-dimensional where $r$ is the rank of the gauge group.  The physics on the CB is the Higgs mechanism where $\vev{\varphi}$ Higgses the gauge group to ${\rm U}(1)^r$.  For non-lagrangian theories we will take the dimensionality of their CBs as the definition of the ``rank" of such theories.  We indicate points on the CB by some complex coordinates $u$ with, locally,  $u\in\C^r$.   We restrict our discussion to rank 1 theories, so henceforth $u\in\C$.

The assumption of unbroken $\cN=2$ supersymmetry constrains the low energy effective action on the CB, with the result that the CB geometry is (rigid) special K\"ahler (SK).  In the rank-1 case this is a rather weak condition locally, and just says that the CB must be a complex manifold with metric (in a special coordinate system) $ds^2 = \Im\t(u)\, du\, d\bar u$ where $\t(u)$ is a holomorphic function with positive imaginary part, but may be identified up to EM duality transformations, $\t \to (a\t+b)/(c\t+d)$ for $\left( \begin{smallmatrix} a&b\\ c&d \end{smallmatrix} \right) \in \SL(2,\Z)$.  (A more detailed discussion of the definition of rank-1 SK geometry is given in appendix \ref{app:details}.)   On any contractible set $\t$ is single-valued since $\SL(2,\Z)$ is discrete.  Furthermore, since the scalar curvature, $R=-|\del_u\t|^2$, is non-positive and $\del_u\t$ is a holomorphic function (i.e., single-valued), any complete SK geometries are flat.  (This generalizes to rank $>1$ SK geometries as well \cite{Freed:1997dp}.)

Interesting SK geometries have singularities.  The singularities are interpreted as vacua for which there are additional massless states in the theory charged under the low energy $\U(1)$ gauge symmetry \cite{Seiberg:1994rs,Seiberg:1994aj}.  In what follows we will refer to the singular SK geometries that occur on Coulomb branches of $\cN=2$ field theories as ``CB geometries".  The basic question we are addressing in this paper is what are the conditions that incomplete SK geometries must satisfy at their singularities for them to be CB geometries.

The simplest CB geometries correspond to those of $\cN=2$ SCFTs.  The scalars $u$ getting vevs on the CB have
\begin{align}\label{Drbound}
\D(u)=r(u)\ge1
\end{align}
by unitarity bounds in the SCFT.  Here $\D$ is the mass scaling dimension and $r$ is the $\U(1)_r$ charge.  (We set our notation and normalizations for the $\cN=2$ superconformal algebra and review its representation theory in appendix \ref{app:repre}.)   The bound \eqref{Drbound} already at rank 1 imposes a restriction on the allowed singular SK geometries, though it is implied by the planarity condition; see appendix \ref{app:details}.

The assumption of the existence of complex coordinates on the CB of definite dimensions satisfying \eqref{Drbound} endows the CB with a faithful $\C^*$ action, $u \mapsto \l^{\D(u)} u$ for $\l\in\C^*$, making it a K\"ahler cone, with the singular point $u=0$ at the tip of the cone corresponding to the superconformal vacuum.  Points with $u\neq0$ introduce a scale $\sim |u|^{1/\D(u)}$ and thus spontaneously break scale invariance.  For rank 1, this implies that the tip is the only singularity, and the CB geometry is that of a flat cone with $\t(u)=\t_0$ a constant; see figure \ref{fig:cone}a.  The geometry of the cone is characterized by a deficit angle $\d = 2\pi(1-\D(u)^{-1})$.

Furthermore, the EM duality monodromy, $M_0\in \SL(2,\Z)$, around the tip of the cone is related to the deficit angle by SK geometry, as described in appendix \ref{app:details}.  The discreteness of $\SL(2,\Z)$ and the unitarity bound \eqref{Drbound} restricts scale-invariant CB cones to a small set of allowed deficit angles, classified long ago by Kodaira \cite{KodairaI, KodairaII} in a different context, and shown in table \ref{Table:Kodaira}.

\begin{table}
\centering
$\begin{array}{|c|l|c|c|c|c|c|}
\hline
\multicolumn{7}{|c|}{\text{\bf Possible scaling behaviors near singularities of a rank 1 CB}}\\
\hline\hline
\text{Name} & \multicolumn{1}{c|}{\text{planar SW curve}} & \ \text{ord}_0(D_{x})\ \ &\ \D(u)\ \ & M_{0} & \text{deficit angle} 
& \t_0 \\
\hline
II^*   &\parbox[b][0.45cm]{4cm}{$\ y^2=x^3+u^5$}             
&10 &6 &ST &5\pi/3 & \ e^{i\pi/3}\ \\
III^*  &\ y^2=x^3+u^3x &9 &4 &S &3\pi/2 & i\\
IV^*  &\ y^2=x^3+u^4 &8 &3 &-(ST)^{-1} &4\pi/3 & e^{i\pi/3}\\
I_0^* &\ y^2=\prod_{i=1}^3\left(x-e_i(\t)\, u\right)
&6 &2 &-I &\pi & \t\\
IV &\ y^2=x^3+u^2 &4 &3/2 &-ST &2\pi/3 & e^{i\pi/3}\\
III &\ y^2=x^3+u x &3 &4/3 &S^{-1} &\pi/2 & i\\
II  &\ y^2=x^3+u &2 &6/5 &(ST)^{-1} &\pi/3 &e^{i\pi/3}\\
\hline
\hline
I^*_n\ \ (n{>}0) &
\parbox[b][0.45cm]{5cm}{
$\ y^2=x^3+ux^2+\L^{-2n}u^{n+3}\ \ $}
& n+6 & 2 & {-T^n} & 2\pi\ \text{(cusp)} 
& i\infty\\
I_n\ \ (n{>}0)    &\ y^2=(x-1)(x^2+\L^{-n}u^n)  
& n     & 1 & {T^n} & 2\pi\ \text{(cusp)} 
& i\infty\\[0.5mm]
\hline
\end{array}$
\caption{\label{Table:Kodaira} Scaling forms of rank 1 planar special K\"ahler singularities, labeled by their Kodaira type (column 1), a representative family of elliptic curves with singularity at $u=0$ (column 2), order of vanishing of the discriminant of the curve at $u=0$ (column 3), mass dimension of $u$ (column 4), a representative of the $SL(2,\Z)$ conjugacy class of the monodromy around $u=0$ (column 5), the deficit angle of the associated conical geometry (column 6), and the value of the low energy $\U(1)$ coupling at the singularity (column 7).  The first seven rows are scale invariant.  The last two rows give infinite series of singularities which have a further dimensionful parameter $\L$ so are not scale invariant; they can be interpreted as IR free theories since $\t_0=i\infty$.}
\end{table}

Each of the geometries listed in table \ref{Table:Kodaira} in fact corresponds to the Coulomb branch geometry of at least one $\cN=2$  SCFT.  For instance, the $I^*_0$ singularity, which depends on an adjustable dimensionless parameter $\t$, is the CB of two different lagrangian theories, the conformal $\SU(2)$ gauge theory with complex gauge coupling $\t = (\th/2\pi)+i(4\pi/g^2)$ and with either 4 massless fundamental hypermultiplets or 1 massless adjoint hypermultiplet \cite{Seiberg:1994aj}.\footnote{The $e_i(\t)$ functions appearing in table \ref{Table:Kodaira} are defined in \cite{Seiberg:1994aj}.}  Similarly, SCFTs corresponding to the other singularities have been found either by RG flows from asymptotically free theories \cite{Argyres:1995jj, Argyres:1995xn} or by tuning exactly marginal couplings to special values (i.e., by S-duality) in conformal gauge theories \cite{Argyres:2007cn, Argyres:2007tq}.  Likewise, the $I_n$ and $I^*_n$ singularities with $n>0$ correspond to the CBs of IR-free $\U(1)$ or $\SU(2)$ gauge theories, respectively, whose 1-loop beta function coefficient is $b_0=n$ (in a specific normalization discussed in section \ref{sec:udefsing} below).\footnote{The mass scale $\L$ appearing in the description of the $I_n$ and $I^*_n$ singularities in table \ref{Table:Kodaira} is the strong-coupling scale (Landau pole) of the corresponding IR-free theories.  The singular geometries only describe the scaling in the vicinity of the IR-free point, i.e., are valid only for $|u| \ll \L^{\D(u)}$.}

Yet this classification is not refined enough: each geometry in the Kodaira classification corresponds to multiple conformal field theories.  Thus, we must go beyond scale-invariant CB geometries if we are to learn more about the corresponding SCFTs.  Studying the \emph{inequivalent deformations} of scale-invariant CB geometries which preserve a SK structure on the CB will allow us to learn about the $\cN=2$ deformations of the corresponding SCFTs, and so gain information about the spectrum of operators of the SCFTs.   Thus distinct deformations of the same scale-invariant CB singularity will correspond to distinct SCFTs.

In weakly-coupled examples, where we have a lagrangian description, such inequivalent mass deformations are familiar: they correspond to different choices of gauge representation for the matter fields such that the gauge coupling remains marginal.  The inequivalent mass deformations of the geometries associated to non-lagrangian theories can, heuristically, be thought of in the same way:  they are strongly coupled rank 1 gauge theories all with the same ``gauge group" (corresponding to the singularity) but with different ``matter content" (corresponding to the different mass deformations).

For instance, there are two different conformal $\SU(2)$ gauge theories, one with 4 massless fundamental hypermultiplets, and one with 1 massless adjoint hypermutliplet.  Nevertheless, they are described in the IR by the same $I_0^*$ conical CB geometry.  For the theory with four fundamental hypermultiplets, there are four independent complex mass deformations breaking the $\SO(8)$ flavor symmetry of the SCFT, while in the theory with a single adjoint hypermultiplet there is only a single mass deformation which breaks the $\Sp(2)$ flavor symmetry.  

Recall from \cite{Seiberg:1994aj} that the effect of these mass deformations on the CB geometry is to split the conical $I_0^*$ singularity into a set of (cusp-like) singularities of $I_n$ Kodaira type.  From the lagrangian point of view this is easy to understand.  In terms of $\cN=1$ superfields, where the hypermultiplet is two $\cN=1$ chiral superfields, $\tQ$ and $Q$, and the vector multiplet contains an adjoint $\cN=1$ scalar chiral multiplet, $\Phi$, the superpotential is of the form
\begin{align}\label{N1W}
\cW = \tr(\tQ\Phi Q) + m \tr(\tQ Q) ,
\end{align}
where $m$ is the mass parameter.  This makes it apparent that the mass term for some components of the hypermultiplets can be cancelled by appropriately tuning the vev of $\Phi$ on the CB.  Thus, upon turning on $m$, vacua with massless hypermultiplets charged under the low energy $\U(1)$ gauge symmetry appear at different positions on the CB.  These massless-hypermultiplet vacua are precisely the ones described by the $I_n$ Kodaira singularities.  

This splitting of the conical singularity into lower order ones deforms the geometry of the CB only in the region of the tip of the cone as pictured in figure \ref{fig:cone}b.   We will call deformations which do not change the geometry asymptotically far from the tip of the cone, \emph{near deformations}.   In the next section we will explain why this is the effect on the CB geometry of deforming a SCFT by a relevant operator, while irrelevant operators give \emph{far deformations} which, by contrast, do not affect the geometry asymptotically close to the tip of the cone.  We will not have much to say about far deformations here since they are difficult to tame using our techniques.  Nevertheless, their classification is an interesting question, especially for IR-free theories.


\section{Operator spectrum and Coulomb branch deformations\label{sec:CBdefs}}

In this section we will elucidate the relationship between the deformations of $\cN=2$ superconformal field theories by local operators and deformations of their moduli spaces.  We start with a general discussion of the effects of RG flows on moduli spaces, then use this to analyse the effects of possible $\cN=2$ supersymmetric local deformations of $\cN=2$ SCFTs.

\subsection{RG flows and moduli space\label{sec:RGflows}}

When a SCFT with a moduli space of vacua is deformed, the moduli space can either be (partially) lifted, or can be deformed.  From the point of view of the low energy theory on the CB, $\cN=2$ supersymmetry disallows any scalar potential for the $\U(1)$ vector multiplets.\footnote{We will show in section \ref{sec:localdefs} that $\cN=2$ Fayet-Iliopoulos terms \cite{Antoniadis:1995vb} --- which can lift CBs --- do not occur as local $\cN=2$ deformations of SCFTs.}  This, together with complex analyticity and the existence of asymptotically undeformed regions of the CB implies that the CB cannot be lifted, but only deformed \cite{Seiberg:1994rs}.  Such asymptotically undeformed regions will be argued below to exist even in theories without any weakly coupled limit.  This persistence of the CB is the main motivation for focussing on their structure:  their deformations will reflect some of the structure of the SCFT.   This is in contrast to Higgs (and mixed) branches where the low energy theory has neutral massless hypermultiplets.  $\cN=2$ supersymmetry does allow potential terms for the scalars in hypermultiplets, e.g., the mass terms in \eqref{N1W}. 

We now give a general discussion of the effects that relevant and irrelevant operator deformations can have on moduli space geometries.  The basic picture that emerges is that the deformation of the moduli space under an RG flow provides a kind of ``map" of the RG flow which contains an imprint of the crossover scale(s) of the flow.  Another way of saying this is that for operators which do not lift the moduli space, it is not enough to specify the local operator deforming the UV fixed point to specify the flow, but one must also specify a particular vacuum in the moduli space.  Specifying a vacuum in the moduli space can be thought of as turning on a specific relevant but non-local operator in the UV.  We start by reviewing some basics about topologies of RG flows between fixed points.  The notion of a \emph{dangerously irrelevant} operator will be important in the next section.

\paragraph{Generalities about RG flows.}

Consider a fixed point, $P$, with a relevant operator, $\cR$, of dimension $\D(\cR) \le 4$, and an irrelevant operator, $\cP$, of dimension $\D(\cP) \ge 4$.  When $\cR$ is turned on at $P$ it gives a flow which ends at a fixed point, $R$.  We'll call $\cP$ a ``safely" irrelevant operator at $P$ if, when turned on at the same time as any relevant $\cR$, it leads to a flow ending at the same fixed point, $R$.  $\cP$ is a ``dangerously" irrelevant operator \cite{Amit:1982az} at $P$ if, when turned on at the same time as $\cR$, it leads to a flow ending at a \emph{different} fixed point, $S$.\footnote{In \cite{Gukov:2015qea} S. Gukov uses a different definition of dangerously irrelevant operator as an operator whose dimension crosses from irrelevant to relevant along an RG flow.  It is not clear to us what is the relationship between these two definitions.  We thank S. Gukov for helpful comments on this point, as well as other points in this section.}

\begin{figure}[ht]
\centering 
\includegraphics[width=.95\textwidth]{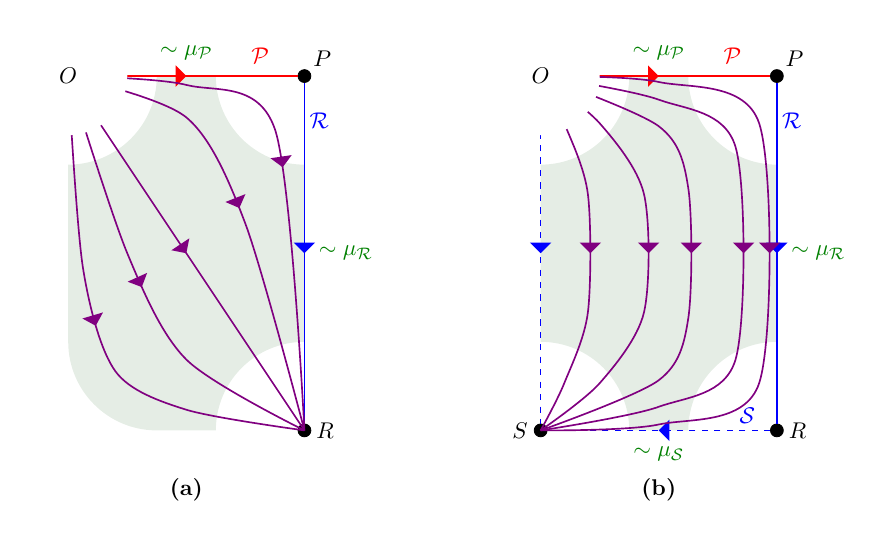}
\caption{{\bf (a)} $\cP$ is ``safely" irrelevant at $P$. {\bf (b)} $\cP$ is dangerously irrelevant at $P$.}
\label{fig:RG1}
\end{figure}

Figure \ref{fig:RG1} illustrates the difference between a safely and a dangerously irrelevant operator.  Here in each case the red flow is the one induced by turning on only the irrelevant operator, $\cP$, at $P$, and the blue flow by turning on only the relevant operator, $\cR$, at $P$.  

To each flow is associated one or more crossover scales describing the order-of-magnitude separation of scales where the effective theory is dominated by one fixed point or another.  The shaded regions in the figures are the crossover regions, and the corresponding crossover scales (at various edges of the crossover regions) are denoted in green.  E.g., to the relevant operator $\cR$ there is a crossover scale $\m_\cR$ such that for energies $E\gg\m_\cR$, we are near the $P$ fixed point, and for $E\ll \m_\cR$ we are near the $R$ fixed point.  In the limit that we send the coupling to $\cR$ to zero, $\m_\cR \to 0$; i.e., as we turn off the $\cR$ deformation, we spend more ``RG time" near $P$.  Of course, if $\cR$ is the only deformation, then this has no observable meaning, since $\m_\cR$ is the only scale in the theory.  Similar comments apply to an irrelevant operator $\cP$ at $P$:  for $E\ll \m_\cP$, the effective theory is near $P$, while in the opposite limit some other fixed point, say ``$O$", governs the flow.  In the limit that we turn off the $\cP$ deformation at $P$, $\m_\cP \to \infty$.

For a safely irrelevant operator, $\cP$ at $P$, when a relevant operator, $\cR$, is also turned on then there will be two corresponding independent crossover scales, $\m_\cP$ and $\m_\cR$.  In the limit that the deformation of $P$ by $\cP$ and $\cR$ is small, there will be a large separation of these scales, $\m_\cP \gg \m_\cR$; equivalently, the RG flow will spend a long ``time" (corresponding to scales between $\m_\cP$ and $\m_\cR$) in the vicinity of $P$.

Assuming smoothness of RG flows in theory space,\footnote{Note that  \cite{Gukov:2015qea} deduces the existence of non-smooth RG flows in some theories.  This paper also treats cases with exactly marginal operators which we ignore here.} we see that to each dangerously irrelevant operator there will be a corresponding relevant operator $\cS$ at $R$.  When a dangerously irrelevant operator, $\cP$, is slightly turned on at $P$ and when the relevant operator $\cR$ is also slightly turned on, then there will be \emph{three} well-separated crossover scales generated, $\m_\cP \gg \m_\cR \gg \m_\cS$.  Only two of these scales are independent (since only two independent couplings have been turned on at $P$):  the larger $\m_\cP$ (compared to $\m_\cR$), the smaller $\m_\cS$.  By dimensional analysis we must have $\m_\cS = \m_\cR\, f(\m_\cR/\m_\cP)$ for some function $f$, and in the limit of large separation of scales $\m_\cR/\m_\cP \to 0$ (where the flow arbitrarily closely approaches the $OP$-$PR$-$RS$ segments) the leading behavior of $f$ gives
\begin{align}\label{discaling}
\m_\cS \sim \m_\cR \left( \frac{\m_\cR}{\m_\cP} \right)^\g
\end{align}
where $\g$ is some positive constant, characteristic of the theory (i.e., of $P$, $\cP$ and $\cR$).

There is a simple counting relation between the number of independent dangerously irrelevant operators at $P$ and the number and kind of relevant operators at $P$ and $R$.  Let $r_\text{UV}$ be the dimension of the manifold of relevant operators at the UV fixed point $P$, and $r_{\text{UV}\to\text{IR}}$ be the dimension of the submanfold which flows to a given IR fixed point $R$.  Similarly, let $r_\text{IR}$ be the dimension of the space of relevant operators at that IR fixed point $R$.  The dimension of the space of dangerously irrelevant operators at $P$ which give flows relevant at $R$, $d_{\text{UV}\to\text{IR}}$, then satisfies the relation
\begin{align}\label{relopcounting}
d_{\text{UV}\to\text{IR}} = 
r_\text{IR} - ( r_\text{UV} - r_{\text{UV}\to\text{IR}} ) .
\end{align}
The logic is that relevant flows from $P$ which do \emph{not} flow to $R$ give an $(r_\text{UV} - r_{\text{UV}\to\text{IR}})$-dimensional space of relevant flows at $R$.  Thus any further relevant flows at $R$ must correspond to a dangerously irrelevant flow at $P$.  (There are implicit assumptions in this argument that the space of relevant flows is always finite-dimensional, that RG flows are smooth, that they are always fixed-point flows, i.e., never flow to limit cycles, and that there are no exactly marginal operators.)  If the initial fixed point $P$ has a moduli space of vacua $\cM_P$, the picture is quite different and \eqref{relopcounting} no longer straightforwardly applies.  We turn to this discussion now.

\paragraph{Generalities about RG flows and moduli spaces of vacua.}  

Now consider a fixed point theory $P$ which has a moduli space of vacua, $\cM_P$, and further suppose that deforming the theory by $\cR$ does not lift the moduli space, but may continuously deform it.  (We will specialize below to the Coulomb branch of an $\cN=2$ theory, but for now keep the discussion general.)

By scale invariance, the geometry of (any branch of) $\cM_P$ is a cone, with a scale-invariant vacuum at the vertex and all other vacua are where some dimensionful field(s) get a vev ``$u$", and so spontaneously break scale invariance.  Call the mass scale associated to $u$, $\m_u$.  Selecting a vacuum in $\cM_P$ is an operation we can perform in the theory $P$; it is akin to deforming it, but by some nonlocal operator, e.g., by fixing the boundary values at spatial infinity for some scalar fields.  Changing the vacuum in $\cM_P$ is an operation that takes an infinite amount of energy (in infinite volume).  So each point in $\cM_P$ is the vacuum of a theory living in a different superselection sector.  The superselection sector above the point $u$ consists of a scale-invariant IR effective theory plus massive stuff of mass $\gtrsim \m_u$, since that is the only scale in the sector.  (In the case of the Coulomb branch of an $\cN=2$ theory, the IR effective theory is a free Maxwell theory for the generic vacuum.)

Now consider deforming $P$ by a relevant operator, $\cR$.  This adds a scale (the crossover scale in the flow generated by $\cR$), $\m_\cR$, well above which the effective theory is arbitrarily close to $P$, since $\cR$ is relevant at $P$.  What happens at scales well below $\m_\cR$?  By assumption the moduli space of vacua is not lifted.  Let's call the moduli space of the deformed theory $\cM_{P+\d_\cR}$, where $P+\d_\cR$ refers to the theory corresponding to the fixed point theory at $P$ deformed by the relevant operator $\d_\cR = (\m_\cR)^{4-\D(\cR)} \int d^4x\, \cR$.  The sector above a vacuum at large $u$, i.e., with $\m_u \gg \m_\cR$, is nearly the same as that sector in the undeformed theory $P$.  The reason is that $\cR$, being a relevant operator, has little effect on the theory on scales $E \gg \m_\cR$.  But the scale symmetry breaking by vev $u$ has happened at these high scales if $\m_u\gg\m_\cR$, so all that is left at scales $\lesssim \m_\cR$ is the (generically free) scale-invariant IR stuff, which is not lifted, by assumption.  Thus relevant operators correspond to deformations which modify the geometry near the $P$ singularity, but do not change it at large distances,\footnote{By $|u|$ we mean the distance on $\cM_{P+\d_\cR}$ measured in units of mass.  This is the usual metric on the moduli space induced by the low energy effective action nlsm terms since free scalars in four dimensions have mass dimension 1.}
\begin{align}\label{neardef}
\cM_{P+\d_\cR} \approx \cM_{P}
\quad\mbox{for}\quad |u| \gg \m_\cR.
\end{align}
We call such deformations \emph{near deformations}.

If instead $P$ is deformed by an irrelevant operator, $\cP$, then there is a crossover scale $\m_\cP$ \emph{below which} we will be close to $P$.  Above $\m_\cP$ the dynamics is controlled by a different fixed point, $O$, and so, by the above argument, it is $O$ which determines the large-$u$ behavior on the moduli space.  Thus irrelevant operators correspond to deformations which modify the geometry of $\cM_P$ at large distances, while leaving the vicinity of the singularity $P$ unchanged,
\begin{align}\label{fardef}
\cM_{P+\d_\cP} \approx \cM_{P}
\quad\mbox{for}\quad |u| \ll \m_\cP.
\end{align}
We wil  call such deformations \emph{far deformations} of the CB.

For $\cP$ a safely irrelevant operator, if we turn on both $\cP$ and $\cR$, we can deform both the near-$P$ and far-$P$ regions.  For large-enough separation of scales, $\m_\cP \gg \m_\cR$, there will be an intermediate region of the deformed moduli space which closely approximates $\cM_P$, as well as a region of small $|u|$ which approximates $\cM_R$.

A dangerously irrelevant operator, $\cP$, turned on at the same time as a relevant operator, $\cR$, will not only deform the distant regions, $|u|\gg \m_\cP$, as above, but also the near region $|u|\lesssim \m_\cS$ corresponding to the relevant deformation of $R$ by $\cS$.  Thus there will be two intermediate regions, $\m_\cP\gg|u|\gg\m_\cR$ where $\cM\approx\cM_P$, and $\m_\cR\gg|u|\gg\m_\cS$ where $\cM\approx\cM_R$, as well as a near region $|u|\ll\m_\cS$ where $\cM \approx \cM_S$.

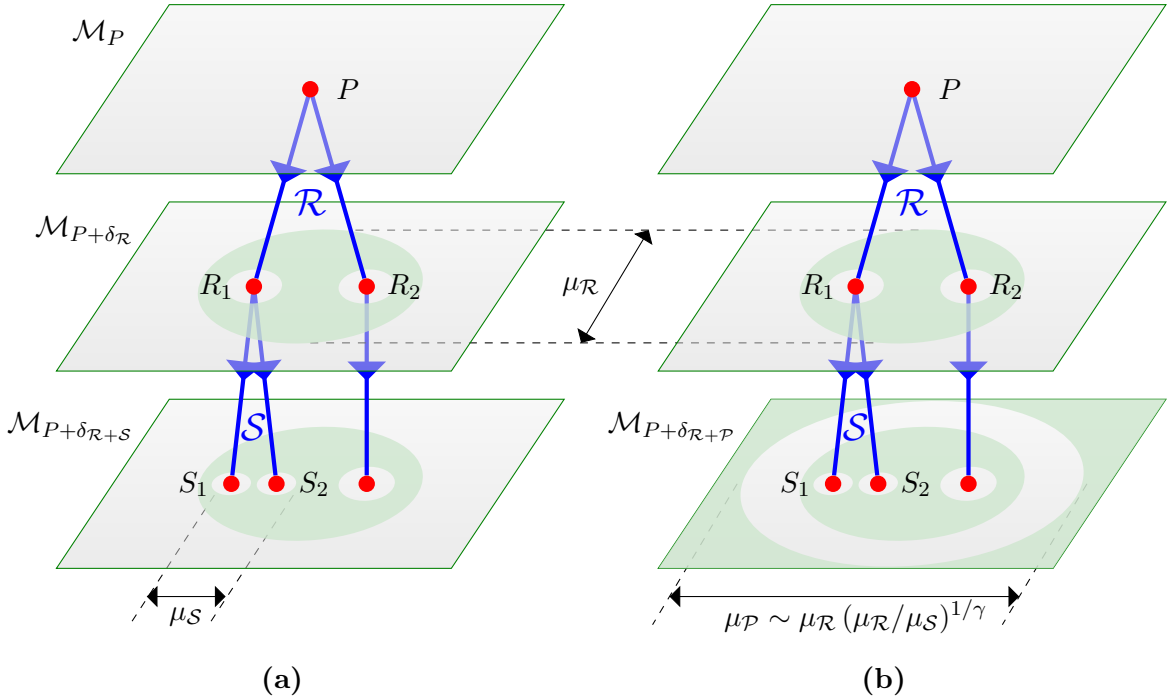
\begin{figure}[ht]
\begin{tikzpicture}[smooth,decoration={markings,
mark=at position .5 with {\arrow{>}}}]
%
%
\begin{scope}[scale=0.75]
\node at (4,-6) {{\bf (a)}};
\begin{scope}[xshift=0cm, yshift=-4.0cm]
\draw[dashed] (2.8,1.3) -- (1.3,-1.0) (4.2,1.3) -- (2.7,-1.0);
\draw[<->] (1.6,-0.5) -- (3.0,-0.5);
\node at (2.3,-0.85) {$\m_\cS$};
\shadedraw[top color=black!05,bottom color=black!15, draw=green!50!black, fill opacity=0.5]
    (0,0) -- (7,0) -- (9,3) -- (2,3) -- cycle;
\fill[fill=green!50!black!20, fill opacity=0.75, even odd rule]
    (4.5,1.5) circle [x radius=2., y radius=1., rotate=5]
    (3.1,1.5) circle [x radius=0.35, y radius=.2, rotate=5]
    (3.9,1.5) circle [x radius=0.35, y radius=.2, rotate=5]
    (5.5,1.5) circle [x radius=0.5, y radius=.3, rotate=5];
\node at (0.25,2.5) {$\cM_{P+\d_{\cR+\cS}}$};
\node[rc] (S1) at (3.1,1.5) {};
\node[rc] (S2) at (3.9,1.5) {};
\node[rc] (R2p) at (5.5,1.5) {};
\draw[ultra thick,blue,postaction={decorate}] (3.5,5.0) to (S1);
\draw[ultra thick,blue,postaction={decorate}] (3.5,5.0) to (S2);
\draw[ultra thick,blue,postaction={decorate}] (5.5,5.0) to (R2p);
\node at (S1) [shift={+(-.5,0)}] {$S_1$};
\node at (S2) [shift={+(.5,0)}] {$S_2$};
\node[font=\Large,color=blue] at (3.5,2.5) {$\cS$};
\end{scope}
\begin{scope}[xshift=0cm, yshift=-0.5cm]
\draw[dashed] (4.5,0.5) -- (14.5,0.5) (5.3,2.5) -- (15.3,2.5);
\draw[<->] (9.3,0.5) -- (10.5,2.5);
\node at (9.3,1.5) {$\m_\cR$};
\shadedraw[top color=black!05,bottom color=black!15, draw=green!50!black, fill opacity=0.5]
    (0,0) -- (7,0) -- (9,3) -- (2,3) -- cycle;
\fill[fill=green!50!black!20, fill opacity=0.75, even odd rule]
    (4.5,1.5) circle [x radius=2., y radius=1., rotate=5]
    (3.5,1.5) circle [x radius=0.5, y radius=.3, rotate=5]
    (5.5,1.5) circle [x radius=0.5, y radius=.3, rotate=5];
\node at (0.5,2.5) {$\cM_{P+\d_\cR}$};
\node[rc] (R1) at (3.5,1.5) {};
\node[rc] (R2) at (5.5,1.5) {};
\node at (R1) [shift={+(-.5,0)}] {$R_1$};
\node at (R2) [shift={+(.5,0)}] {$R_2$};
\draw[ultra thick,blue,postaction={decorate}] (4.5,5) to (R1);
\draw[ultra thick,blue,postaction={decorate}] (4.5,5) to (R2);
\node[font=\Large,color=blue] at (4.5,3) {$\cR$};
\end{scope}
\begin{scope}[xshift=0cm, yshift=3cm]
\shadedraw[top color=black!05,bottom color=black!15, draw=green!50!black, fill opacity=0.5]
    (0,0) -- (7,0) -- (9,3) -- (2,3) -- cycle;
\node at (0.75,2.5) {$\cM_{P}$};
\node[rc] (P) at (4.5,1.5) {};
\node at (P) [shift={+(.5,0)}] {$P$};
\end{scope}
\end{scope}
%
%
\begin{scope}[xshift=8cm,scale=0.75]
\node at (4,-6) {{\bf (b)}};
\begin{scope}[xshift=0cm, yshift=-4.0cm]
\draw[dashed] (1.4,1.5) -- (-0.1,-1.0) (7.6,1.5) -- (6.1,-1.0);
\draw[<->] (0.2,-0.5) -- (6.4,-0.5);
\node at (3.5,-0.85) {$\m_\cP\sim \m_\cR\, (\m_\cR/\m_\cS)^{1/\g}$};
\shadedraw[top color=black!05,bottom color=black!15, draw=green!50!black, fill opacity=0.5]
    (0,0) -- (7,0) -- (9,3) -- (2,3) -- cycle;
\fill[fill=green!50!black!20, fill opacity=0.75, even odd rule]
    (0,0) -- (7,0) -- (9,3) -- (2,3) -- cycle
    (4.5,1.5) circle [x radius=3.05, y radius=1.45, rotate=5]
    (4.5,1.5) circle [x radius=2., y radius=1., rotate=5]
    (3.1,1.5) circle [x radius=0.35, y radius=.2, rotate=5]
    (3.9,1.5) circle [x radius=0.35, y radius=.2, rotate=5]
    (5.5,1.5) circle [x radius=0.5, y radius=.3, rotate=5];
\node at (0.25,2.5) {$\cM_{P+\d_{\cR+\cP}}$};
\node[rc] (S1) at (3.1,1.5) {};
\node[rc] (S2) at (3.9,1.5) {};
\node[rc] (R2p) at (5.5,1.5) {};
\draw[ultra thick,blue,postaction={decorate}] (3.5,5.0) to (S1);
\draw[ultra thick,blue,postaction={decorate}] (3.5,5.0) to (S2);
\draw[ultra thick,blue,postaction={decorate}] (5.5,5.0) to (R2p);
\node at (S1) [shift={+(-.5,0)}] {$S_1$};
\node at (S2) [shift={+(.5,0)}] {$S_2$};
\node[font=\Large,color=blue] at (3.5,2.5) {$\cS$};
\end{scope}
\begin{scope}[xshift=0cm, yshift=-0.5cm]
\shadedraw[top color=black!05,bottom color=black!15, draw=green!50!black, fill opacity=0.5]
    (0,0) -- (7,0) -- (9,3) -- (2,3) -- cycle;
\fill[fill=green!50!black!20, fill opacity=0.75, even odd rule]
    (4.5,1.5) circle [x radius=2., y radius=1., rotate=5]
    (3.5,1.5) circle [x radius=0.5, y radius=.3, rotate=5]
    (5.5,1.5) circle [x radius=0.5, y radius=.3, rotate=5];
\node[rc] (R1) at (3.5,1.5) {};
\node[rc] (R2) at (5.5,1.5) {};
\node at (R1) [shift={+(-.5,0)}] {$R_1$};
\node at (R2) [shift={+(.5,0)}] {$R_2$};
\draw[ultra thick,blue,postaction={decorate}] (4.5,5) to (R1);
\draw[ultra thick,blue,postaction={decorate}] (4.5,5) to (R2);
\node[font=\Large,color=blue] at (4.5,3) {$\cR$};
\end{scope}
\begin{scope}[xshift=0cm, yshift=3cm]
\shadedraw[top color=black!05,bottom color=black!15, draw=green!50!black, fill opacity=0.5]
    (0,0) -- (7,0) -- (9,3) -- (2,3) -- cycle;
\node[rc] (P) at (4.5,1.5) {};
\node at (P) [shift={+(.5,0)}] {$P$};
\end{scope}
\end{scope}
\end{tikzpicture}
\caption{{\bf (a)} $\cS$ is a relevant operator at $P$.  {\bf (b)} $\cS$ is related to a dangerously irrelevant operator, $\cP$, at $P$.}
\label{fig:RG2}
\end{figure}

For concreteness, consider the case where a relevant deformation at a fixed point $P$ results in a near deformation of the moduli space which splits the conical singularity into two or more other singularities.  (We will consider other possible behaviors later.)  
Call the coordinates of the singularities $u_i$, $i=1,2,\ldots$, and name the IR fixed point theories at those points $R_i$.  The typical separation of the $u_i$ singularities is $\m_\cR$, as this is the only scale in the problem.   From the above discussion, for points $u\in\cM_{P+\d_\cR}$ close enough to $u_i$, the low energy action on the moduli space will be governed by the $R_i$ fixed point.  The moduli space of $R_i$ is (by scale invariance) a cone $\cM_{R_i}$, so
\begin{align}\label{}
\cM_{P+\d_\cR} \approx \cM_{R_i}
\quad\mbox{for}\quad |u-u_i| \ll \m_\cR.
\end{align}
$\cM_P$ and its deformation $\cM_{P+\d_\cR}$ are illustrated on the top two lines of figure \ref{fig:RG2}.

Now consider a situation where the $R_1$ fixed point theory (after the relevant $\cR$ deformation of the $P$ theory) itself has a relevant deformation $\cS$.  This deformation either is or is not related to a dangerously irrelevant operator.  In the latter case, shown in figure \ref{fig:RG2}(a), $\cS$ must also be a relevant operator at $P$.  
The green shaded regions on the various deformed CBs represent the crossovers between areas where the geometry is (close to) the scaling geometry determined by a single fixed point.

By contrast, in the case where $\cS$ is turned on by a dangerously irrelevant operator, $\cP$, in $P$, the geometry of the moduli space is deformed at large vevs, as shown in figure \ref{fig:RG2}(b).  The scale, $\m_\cP$, at which this deformation becomes apparent diverges to infinity as $\m_\cS\to0$.  Thus there is a clear qualitative difference in the CB geometry associated with relevant deformations in the IR that flow from relevant versus (necessarily dangerously) irrelevant operators in the UV.

As mentioned above, the counting relation \eqref{relopcounting} relating the dimensions of spaces of relavant opertors at the UV and IR fixed points to the dimension of the space of dangerously irrelevant opertors no longer makes sense when there is a moduli space of vacua, since, as we have emphasized, the notion of a single IR fixed point does not make sense.  We will discuss how to properly interpret this kind of relation in the case of rank-1 CBs in section \ref{ssec:sings.1}.

\subsection{Local deformations of $\cN=2$ SCFTs\label{sec:localdefs}}

The possible local $\cN=2$ supersymmetric deformations of a SCFT are built as integrals over space-time of some Lorentz scalar descendant, $\d$, of a superconformal primary field $X$.  To deform the theory it should not be a total derivative, so $\d$ can only be a descendant formed by acting with some combination of the eight $\cN=2$ supercharges, $Q$ and $\tQ$, on $X$,
\begin{align}\label{}
\d_{n,m}\sim Q^n \tQ^m X
\end{align}
for $n,m\in\{0,1,\ldots,4\}$.  To be $\cN=2$ supersymmetric, $Q$ or $\tQ$ acting on it must annihilate it or give a total derivative.  This can happen either by virtue of the supersymmetry algebra or if the primary $X$ satisfies a null state condition by virtue of it being in a shortened representation of the superconformal algebra.  The unitary, positive energy representations of the $\cN=2$ superconformal algebra and their associated null states are reviewed in appendix \ref{app:repre}.  Then, as in the $\cN=1$ case \cite{Green:2010da}, it is a straightforward exercise to classify the allowed deformations.  This argument is given in appendix \ref{app:defo}, with the resulting 6 types (and their conjugates) shown in table \ref{N2deftable}.  In every case the primary $X$ turns out to be a Lorentz scalar, and we list its $\SU(2)_R$-spin $R$, $\U(1)_R$-charge $r$, and dimension $\D$.  In the table, $R_\d$, $r_\d$, and $\D_\d$ are the $\SU(2)_R$-spin, $\U(1)_R$-charge, and dimension of the deformation operator descendant.  The conjugates $\bar{\d_{n,m}}=\d_{m,n}$, have the same properties but with the sign of the $\U(1)_R$ charges reversed.  Note that $\d_{0,0}$ is just the identity operator.  
\begin{table}[ht]
\centering
$\begin{array}{|l|lrrl|lrr|}\hline
\multicolumn{8}{|c|}{\text{\bf Local $\cN=2$ deformations of SCFTs}}\\ \hline\hline
\text{Deformation type} & 
\multicolumn{4}{c|}{\text{Primary operator rep.\ \& charges}}
&\multicolumn{3}{c|}{\text{Deformation op.\ charges}}
\\
\parbox[b][0.45cm]{3cm}{$\d_{n,m} \sim Q^n \tQ^m X$}  
& X\in\ & R &\qquad r &\quad \ \ \ \D 
&\ R_\d &\qquad r_\d &\qquad\qquad \D_\d\\[0.5mm]
\hline
\d_{0,0} := X & \hat\cB_0 & 0 & 0 &\quad \ \ \ 0 &\ 0 & 0 & 0\\
\d_{0,2} := (\tQ^2)^1_0 X & \hat\cB_1 
& 1 & 0 &\quad \ \ \ 2  
&\ 0 & {-}1 & 3\\
\d_{0,4} := \tQ^4 X & \cE_{r\,(0,0)} 
& 0 & {>}1 &\quad \ \ \ r   
&\ 0 & r{-}2 & \D{+}2 > 3\\
\d_{2,2} := (Q^2)^1_0 (\tQ^2)^1_0 X\  & \hat\cB_R 
& {\ge}2 & 0 &\quad \ \ \ 2R   
&\ R{-}2 & 0 & \D{+}2 \ge_\Z 6\\
\d_{2,4} := (Q^2)^1_0 \tQ^4 X & \cB_{R,r\,(0,0)} 
& {\ge}1 & {>}1 &\quad \ \ \ 2R{+}r   
&\ R{-}1 &\ r{-}1 & \D{+}3 > 6\\
\d_{4,4} := Q^4 \tQ^4 X & \cA^\D_{R,r\,(0,0)} \ 
& \ {\ge}0 & \ &\quad > 2R{+}2{+}|r| \ 
&\ R & r & \D{+}4 > 6\\[0.2mm]
\hline
\end{array}$
\caption{The quantum numbers $(R_\d, r_\d, \D_\d)$ of local operators preserving $\cN=2$ supersymmetry that can deform an $\cN=2$ SCFT, along with the representation and quantum numbers $(R,r,\D)$ of the primary field, $X$ they are descended from.\label{N2deftable}}
\end{table}

We have also listed the type of superconformal representation the primary is in, using the naming scheme of \cite{Dolan:2002zh}, whose definitions and properties are listed in table \ref{N2table} in appendix \ref{app:repre}.  We see that all deformations up to and including dimension 6 are in protected (short) multiplets.  In particular all relevant and marginal operator primaries are in $\cBh_R$ and $\cB_{R,r\,(0,0)}$ representations.  The Higgs, Coulomb, and mixed branch scalar moduli are primaries of precisely these representations;  see appendix \ref{app:repre}. 

In our analysis we are particularly interested in relevant operators. From table \ref{N2deftable}, the relevant operators are the $R=1$, $\d_{0,2}=\tQ^2 \cBh_1$ deformations of dimension 3, and the $R=0$, $\d_{0,4}=\tQ^4 \cE_{r(0,0)}$ deformations with $1<r<2$ which have dimensions $3<\D<4$.  ($\cE_{r(0,0)}$ is another name for a $\cB_{R,r\,(0,0)}$ multiplet with $R=0$.)  This explains the pattern of dimensions of deformation parameters of scale invariant Coulomb branch geometries observed in the literature.\footnote{See however \cite{Gaiotto:2010jf} for an example showing that the assignment of these scaling dimensions may be ambiguous.}  There we see either deformation parameters with dimensions $0<\D<1$, or parameters with dimensions $\D\in\N$.  The former correspond to the $\d_{0,4}$ deformations and the latter to the $\d_{0,2}$ deformations.  

Note that all relevant deformations have $R_\d=0$, so do not break the $\SU(2)_R$ symmetry.  This means that the Fayet-Iliopoulos term which is allowed by $\cN=2$ supersymmetry (and has $R_\d=1$) \cite{Antoniadis:1995vb} is not allowed as a local $\cN=2$ deformation of a unitary $\cN=2$ superconformal theory.

We now take a closer look at the effects of the allowed relevant deformation operators.

\paragraph{Relevant deformations: mass terms.}

The $R=1$, $\d_{0,2}$ deformations necessarily transform in the adjoint representation of the flavor symmetry algebra, $F$, since their primary operator is in a $\cBh_1$ multiplet which contains conserved flavor currents, $J_\m^A$, which by definition carry an adjoint index, $A=1,\ldots,\text{dim}(F)$.  Thus the general $R=1$, $\d_{0,2}$ deformation has the form\footnote{The supercharges are understood to be acting to the right by (anti)commutators and are contracted with $X$ to Lorentz and $\SU(2)_R$ singlets in a particular way; see appendix \ref{app:defo} for the details.} 
\begin{align}\label{CFTmassterm}
m_A \int d^4x\, \tQ^2 X^A + \text{c.c.},\qquad\text{for}\ X^A\in\cBh_1 .
\end{align}
Although there are dim$(F)$ dimension-1 complex parameters, $m_A$, in \eqref{CFTmassterm}, since they are related by the $F$ global symmetry, they give rise to only rank$(F)$ inequivalent deformations of the SCFT.  This is because generic values of the $m_A$ break $F\to \U(1)^{\text{rank}(F)}\rtimes \text{Weyl}(F)$, where Weyl$(F)$ is the Weyl group of $F$.\footnote{The Weyl group is a finite group generated by real reflections that is a certain subgroup of the isometry group of the root system of $F$; Weyl groups for simple Lie algebras are listed in tables, e.g., \cite{mckay1981tables, humphreys1990coxeter}.}  These inequivalent deformations are labeled by a generating set of the rank$(F)$ algebraically independent adjoint Casimirs formed from the $m_A$, which are Weyl$(F)$-invariant polynomials.  A generating set can be taken to be homogeneous polynomials in the $m_A$.  Although there is no unique (canonical) generating set, the set of the degrees of the generators is unique, and uniquely determines Weyl$(F)$.  

Upon weakly gauging the flavor symmetry of the SCFT (after adding free massless hypermultplets transforming under $F$, if necessary, to make the $F$ coupling IR-free), the $m_A$ become vevs of the $F$-vector multiplet scalar, enlarging the Coulomb branch.   Since Coulomb branches are complex varieties, we learn upon taking the flavor gauge coupling to zero that the Coulomb branch of the deformed SCFT can only depend holomorphically on the $m_A$.  Furthermore, as explained above, the broken flavor symmetry implies that the deformed CB can only depend on Weyl$(F)$-invariant combinations of the $m_A$, implying that there will be rank$(F)$ deformations parameters of integer dimensions corresponding to the degrees of the adjoint Casimirs of $F$.\footnote{This argument also implies \cite{Argyres:1996eh} that since the the $m_A$ enter the deformed SCFT in the same way as CB vevs, they must also satisfy the CB ``D-term" constraint
\begin{align}\label{CBDterm}
f^{ABC}m_A \bar m_B = 0 ,
\end{align}
where $f^{ABC}$ are the structure constants of $F$.  This means that although turning on an arbitrary complex $m_A \tQ^2 X_A$ deformation preserves $\cN=2$ supersymmetry, adding its complex conjugate breaks the supersymmetry unless \eqref{CBDterm} is satisfied.  It is interesting to understand how this follows from the SCFT operator algebra.  
The $\cBh_1^A(x) \cBh_1^B(y) \sim |x-y|^{-4} \d^{AB} + |x-y|^{-2} f^{AB}_C \cBh_1^C(y) + \cdots$ OPE implies that their descendants' OPE has the contact term $\tQ_x^2\cBh_1^A(x) Q_y^2\cBh_1^B(y) \sim \cBh_1^C(y) f^{AB}_C \d^4(x-y)$, which in turn implies there must be a counterterm of the form $\int d^4x\, \bar m_A m_B f^{AB}_C\cBh_1^C(x)$ which breaks supersymmetry unless \eqref{CBDterm} is satisfied.  We thank Yifan Wang \cite{yifanPC} for explaining this to us.}

We will now argue that the effect of these mass operators is to split the conical singularity on the CB into multiple Kodaira singularities.  Being relevant operators, they can only lead to near deformations of the CB.  This deformation can do one of three things: (i) remove (smooth out) the singularity, (ii) not split the singularity, (iii) or split it.  

Possibility (i) is ruled out because a near deformation cannot change the EM duality monodromy at infinity on the CB and all Kodaira singularities have nontrivial monodromy (and are, in fact, characterized by them), whereas a removal of the singularity would give a trivial monodromy at infinity.  A geometrical argument leading to the same conclusion is that since the Kodaira singularities all have positive deficit angles, any smoothing of them would necessarily create regions of positive scalar curvature, contradicting the non-positivity of the scalar curvature on SK manifolds. 

Possibility (ii) is similarly constrained.  If the singularity does not split, but changes Kodaira type, then the monodromy constraint is violated.  If the singularity type stays the same but the geometry is modified in a region around the singularity, the non-positivity of the curvature will be violated somewhere in the deformation region.  So  the only possibility is that the geometry is changed by at most an overall shift of the position of the tip of the cone, which has no effect on the CB geometry.  Though a logical possibility, such an overall shift is unnatural since it implies a decoupling of the relevant mass deformation operator from the low energy theory on the Coulomb branch which is not enforced by a symmetry.  We will assume that this kind of ``invisible" relevant deformation does not occur.  

The possible exception to this is when the conformal theory has a free (gauge singlet) vector multiplet.  This multiplet then describes a $\U(1)$ gauge factor decoupled from the rest of the theory.  It contributes a flat cartesian factor $\C$ to the CB geometry.  In the rank-1 case this would be the whole CB, which would therefore have no singularity and trivial EM monodromy.  (Though it is not listed in table \ref{Table:Kodaira}, it is called the $I_0$, or regular fiber, in the Kodaira classification.)  However, there is no allowed mass term in a free $\U(1)$ pure gauge theory.

Note that this discussion also applies to deformations of the non-scale-invariant $I_n$ and $I^*_n$ Kodaira singularities which correspond to IR-free $\U(1)$ and $\SU(2)$ gauge theories, respectively, with massless charged hypermultiplets (a more detailed discussion of the $I_n$ and $I_n^*$ singularities will be given below).  But now there is an exception to the argument eliminating possibility (ii).  In the case of the $\U(1)$ theories there is a free gauge-singlet vector multiplet which decouples in the IR and whose vev therefore has dimension 1 and can be shifted by a mass term for a $\U(1)$ factor of the flavor symmetry.   This can be restated as saying that a mass term \eqref{CFTmassterm} with mass parameter $m$ corresponding to a $\U(1)$ factor of the flavor symmetry may deform the CB geometry by an overall shift of a CB parameter $u$ with $\D(u)=1$.  Indeed, only in this case does $m$ carry the same dimension, R-charges, and flavor quantum numbers as does $u$. 

So, by elimination, the effect of a mass term \eqref{CFTmassterm} must be possibility (iii), that is to split a scale-invariant Kodaira singularity into two or more Kodaira singularities.  This splitting is tightly constrained by the SK geometry of the CB, as we discuss in detail in the next section.

\paragraph{Relevant deformations: chiral terms.}

The second class of $\cN=2$ supersymmetric relevant local deformations of a SCFT are chiral terms (i.e., integrals over a chiral half of $\cN=2$ superspace) of the form
\begin{align}\label{chiraldef}
\m \int d^4x\, \tQ^4 X + \text{h.c.}
\qquad \text{with}\qquad X\in\cE_{r\,(0,0)}\quad\text{for}\quad 1<r<2.
\end{align} 
Here recall that $\cE_{r\,(0,0)}$ are the short superconformal representations with dimension $\D=r$ scalar primaries which can get vevs on the Coulomb branch.
The complex parameter $\m$ has mass dimension and $\U(1)_R$ charge $2-r$.  Furthermore, $\m$ is a flavor singlet since, by $\SU(2)_R$ conservation, $\cE_r$ cannot appear on the right side of the $\cE_r \cBh_1$ OPE.

The same arguments as for the mass terms imply that the effect of these terms is also to split singularities.  As before, an exception occurs when a Coulomb branch multiplet decouples.  This happens in the $r\to1$ limit, where \cite{Dolan:2002zh} 
\begin{align}\label{}
\lim_{r\to1} \cE_{r\,(0,0)} = \cD_{0\,(0,0)}\oplus \cBh_1.
\end{align}
The $\cD$ multiplet is the short multiplet containing the $\cE$ primary and is a free vector multiplet, and the $\cBh_1$ multiplet's primary is the second level descendant $(\tQ^2)^1_0 \cE$; it is necessarily a flavor scalar since $\cE$ is, so must generate a $\U(1)$ factor of the flavor symmetry.  Thus in this limit $\tQ^4 X = [(\tQ^2)^1_0 (\tQ^2)^1_0 X]^0 = (\tQ^2)^1_0 X'$ with $X'\in\cBh_1$ (where the indices denote spins and R-spins following the notation introduced in appendix \ref{app:defo}).  This shows that in this limit the chiral term becomes a mass term for a $\U(1)$ flavor generator.  Since it is also accompanied by a free vector multiplet it is precisely the mass term which shifts but does not deform the $I_n$ singularities.\footnote{This argument also shows that \eqref{CFTmassterm} with $X$ a $\U(1)$ flavor generator and $m$ the CB vev of a free $\U(1)$ vector multiplet is the only allowed relevant coupling between a SCFT and $\U(1)$ multiplets; answering a question posed in \cite{Argyres:2012fu}.}

\paragraph{Marginal deformations.}

The possible marginal operators are of the form 
$\d_{0,4}=\tQ^4 \cE_{2(0,0)}$ (with $R_\d=r_\d=0$).  By the arguments of \cite{Green:2010da} these can only be either exactly marginal or marginally irrelevant at a SCFT without free fields.  Furthermore, \cite{Green:2010da} show that the set of exactly marginal deformations is the K\"ahler quotient of the space of marginal couplings by the $\cN=1$ flavor symmetry, $F^{(\cN=1)}$.  If we embed the $\cN=1$ in the $\cN=2$ algebra by $(Q,\tQ)^{(\cN=1)} = (Q_1,\tQ_2)$ then the $\U(1)_f$ generated by $r-R_3$ commutes with the $\cN=1$ $\U(1)^{(\cN=1)}_R$ symmetry, so the $\cN=1$ flavor symmetry obeys $F^{(\cN=1)} \supset F \times \U(1)_f$.  Alternatively, if we embed the $\cN=1$ in the $\cN=2$ algebra by $(Q',\tQ')^{(\cN=1)} = (Q_2,-\tQ_1)$ then the $\U(1)_{f'}$ generated by $r+R_3$ commutes with $\U(1)^{(\cN=1)}_R$, so the flavor symmetry for this $\cN=1$ theory obeys $F'^{(\cN=1)} \supset F \times \U(1)_{f'}$.  Therefore an $\cN=2$ marginal operator is neutral under all $\cN=1$ flavor symmetries only if it is neutral under $F$ and has $r=R_3=0$ and therefore is $\U(1)_R\times\SU(2)_R$ neutral.

The $\d_{0,4}$ marginal deformation is $\U(2)_R$ neutral, so is exactly marginal when it is neutral under the ($\cN=2$) flavor algebra $F$.   But this operator is, in fact, flavor neutral because $\cE_{r(0,0)}$ cannot appear in the $\cBh_1 \times \cE_{r(0,0)}$ OPE by R-spin conservation.  
Thus the only possible exactly marginal $\cN=2$ operators are the fourth-level descendant of dimension-2 CB operators, $\tQ^4 \cE_{2\,(0,0)}$, as stated in \cite{Beem:2014zpa}.  


\section{Physical deformations}\label{sec:sings}

In this section we describe the constraints that $\cN=2$ supersymmetry puts on possible near deformations of scale-invariant Kodaira singularities.  The special K\"ahler (SK) conditions that the $\cN=2$ CB geometries satisfy are not apparent from the relation of the CB chiral ring to the local SCFT deformations discussed in the last section.  These SK conditions put further constraints on possible near deformations of the CB. We will show that, at least in the rank-1 case, these SK conditions are weak: the problem we face is that they present no obstruction to ``turning on" two SK-preserving near deformations to obtain a larger SK near deformation.  Indeed, each Kodaira singularity has a unique \emph{maximal SK deformation} (to be defined below) which was already constructed in \cite{Argyres:1995xn,Minahan:1996fg,Minahan:1996cj}.  So the question becomes: when is a sub-maximal deformation (a restriction of the maximal deformation to fewer parameters) a physically distinct deformation?  This can only be answered by going beyond the SK conditions and using other physical conditions.  The main purpose of this section is to motivate a conjecture for the key physical condition that deformations of CBs must satisfy. 

\subsection{Special K\"ahler constraints on splitting of singularities\label{SKsplits}}

As argued in the last section, relevant local $\cN=2$ deformations of SCFTs give near deformations of a Kodaira singularity which splits the singularity (with the exception of a $\U(1)$-flavor mass deformation of an $I_n$ singularity which simply shifts the singularity position). Each singularity which follows from the deformation should be interpreted as an IR fixed point, thus the geometry around the singularities should be locally scale-invariant. In other words the split of a scale-invariant singularity can only result in singularities which themselves belong to the Kodaira classification reported in table \ref{Table:Kodaira}. We will call the \emph{pattern of the deformation} the list of Kodaira types of the singularities that result from the splitting.  Examples of these deformation patterns are shown in the third column of table \ref{table:theories}.  For example, the second deformation listed there splits the $II^*$ singularity into six $I_1$ singularities and one $I_4$ singularity (whose physical interpretation will be given below).

\begin{figure}[tbp]
\centering
\begin{tikzpicture}[decoration={markings,
mark=at position .5 with {\arrow{>}}}]
\clip (0,0) rectangle (15,4);
\fill[color=black!05] (0,0) rectangle (6,4);
\node[Bl] (or) at (3,3) {};
\node[R] (br) at (3,2) {};
\node[red] at (3,2.35) {$u{=}0$};
\draw[thick,blue,postaction={decorate}] (or) .. controls (0,0) and (6,0) .. (or);
\node[blue] at (4.25,1) {$K_0$};
\node[single arrow, draw, black] at (7.5,2) {{\small deformation}};
\fill[color=black!05] (9,0) rectangle (15,4);
\node[Bl] (or1) at (12,3) {};
\node[R] (br1) at (10.5,2) {};
\node[red] at (10.5,2.35) {$u_1$};
\node[R] (br2) at (12,1.5) {};
\node[red] at (12,1.85) {$u_2$};
\node[R] (br3) at (13.5,2) {};
\node[red] at (13.5,2.35) {$u_3$};
\draw[thick,blue,postaction={decorate}] (or1) .. controls (8,3) and (10,-1) .. (or1);
\node[blue] at (10,3) {$K_1$};
\draw[thick,blue,postaction={decorate}] (or1) .. controls (10.5,0) and (13.5,0) .. (or1);
\node[blue] at (11.3,1) {$K_2$};
\draw[thick,blue,postaction={decorate}] (or1) .. controls (14,-1) and (16,3) .. (or1);
\node[blue] at (14,3) {$K_3$};
\end{tikzpicture}
\caption{Singularities and their monodromies on the CB before and after a relevant deformation.  (This is a ``top down" view of figure \ref{fig:cone}.)  The solid points with coordinates $u_a$ are the singularities.  The $K_a\in SL(2,\Z)$ are EM duality monodromies associated to the closed paths looping around these singularities, and satisfy $K_0 = K_3 K_2 K_1$.}
\label{fig:monod}
\end{figure}
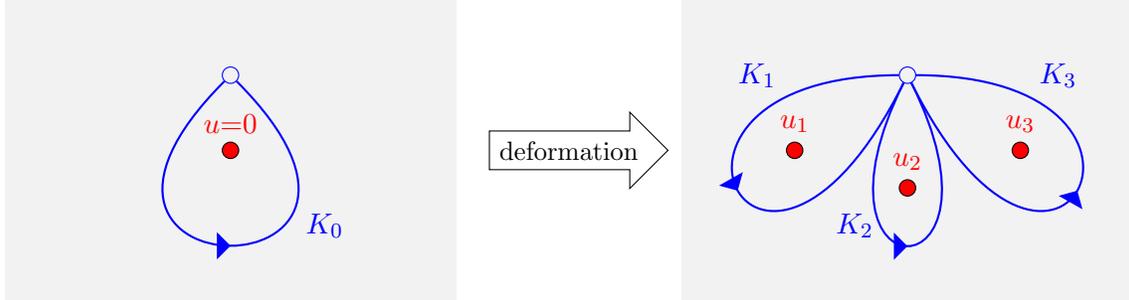

The SK condition constrains this splitting in a number of ways.  Since the deformation is a near deformation (i.e., does not affect the geometry asymptotically far from the singularity), the total electric-magnetic (EM) duality $\SL(2,\Z)$ monodromy around the singularities cannot change.  This means that the product of the monodromies around the split singularities must equal the monodromy around the original singularity, as shown in figure \ref{fig:monod}.  Note that the Kodaira type of the singularity does not determine its EM monodromy, but only its conjugacy class in $\SL(2,\Z)$, shown in table \ref{Table:Kodaira}.

Furthermore, the positions of the singularities on the CB are given by the zeros of a polynomial in $u$, the discriminant of the SW curve, $D_x$, discussed in appendix \ref{app:details}.  A near deformation cannot change the order of the polynomial since that would involve changing the asymptotic behavior of the CB geometry.  So deformations of singularities must proceed by splitting zeros of $D_x$ of multiplicity greater than one into multiple zeros of lower multiplicity.  Each Kodaira singularity has a characteristic order of vanishing of $D_x$, listed in table \ref{Table:Kodaira}.  In this way we get a hierarchy of Kodaira singularities, ordered by decreasing order of vanishing of $D_x$.  For instance, the largest value of ord$_0(D_x)$ for a rank-1 scale-invariant singularity is 10 which corresponds to the $II^*$ singularity.  So upon splitting this singularity only Kodaira singularities with ord$_0(D_x)<10$ can appear, and all possibilities can easily be listed: $II^* \to \{III^*,I_1\}$, $II^*\to\{I^*_3,I_1\}$, $II^*\to\{I_9,I_1\}$, $II^*\to\{IV^*,I_2\}$, etc.  This limits near deformations to a finite but large number, 383, of distinct possible deformation patterns.   Not all of these patterns can be realized in such a way as to satisfy the monodromy constraint explained above.  It is not an easy problem to determine which of the deformation patterns are disallowed by this monodromy constraint, but it is not hard to see \cite{Argyres:2015gha} that at least many dozens of these patterns are consistent with it.  

A {\it maximal deformation} splits a Kodaira singularity into ord($D_x$) $I_1$ singularities, and is always realizable.  Since the order of vanishing of $D_x$ of an $I_1$ singularity is 1, no further splitting is possible.  The set of relevant operators which implement the maximal deformations for each of the singularities in the Kodaira classification, along with the explicit form of their SW curves and 1-forms were constructed in \cite{Argyres:1995xn,Minahan:1996fg,Minahan:1996cj}.  Every sub-maximal deformation pattern can obviously be extended by further splittings to the maximal deformation pattern.  

This raises the question: what physical effect can prevent singularities from splitting?  As already mentioned, each singularity (after deformation) is an IR fixed point.  Saying it cannot be further split means either
\begin{itemize}
\item[(i)] as a CFT it has no further relevant deformation operators (other than CB shift operators), or
\item[(ii)] it does have relevant deformation operators, but they cannot be turned on by relevant deformations of the UV fixed point.
\end{itemize}
Note that possibility (ii) means that there must be dangerously irrelevant operators in the UV associated to the ``extra" relevant operators at the IR fixed point, as explained in the last section.

We will now argue that possibility (ii) does not occur.
Upon explicitly constructing the geometries of SK deformations of the Kodaira cones, we find \cite{Argyres:2015gha} that there are no SK obstructions to turning on near CB deformations simultaneously.  Concretely, what this means is the following:  Suppose that a near deformation of a Kodaira singularity with parameters $\{m_i\}$ is turned on, and splits the singularity into some IR singularities.  If one (or more) of the resulting IR Kodaira singularities has further deformation parameters, $\{m'_a\}$, then we find that we can always construct a SK near deformation of the original singularity with the enlarged set $\{m_i\}\cup\{m'_a\}$ of deformation parameters.  

This statement is equivalent to the following property of rank-1 SK deformations:  Every submaximal deformation can be realized simply by restricting the vector space of \emph{linear} deformation parameters --- e.g. mass parameters with $\D(m_i)=\D(m'_a)=1$ or chiral term parameters with $3<\D(\m)<4$ --- to a linear subspace.  It is easy to see (see appendix \ref{app:details} for the details) that such a restriction of a SK deformation will give another SK deformation, but it is not obvious that \emph{every} submaximal SK deformation is found in this way.  Indeed, we do not have a proof of this property, but only observe it to be true of all the rank-1 SK geometries that we have been able to construct.  The details of these constructions are given in \cite{Argyres:2015gha} and thus are evidence for the validity of this property.

Consider a conformal theory $P$ deformed by a relevant operator $\cR$ with parameters $\{m_i\}$ which leads to the splitting of the initial singularity of theory $P$. Say that one of the resultant singularities is associated to an IR fixed point $R_1$ at which a further relevant deformation $\cS$ with parameters $\{m'_a\}$ is turned on.  The above property then says that there is a near deformation of the CB of $P$ by $\{m_i\}\cup\{m'_a\}$ which leads to the same near deformation of the $R_1$ CB geometry as would be obtained by turning on $\cS$ at $R_1$. To exclude possibility (ii) we need to show that $\cS$ cannot be a dangerously irrelevant operator. Dangerously irrelevant operators, as we discussed in section \ref{sec:RGflows}, deform the Coulomb branch in a very distinctive manner, that is by deforming at the same time the near and the far region. There is nothing that ``forces" the near deformation of $R_1$ by $\cS$ to be accompanied by a far deformation, and therefore $\cS$ cannot associated to a dangerously irrelevant operator at $P$.  In other words, the absence of obstructions to extending near deformations of SK geometries leads to the picture of $\cN=2$ RG flows given in figure \ref{fig:RG2}a and not the one given in figure \ref{fig:RG2}b.  This is evidence for our main ``safely irrelevant" conjecture:

\begin{center}
\emph{$\cN=2$ conformal theories do not have dangerously irrelevant operators.}
\end{center}

\noindent We will see that this conjecture dramatically constrains the possible physical deformations of the scale-invariant Kodaira singularities, nearly restricting it to the set of conformal field theories already known from RG flows and S-dualities of lagrangian theories.  This in itself can be interpreted as a kind of evidence for the correctness of the conjecture.  It would be interesting to analyze higher-rank lagrangian and class $\cS$ theories to test this conjecture.\footnote{Note that examples of $\cN=1$ flows with dangerously irrelevant operators are found in \cite{Gukov:2015qea} while the analysis of $\N=2$ flows is consistent with our safely irrelevant conjecture.}

Accepting this conjecture, we are constrained to possibility (i) above, which says that we should only consider as physical a deformation which is ``complete" or ``non-extendable", i.e., only if all of its resultant IR singularities are theories that admit no further deformations.  The IR singularities can only be one of the Kodaira singularities shown in table \ref{Table:Kodaira}.  Which of these can correspond to theories which admit no further deformation?

\subsection{Undeformable singularities\label{froSing}}

In section \ref{sec:localdefs} we showed that a relevant local $\cN=2$ supersymmetric deformation of a SCFT splits singularities unless it is a mass operator of the form \eqref{CFTmassterm} for an overall $\U(1)$ flavor symmetry factor in an IR free theory described by an $I^*_n$ singularity with $n>1$.  So other than this case, a singularity is undeformable only if it corresponds to a theory which admits no relevant operators, either mass terms \eqref{CFTmassterm} or chiral terms \eqref{chiraldef}.  We will now study which of the Kodaira singularities listed in table \ref{Table:Kodaira} can be undeformable.

\subsubsection*{Types $II$, $III$ and $IV$}

The type $II$, $III$, and $IV$ singularities have CB parameters with dimension $1<\D(u)<2$.  Their corresponding SCFTs therefore have a superconformal scalar primary field of type $\cE_{r\,(0,0)}$ with $1<r<2$ whose vev is $u$.  These theories necessarily have a chiral term deformation of the form \eqref{chiraldef} and therefore cannot be undeformable.

\subsubsection*{Type $I_0^*$}

The type $I_0^*$ singularity has CB parameter of dimension 2, and so has a superconformal scalar primary field of type $\cE_{2\,(0,0)}$.  When the $I^*_0$ singularity is not weakly coupled, i.e., when its parameter $\t\neq i\infty$, its CFT will have an exactly marginal chiral term deformation by the discussion in \ref{sec:localdefs}.  With the additional assumption that turning on this marginal deformation changes the $\t$ parameter of the $I_0^*$ singularity, i.e., that $\t$ is not constant on the conformal manifold,  then holomorphy of $\t$ implies that $\t$ must cover some multiple of $\SL(2,\Z)$ fundamental domains, and therefore will include the $\t=i\infty$ weak coupling point.  But a weakly coupled rank-1 theory with a dimension-2 CB parameter is naturally interpreted as a  scale invariant $\SU(2)$ gauge theory, i.e., either the one with four fundamental hypermultiplets or the one with one adjoint hypermultiplet. These have an $\SO(8)$ and an $\Sp(2)$ flavor symmetry respectively, and so in both cases they have mass deformations.  This singularity therefore cannot be undeformable.

There are two ways around the above conclusion.  One is to allow gauging of discrete symmetries.  Indeed, examples of such frozen $I_0^*$ singularities arise as the low energy desription of a free $\U(1)$ gauge theory for which a particular $\Z_2$ symmetry (which acts on the coupling as well as the CB) is gauged.  This possibility is described in detail in \cite{Argyres:2016yzz}.

A different and more exotic possibility is that of a non-lagrangian frozen $I_0^*$ singularity. This could for instance arise by weakly gauging an $\SU(2)$ subgroup of the flavor group of a putative rank-0 SCFT, ``$X_0$''.  In order for the singularity to be scale invariant, the $X_0$ contribution to the $\SU(2)$ beta function should be such that the beta function vanishes.  While in order for the singularity to be frozen, there must be no commutant of the $\SU(2)$ in the $X_0$ flavor group.  No such rank-0 SCFT is known to exist.

Even if we assume there are no rank-0 SCFTs, because of the existence of the free frozen discretely gauged $I^*_0$ theories, we must include the possibility of frozen $I_0^*$ singularities in our classification of possible rank-1 CB geometries and so they appear in table \ref{table:theories}.

\subsubsection*{Types $II^*$, $III^*$ and $IV^*$ }

These strongly coupled theories do not have relevant chiral term deformations since their CB parameters all have $\D(u)>2$.  It is conceivable that ``frozen" SCFTs with CBs of types $II^*$, $III^*$, or $IV^*$ exist which do not admit any relevant deformation and, in particular, have no flavor symmetry.  Such conjectured non-lagrangian frozen singularities are speculative, as we have no evidence either for or against their existence.  We know of no argument showing that all SCFTs corresponding to these singularities must have non-trivial flavor symmetries, or, equivalently, necessarily admit mass term deformations. In fact there is at least one known W-algebra, namely the W(2,7) in \cite{Blumenhagen:1990jv}, which could be consistently interpreted as 2d chiral algebra associated to a frozen $IV^*$ singularity\footnote{We thank Madalena Lemos for pointing this out to us.} Perhaps bootstrap arguments could also address this question.  On the other hand, we know of no construction (e.g., using string techniques) that imply the existence of $\cN=2$ theories with these types of CBs and empty flavor symmetry.

More recently, it has been pointed out in \cite{Argyres:2016yzz} that frozen $II^*$, $III^*$, and $IV^*$ singularities can arise upon discretely gauging certain special (non-flavor) symmetries of free vector multiplets.  

So, we must include the possibility of frozen $II^*$, $III^*$, and $IV^*$ singularities in our classification of possible rank-1 CB geometries and they are shown in table \ref{table:theories}.

\subsubsection*{Types $I_n$, $n>0$}

The $I_n$ singularities with $n>0$ have a CB parameter of dimension $\D(u)=1$ and at the singularity the low energy gauge coupling on the CB becomes free: $\lim_{u\to 0}\t(u) = i\infty$.  They therefore are naturally interpreted as IR free $\U(1)$ gauge theories.   The $I_n$ singularity has a ${T^n}$ EM-duality monodromy (see table \ref{Table:Kodaira}) and therefore corresponds to a $\U(1)$ theory with beta function with one-loop coefficient proportional to $n$ \cite{Seiberg:1994rs,Seiberg:1994aj}.   Choose the normalization of the $\U(1)$ gauge coupling so that a hypermultiplet with charge 1 contributes 1 to the one-loop beta function coefficient.  Then a $\U(1)$ gauge theory with $N$ massless charged hypermultiplets of (electric) charges $Q_a$, $a=1,\ldots,N$, gives an $I_n$ singularity with $n=\sum_{a=1}^N Q_a^2$.  

For instance, an $I_n$ singularity could be realized by a $\U(1)$ gauge theory with $n$ charge-1 hypermultiplets, or by one with a single charge-$\sqrt{n}$ hypermultiplet.  In the first case, the theory would have a $\U(n)$ flavor symmetry and thus $n$ distinct mass deformations, while in the second case the theory has a $\U(1)$ flavor symmetry and a single mass term.  (Electric and magnetic charges which are multiples of square roots of integers are allowed by the Dirac quantization condition; see the discussion of charge normalizations in appendix \ref{app:details} and section \ref{ssec:DQC}.)  

As discussed in section \ref{sec:localdefs}, a mass term deformation of an IR free $\U(1)$ gauge theory where the mass is for an overall $\U(1)$ flavor symmetry factor is the only kind of mass deformation which does not split the singularity, but instead only shifts its position on the CB.  This is easy to see since this deformation gives a common mass to all the hypermultiplets, and so can be undone by a shift in $u$ as in the discussion after \eqref{N1W}.  This shift deformation will be the \emph{only} deformation of the $\U(1)$ theory if it has only a single massless hypermultiplet, for if there are two or more hypermultiplets, then there will be additional deformations which change the relative masses of the hypermultiplets and split the singularity.  Therefore, for each $n>0$, there is one theory with an undeformable $I_n$ singularity, namely, the $\U(1)$ gauge theory with a single massless hypermultiplet of EM duality invariant charge $Q=\sqrt n$.

As in the discussion of the physical interpretation of the $I^*_0$ singularity above, there are conceivable alternate interpretations of frozen or undeformable $I_n$ singularities other than IR free $\U(1)$ gauge theories.   One possibility is that they are IR free theories but with some global discrete symmetry gauged.   We will present an argument immediately below suggesting that there is an obstruction to gauging a discrete subgroup of the flavor symmetry of an $\cN=2$ SCFT.  An alternative is to gauge other, non-flavor, discrete symmetry groups, as described in \cite{Argyres:2016yzz}.  However, as discussed there, frozen $I_n$ theories do not arise in this way.

Another possibility is that a frozen $I_n$ singularity comes from gauging a $\U(1)$ flavor symmetry of a rank-0 SCFT ``$X_0$''.  For this to produce an $I_n$ singularity, $X_0$ would have to contribute $n$ to the $\U(1)$ beta function coefficient, and the commutant of $\U(1)$ in the flavor symmetry algebra of $X_0$ would have to be empty (thus restricting the flavor symmetry of $X_0$ to be either $\U(1)$ or $\SU(2)$).  Again, no such rank-0 SCFT is known to exist.

So, to make progress, we make two additional assumptions, mentioned in the introduction: 
\begin{quote}
{\bf No rank 0 theories:}\ interacting $\cN=2$ SCFTs with no CB do not occur, and\\[2mm]
{\bf No discretely gauged flavor:}\ $\cN=2$ supersymmetric gaugings of discrete nonabelian flavor symmetries do not occur.\footnote{However, supersymmetric gaugings of certain discrete symmetries which include flavor \emph{outer} automorphisms are allowed, and are discussed in \cite{Argyres:2016yzz}.}
\end{quote}
Then all undeformable $I_n$ singularities have the interpretation as IR free $\U(1)$ gauge theories with a single massless hypermultiplet of charge $Q=\sqrt n$.  Our classification of possible rank-1 CB geometries shown in table \ref{table:theories} results from this assumption.  Clearly, the first assumption could be weakened, since to justify our classification we only need to assume the non-existence of certain very special rank-0 theories.  We now turn briefly to a discussion of the second assumption.

\subsubsection*{Gauging discrete flavor symmetries}

A way one might imagine constructing frozen or undeformable singularities is by starting with a deformable theory and gauging a discrete global symmetry to form a new theory.  The discretely gauged theory will have the same number of degrees of freedom (and hence the same rank) as the original theory, but only combinations of operators neutral under the discrete group will be allowed.  In particular, gauging a large enough discrete symmetry may disallow all of the relevant operators, leaving a frozen theory.  

We will now present some arguments indicating that this does not occur in $\cN=2$ supersymmetric theories when the discrete symmetry is part of the flavor symmetry of the theory (i.e., its generators commute with the R-symmetries and the outer-automorphisms of the flavor algebra).  Since these arguments are less than rigorous, it is important to point out that if the conclusion is wrong and discretely gauged frozen singularities are allowed, then our classification of safe deformations given in table \ref{table:theories} will have missed some possibilities.

Consider an SCFT with a continuous flavor symmetry algebra $F$ and, therefore, with rank$(F)$ mass deformations.  We can gauge a discrete subgroup of $F$, which can be taken as a  subgroup of the flavor inner automorphisms, Inn$(F)$, which act by conjugation by elements of the maximal torus of $F$.\footnote{Inn$(F)\simeq F/Z(F)$ where $Z(F)$ is the center of $F$.  But gauging generators in the center of $F$ cannot decrease the rank of $F$, so we can ignore the center for our purposes.}  Finite abelian subgroups $Z\subset\text{Inn}(F)$ fix a subalgebra $F_Z \subset F$ of the same rank: rank$F_Z = \text{rank}F$; see, e.g., ch.\ 11 of \cite{fuchs2003symmetries} for a discussion.  However, it is not too hard to see that non-abelian subgroups, $\G\subset\text{Inn}(F)$, can always be found that fix only the trivial subalgebra, $F_\G=0$.\footnote{For example, if $F=\SU(2)$, then Inn$(F)={\rm SU}(2)$ acting by conjugation on $F$.  The discrete $\Z_2\subset {\rm SU}(2)$ generated by the Pauli matrix $\s_1$ fixes $F_{\Z_2} \simeq \U(1)$, but the 8-element nonabelian binary dihedral subgroup $D_2 \subset {\rm SU}(2)$ generated by $\s_1$ and $\s_2$ fixes $F_{D_2}=0$.}  Thus gauging such a nonabelian discrete subgroup $\G\subset\text{Inn}(F)$ would eliminate the flavor symmetry and thus the mass deformations, giving a frozen theory.

This kind of construction allows us to construct frozen versions of all our lagrangian theories including the IR free ones, giving many new safe deformation patterns.  It is possible that these deformation patterns are actually realized among rank-1 $\cN=2$ SCFTs by starting with discretely gauged versions of lagrangian SCFTs then flowing with relevant operators or tuning marginal operators to arrive at new strongly coupled SCFTs.  In this sense, the classification of CB geometries that we present here is insensitive to the existence of theories with discretely gauged flavor symmetries:  for each deformed CB geometry that we constructed in table \ref{table:theories}, one would then have to ask which ``daughter'' geometries could arise from it by gauging a discrete subgroup of the flavor group of the UV theory.

An alternative to accepting this multiplication of the number of inequivalent lagrangian field theories is if there is some obstruction to the discrete gaugings.   For instance, one can try to realize gauging an inner automorphism of the flavor symmetry by constucting it as the low energy limit of a larger theory without discrete gaugings.   Take the original theory with flavor symmetry $F$, and weakly gauge $F$ (if necessary, add additional free hypermultiplets and let them transform under $F$ to make the $F$ gauge coupling IR free).  Then one can engineer a low energy theory with a subgroup $\G\subset F$ gauged by adding a Higgs field with vev invariant under $\G$ and integrating it out.  However, in $\cN=2$ theories this does not work since the Higgs will be part of a hypermultiplet or vector multiplet and its vev cannot be fixed to a specific value; instead the theory will have a whole moduli space of allowed vevs and accompanying massless degrees of freedom, and so will not flow in the IR to the desired theory with $\G\subset F$ discretely gauged.  But, this argument may not be convincing since one can imagine engineering the theory as above but without demanding $\cN=2$ supersymmetry at the Higgs scale; the $\cN=2$ supersymmetry would then be an accidental symmetry in the IR.

The above discussion does not apply to flavor outer automorphisms.   However, outer automorphism groups for reductive Lie algebras are relatively small and gauging them can reduce the rank of $F$ but not freeze it completely \cite{fuchs2003symmetries}.  Thus, though interesting, even if gauging outer automorphisms of $F$ by themselves is allowed in $\cN=2$ theories, it does not affect our above classification of undeformable theories.

Finally note that, following the work of \cite{Garcia-Etxebarria:2015wns}, it was realized that one can gauge certain $\Z_m$ symmetries whose generator involves a combination of a flavor outer automorphism transformation, a $\U(1)_R$ symmetry transformation, and an $SL(2,\Z)$ EM-duality transformation \cite{Argyres:2016yzz}.  This observation does in fact give additional possibilities for frozen $I_n^*$ singularities, to which we now turn. 

\subsubsection*{Types $I^*_n$, $n>0$}

The $I^*_n$ singularities with $n>0$ have a CB parameter of dimension $\D(u)=2$ and at the singularity the low energy gauge coupling on the CB becomes free: $\lim_{u\to 0}\t(u) = i\infty$.  They therefore are naturally interpreted as IR free $\SU(2)$ gauge theories.   For such a theory to be undeformable it must have no relevant $\cN=2$ deformations.  Since $\D(u)=2$, there are no relevant chiral term deformations \eqref{chiraldef}.  So the possible relevant deformations are mass terms of the type \eqref{CFTmassterm} which are associated to flavor symmetries.  An undeformable $I^*_n$ singularity must therefore correspond to an IR free $\SU(2)$ gauge theory with no flavor symmetry.  We will now show that, perhaps surprisingly, such frozen IR free theories exist, and imply that there is at least one frozen theory giving an $I^*_n$ singularity for each $n>0$.

Consider an $\cN=2$ $\SU(2)$ gauge theory with massless half-hypermultiplets (which have the field content of an $\cN=1$ chiral multiplet) transforming in a reducible representation $\oplus_{i=1}^N \bR_i$ where $\bR$ denotes the irreducible $\SU(2)$ representation of dimension $R$.  Then the coefficient of the one-loop beta function is proportional to 
\begin{align}\label{su2b0}
b_0 =- T({\bf 3})+ \frac12 \sum_{i=1}^N T(\bR_i)
\end{align}
where $T(\bR)$ is the Dynkin index of the representation $\bR$.   Choose a normalization of $\SU(2)$ generators where the $t_3$ generator in irrep $\bR$ is
\begin{align}\label{su2norm}
t_3(\bR) := \frac12 \begin{pmatrix} R-1 & & & & \\
& R-3 & & & \\ & & \ddots & & \\ & & & -R+3 & \\ & & & & -R +1
\end{pmatrix} .
\end{align}
Then a standard normalization for the Dynkin index is
\begin{align}\label{su2dynkin}
T(\bR) := 2\, {\tr}_\bR(t_3^2) = \frac{1}{6}(R-1)R(R+1),
\end{align}
so $T({\bf 2})=1$, $T({\bf 3})=4$, $T({\bf 4})=10$, etc.  Therefore, in this normalization, the theory is IR free whenever $\sum_{i=1}^N T(\bR_i) > 8$.

Even-$R$ $\SU(2)$ representations are symplectic (pseudo-real) while those with odd $R$ are orthogonal (real).  An $\cN=2$ gauge theory can be consistently coupled in an $\cN=2$ supersymmetric way to any number $N$ (even or odd) of half-hypermultiplets in a given symplectic irreducible representation, giving an $\SO(N)$ flavor symmetry factor.  For an orthogonal representation, only an even number, $2N$, of half-hypers can be coupled, giving an $\Sp(2N)$ flavor factor \cite{McOrist:2013bga}.\footnote{An easy way to see this is to view the gauge theory as a gauged linear sigma model with the complex scalars of the half-hypers forming the target space.  Then as long as the target space has a hyperk\"ahler structure with respect to which there is a triholomorphic $\SU(2)$ isometry, then this isometry can be gauged in an $\cN=2$ supersymmetric way \cite{Sierra:1983uh, Hull:1985pq, Hitchin:1986ea}.  For example, if there are $N$ half-hypers in the $\bf 2r$ of $\SU(2)$, then before gauging there are $4rN$ free real scalars, so the target space has an $\SO(4rN)\supset\SU(2)_R\times\Sp(2Nr)$ isometry under which they transform as the ${\bf 4rN}=({\bf2},{\bf 2Nr})$.   The $\SU(2)_R$ factor is the R-symmetry which ensures the triholomorphic nature of the $\Sp(2Nr)$ flavor symmetry.   Now gauge the following $\SU(2)$ subalgebra of the $\Sp(2Nr)$ factor: $\Sp(2Nr) \supset \Sp(2r)\times\SO(N) \supset \SU(2)_\text{gauge} \times \SO(N)$ such that the $\bf 2Nr$ decomposes as ${\bf 2Nr} = ({\bf 2r}, {\bf N})=({\bf 2r}, {\bf N})$.  The last factor is the $\SO(N)$ global flavor symmetry.  A similar argument gives the $\Sp(2N)$ flavor symmetry for $2N$ half-hypers in an orthogonal $\boldsymbol{2r{+}1}$ irrep, since $\SO(2(2r{+}1)2N) \supset \SU(2)_R \times \SU(2)_\text{gauge}\times \Sp(2N)$ under which $\boldsymbol{4(2r{+}1)}{\bf N}=({\bf 2},\boldsymbol{2r{+}1},{\bf 2N})$.}  Also, these theories have to be free of the $\Z_2$ global anomaly \cite{Witten:1982fp} which is the case if the total Dynkin index of all the Weyl fermions in symplectic representations is even in the normalization of \eqref{su2dynkin}.  It is not too hard to check that with these constraints on the representation content, the one-loop beta function coefficient $b_0$ defined in \eqref{su2b0} is an integer.

A single half-hypermultiplet in any symplectic irrep will contribute an $\SO(1)$ --- i.e., empty --- flavor factor.  Thus IR free $\SU(2)$ gauge theories with only  this kind of matter content will have no flavor symmetry, and thus will have no $\cN=2$ relevant deformations.  Since $T({\bf 2r}) > 8$ for $r>1$, all such theories will be IR free except for the one with a single half-hypermultiplet in the $\bf 2$ irrep.  The index of irreps $\bf R$ with $R=4k$ are even while those with $R=4k+2$ are odd.  Therefore the $\Z_2$ anomaly-free IR free frozen $\SU(2)$ gauge theories are ones with at most a single half-hypermultiplet in each even-dimensional $\SU(2)$ irrep and with an even total number of half-hypermultiplets in irreps whose dimensions are 2 mod 4.  Thus the list of frozen $\SU(2)$ theories with smallest one-loop beta functions have half-hypermultiplets in representations
\begin{align}\label{frozensu2s}
&{\bf 4} & b_0 =&\ 1 \nonumber\\
&{\bf 2}\oplus{\bf 6} & b_0 =&\ 14 \nonumber\\
&{\bf 2}\oplus{\bf 4}\oplus{\bf 6} & b_0 =&\ 19 \\
&{\bf 8} & b_0 =&\ 38 \nonumber\\
& \ldots & \ldots &\nonumber
\end{align}
It will turn out that only the first theory in this list will play a role in the classification of rank-1 CB deformations.\footnote{Upon breaking $\cN=2$ to $\cN=1$ by giving a mass to the adjoint chiral multiplet in the $\SU(2)$ vector multiplet, the theory with a single half-hypermutiplet in the $\bf 4$ (or spin-3/2) irrep flows to the interesting theory discussed in \cite{Intriligator:1994rx,Poppitz:2009kz}.}   To our knowledge the existence of such frozen IR free gauge theories has not been noted before.

The next question is which $I^*_n$ singularities correspond to each of these frozen theories.  This is basically a question of determining charge normalizations, that is, the normalization of the $\SU(2)$ gauge generators relative to that of the generator of the low energy $\U(1)$ gauge symmetry on the CB.

Note first that the order of vanishing of the discriminant of an $I^*_n$ singularity at $u=0$ is $n+6$, and so its maximal deformation splits it into $n+6$ $I_1$ singularities.  It is straightforward to write this deformation as the general complex deformation of the SW curve for the $I^*_n$ singularity listed in table \ref{Table:Kodaira}, and check that it has $n+4$ independent deformation parameters with dimensions $\{2, 4, \ldots, 2n+6; n+4\}$, see \cite{Argyres:2010py}.  These are the dimensions of the adjoint Casimirs of $\SO(2n+8)$ so this must be the flavor symmetry of the maximal deformation of the $I^*_n$ singularity, and the corresponding IR free $\SU(2)$ gauge theory is the one with $2n+8$ half-hypermultiplets in the $\bf 2$ irrep.  This theory has one-loop beta function coefficient $b_0=n$ in our normalization \eqref{su2b0}.

Now, as discussed above, an $I_1$ singularity has the unique interpretation as an IR free $\U(1)$ gauge theory with a single massless hypermultiplet with EM duality invariant charge $Q=1$ (in the charge normalization explained in appendix \ref{app:details}).  On the other hand, since the $I_1$ singularities arise from a general mass deformation of an $\SU(2)$ theory with hypermultiplets in the $\bf 2$ irrep (which have $t_3$ eigenvalues $\pm\frac12$), it follows upon higgsing $\SU(2)\to\U(1)$ on the CB that $Q=2 t_3$.  (More precisely, $Q$ can be twice any of the eigenvalues of $t_3$.)

We derived this connection between our $\U(1)$ and $\SU(2)$ charge normalizations by identifying the maximal deformation of an $I_n^*$ singularity with a certain $\SU(2)$ gauge theory.  More generally, any allowed $\SU(2)$ representation for massless hypermultiplets with $b_0=n$ will give an IR free theory whose CB is described by an $I^*_n$ singularity.  The different possible choices of hypermultiplet representation with $b_0=n$ will have different mass deformations and so will give rise to different deformations of the $I^*_n$ singularity --- indeed, this is a lagrangian illustration of our basic motivation for studying CB deformations in the first place. 

However, this analysis depended on our initial choice of normalization of the $\SU(2)$ generators \eqref{su2norm}, which was arbitrary.  If we rescale the generators by a factor $a$ so that $t_3 \to a t_3$, then $\U(1)$ charges on the CB will be rescaled by the same factor, and the one-loop beta function coefficient (which is proportional to a sum of squares of charges) will be rescaled by $a^2$.  Thus an $I_n^*$ singularity will correspond to such a charge-rescaled $\SU(2)$ theory if
\begin{align}\label{I*n-su2}
n = a^2 b_0,
\end{align}
and will have $\U(1)$ charges on the CB given by 
\begin{align}\label{u1su2norm}
Q=2 a t_3 .
\end{align}
To be clear: we are keeping our definitions of $b_0$, $t_3$, and $T(\bR)$ fixed to be \eqref{su2b0}, \eqref{su2norm}, and\eqref{su2dynkin} in this discussion, and are only letting the positive parameter $a$ vary.  

But not every value of $a$ is allowed, because the set of allowed $\U(1)$ charges on the CB is constrained by the Dirac quantization condition to be (integer multiples of) square roots of positive integers, $Q^2 \in \N$; see the discussion in appendix \ref{app:details}.   Combining this condition with \eqref{su2norm} and \eqref{u1su2norm} we arrive at the following possibilities:
\begin{align}\label{a2cases}
a^2 &\in \tfrac14\, \N & &\text{if all hypermultiplets are in orthogonal irreps,}
\nonumber\\
a^2 &\in \N & &\text{otherwise.}
\end{align} 
The first case is allowed since the orthogonal irreps are odd-dimensional, and so the eigenvalues of $t_3$ are integers by \eqref{su2norm}.  Note that this is the case where the global form of the gauge group is allowed to be ${\rm SO}(3) \cong {\rm SU}(2)/\Z_2$.  In the second case in \eqref{a2cases} the global form of the gauge group must be ${\rm SU}(2)$.  The extra freedom in the normalization of the charges summarized in \eqref{I*n-su2}, \eqref{u1su2norm}, and \eqref{a2cases} will play an important role in the next section and it will be exploited in detail in \cite{Argyres:2015ccharges}.  For now we focus on its implications for the question of frozen $I^*_n$ singularities.

We have seen that a frozen $\SU(2)$ theory has a half-hypermutiplet in a symplectic representation.  Therefore $a^2\in\N$ by \eqref{a2cases}.   Consider the first frozen $\SU(2)$ theory in \eqref{frozensu2s} with $b_0=1$ which has one half-hypermultiplet in the $\bf 4$.  We see that it can give an $I_n^*$ singularity if we take $a = \sqrt{n}$, in which case the charges on the CB will be quantized in units of $Q=\sqrt{n}$.  Next, the $b_0=14$ theory with half-hypermultiplets in ${\bf2}\oplus{\bf6}$ can give rise to $I^*_{14n}$ singularities with $a=Q=\sqrt{n}$.  And so forth.  Thus, for example, the $I^*_{38}$ singularity corresponds to three different frozen $\SU(2)$ theories: one with a half-hypermultiplet in the $\bf8$ and charges normalized to $Q=1$; one with half-hypermultiplets in ${\bf2}\oplus{\bf4}\oplus{\bf6}$ and charges normalized to $Q=\sqrt2$; and one with a half-hypermultiplet in $\bf4$ with charges normalized to $Q=\sqrt{38}$.

As discussed in the last subsection, possible deformations of the scale-invariant Kodaira singularities can be classified by their deformation patterns which are constrained by the requirement that the sum of the orders of vanishing of the discriminants, ord$_0(D_x)$, of the singularities does not change upon deformation.  Since the largest ord$_0(D_x)$ for a scale-invariant singularity is 10, the only frozen $I_n^*$ singularties that can appear are those with $n\in\{1,2,3\}$ since ord$_0(D_x)$ of $I_n^*$ is $6+n$.  These can only correspond to the frozen $\SU(2)$ theory with a half-hypermultiplet in the $\bf4$.  Thus, such a frozen IR-free $\SU(2)$ gauge theory can only possibly appear as one the following singularities in our classification (table \ref{table:theories})
\begin{align}\label{IRFlist}
I^*_{1\ Q{=}1}, \quad
I^*_{2\ Q{=}\sqrt2}, \quad
I^*_{3\ Q{=}\sqrt3}.
\end{align}
Here we have indicated the unit of charge quantization, $Q$, for each singularity realized in this way.  In table \ref{table:theories}, these $Q=\sqrt n$ subscripts are shown whenever they are different from $Q=1$.

Frozen $I^*_n$ singularities can be realized in a different way.  As shown in \cite{Argyres:2016yzz}, the following ones arise from discretely gauging certain non-flavor symmetries of free vector multiplets:
\begin{align}\label{FDGlist}
I^*_{0\ Q{=}1}, \quad
I^*_{1\ Q{=}\sqrt2}, \quad
I^*_{2\ Q{=}1}.
\end{align}

Finally, one could also imagine the existence of non-lagrangian frozen $I^*_n$ singularities arising from gauging an $\SU(2)$ sub-algebra of the flavor symmetry of an interacting rank-0 SCFT.  In classifying all possible rank-1 planar CB geometries in table \ref{table:theories}, we allowed for this possibility by listing all geometries with frozen $I^*_n$ singularities without putting any restriction on what values of the chrage quantization unit, $Q$, were realized.  In the resulting list there are two entries (numbers 6 and 8) for which the frozen $I^*_n$ singularities are $I^*_{1\ Q{=}\sqrt2}$ and $I^*_{2\ Q{=}1}$.  These do not appear in either \eqref{IRFlist} or \eqref{FDGlist}, so, if they exist, must correspond to more exotic non-lagrangian frozen theories of the type just described. With the \emph{no rank-0} assumption we can discard such possibility.

\begin{center}
\rule[1mm]{2cm}{.4pt}\hspace{1cm}$\circ$\hspace{1cm} \rule[1mm]{2cm}{.4pt}
\end{center}

In summary, the safely irrelevant conjecture together with constraints from the allowed form of relevant deformations of $\cN=2$ SCFTs implies that under generic deformations, scale invariant CB singularities must split into a collection of undeformable $I_n$ and $I^*_n$ singularities, and perhaps also frozen $II^*$, $III^*$, and $IV^*$ singularities.   Restricting to deformation patterns only involving these singularities reduces the number of possible deformation patterns from 383 to 122.   But the fact that these generic IR singularities must all be undeformable allows us to determine the charges of light BPS states on the CB, and thereby further dramatically reduce the number of consistent geometries using the Dirac quantization condition.

Before we turn to the Dirac quantization constraint, however, we pause to discuss some of the implications of the assumed lack of dangerously irrelevant operators for the counting of relevant operators and for the appearance of flavor symmetries in the presence of a moduli space of vacua.

\subsection{Operator counting on moduli spaces\label{ssec:sings.1}}

The operator counting relation \eqref{relopcounting} must be modified in the presence of a CB of vacua.  In the absence of dangerously irrelevant operators at the UV fixed point we have $d_{\text{UV}\to\text{IR}}=0$ and \eqref{relopcounting} gives as the number of independent relevant operators in the IR
\begin{align}\label{relop2}
r_\text{IR} = r_\text{UV} - r_{\text{UV}\to\text{IR}},
\end{align}
where $r_\text{UV}$ is the dimension of the space of relevant operators in the UV and $r_{\text{UV}\to\text{IR}}$ is the dimension of the subspace of those operators which flow to the given IR fixed point.  But, in the presence of a CB there is not a single IR fixed point, but rather a whole CB of IR fixed points.   So we should re-interpret \eqref{relop2} by replacing the notion of an IR fixed point with the whole CB geometry.  In particular, we should interpret $r_\text{IR}$ as the dimension of the space of relevant operators of the theory with a given non-scale-invariant CB.  

Then, since every distinct relevant operator (including the choice of its crossover scale) leads to a different CB, the subspace of operators which flows to a given CB geometry is just a point, so its dimension vanishes, $r_{\text{UV}\to\text{IR}}=0$.  Thus, in the absence of dangerously irrelevant operators at a UV fixed point, we have from \eqref{relop2} that the number of relevant operators in the IR does not change from the number in the UV:
\begin{align}\label{relop3}
r_\text{IR} = r_\text{UV}.
\end{align}
For mass terms this agrees with the local operator analysis of the last section, since mass terms transform in the adjoint of the flavor symmetry, and so leave an unbroken $\U(1)^{\text{rank}(F)}$ flavor symmetry in the IR.  Therefore the number, rank$(F)$, of independent flavor deformations remains the same.  When the UV theory has chiral terms, \eqref{relop3} implies that there must be a corresponding relevant deformation of the IR CB geometry. 

As discussed in section \ref{sec:CBdefs} local relevant deformations of a theory with a CB show up as near deformations of the CB geometry, i.e., ones which do not change the asymptotic geometry at large CB distances.  By the discussion of the last subsection, a generic deformed CB will have some number of disctinct singularities which cannot be further split.  So any further  deformation can only shift their relative positions.  In the vicinity of a singularity, by the discussion of local $\cN=2$ deformations in section \ref{sec:CBdefs}, a shifting deformation is a relevant mass term for a $\U(1)$ flavor symmetry at an $I_n$-type singularity.  For the other Kodaira types of singularity, a CB shift does not correspond to any local relevant operator.  

Since the safely irrelevant conjecture implies that there must be on the generically deformed CB only either ``frozen" singularities with no flavor symmetry or undeformable $I_n$ type singularities with $\U(1)$ flavor symmetry, it follows that there must be at least rank$(F)$ of the $I_n$ type singularities to account for the $\U(1)^{\text{rank}(F)}$ generically unbroken flavor symmetry.  If there are only $I_n$ type singularities, then there must be at least rank$(F)+1$ of them since one overall $\U(1)$ flavor symmetry is a gauge redundancy, corresponding to an unobservable shift of the entire CB; it is only the relative shifts of the singularities which are physical.  The number of $I_n$ singularities may exceed these lower bounds for a couple of reasons.   First, the global electric and magnetic parts of the low energy $\U(1)$ gauge invariance of the Coulomb branch can mix with the flavor $\U(1)$'s \cite{Seiberg:1994aj} giving up to two\footnote{This  number depends on the (electric, magnetic) charge vectors of the light hypermultplets at the singularities.  If these vectors are all commensurate then only one linear combination of the global electric and magnetic $\U(1)$'s acts on the CB.  If some are incommensurate then both $\U(1)$'s act independently on the CB.} additional $I_n$ singularities.  Second, the flavor $\U(1)$'s of the different singularities might simply not be independent of one another.   For example, the $\U(1)$ flavor symmetries of two distinct singularities might act identically in the theory and so be the same $\U(1)$ symmetry. 

There are non-generic deformations of the CB where some simple (i.e., non-abelian) factors of the flavor symmetry remain unbroken.  As discussed earlier, mass term deformations which break non-abelian flavor symmetries necessarily split singularities.  These splittable singularities thus each carry the action of one or more simple factors of the partially broken flavor symmetry.  As in the $\U(1)$ case in the previous paragraph, it is possible that there be more than one singularity that carries the same simple flavor symmetry factor, but that they do not act independently.

We have emphasized above (and in section \ref{sec:CBdefs}) a point of view which interprets the whole moduli space of a deformed theory as the appropriate object to replace the notion of an IR fixed point when analyzing RG flows.  An RG flow in this picture is thus a deformation of a scale invariant moduli space into a non-scale-invariant one which retains the ``imprint" of the crossover scales of the RG flow.  One could, instead, interpret RG flows in the more traditional way as a path between two fixed points, where now the UV fixed point is the conformal vacuum (superselection sector) at the tip of the conical (scale invariant) moduli space, and the IR fixed point is a particular choice of vacuum on the deformed moduli space.  Thus, in this picture we specify not only the UV theory and some local relevant deformation operator, but also a particular ``relevant" nonlocal deformation operator which picks out the particular vacuum superselection sector.  We have avoided the latter point of view because it makes it difficult to clearly distinguish between safely and dangerously irrelevant operators.

For example, in a theory with a CB vev $u$ of scaling dimension $\D(u)=r$ with $1<r<2$, there is a local relevant chiral term of the form \eqref{chiraldef} where $\vev{X}=u$  for $X\in\cE_{r\,(0,0)}$.  This theory therefore also has a local irrelevant operator
\begin{align}\label{}
\d_\cP = \n \int d^4x\, \tQ^4 X^2 + \text{h.c.} 
\qquad \text{with}\qquad
X^2 \in \cE_{2r\,(0,0)}.
\end{align}
If one interprets a scale invariant (conical) moduli space in terms of  RG flows from the conformal vacuum at the vertex of the cone to a given IR vacuum point away from the vertex, then the above irrelevant operator looks dangerously irrelevant: upon turning it on and turning on the nonlocal operator that sets $\vev{X}=u$, the above operator induces the operator
\begin{align}\label{}
\d_\cS= 2u\n \int d^4x\, \tQ^4 X + \text{h.c.}
\end{align}
which is a local relevant chiral term \eqref{chiraldef} with $\m = 2u\n$.  But this induced ``relevant" operator, associated with a crossover scale increasing as $|u|$, gives a far deformation of the CB geometry as in \eqref{fardef}.  So from the point of view of its effects on the CB developed in the last section, it should simply be understood as the typical effect of any irrelevant operator and not as the sign of a dangerously irrelevant operator.

It is to avoid this type of confusion of the action of the spacetime nonlocal ``choose-a-vev" operation with the flows generated by local operators that we keep track of the change of the whole moduli space geometry.  This infomation makes the distinction between safely and dangerously irrelevant operators apparent, as illustrated in figure \ref{fig:RG2}.

\subsection{Action of the flavor symmetry on families of deformed CBs\label{ssec:sings.2}}

Upon deforming the SCFT by mass terms the flavor symmetry is broken to $\U(1)^{\text{rank}(F)}\rtimes \text{Weyl}(F)$, so there are rank$(F)$ linearly independent complex mass parameters, $m_i$, that the family of deformed CB geometries can depend on.  The discrete Weyl symmetry further restricts these parameters to appear in the CB geometry only in Weyl$(F)$-invariant polynomial combinations, $M_a(m_i)$.  Thus there will be rank$(F)$ algebraically independent deformation parameters, $\{M_a\}$, of integer dimensions given by their degrees as polynomials in the linear masses.

Although there is no unique way of choosing a basis of the $M_a$, the set of their dimensions, $\{\D(M_a)\}$, is unique, and, furthermore, these dimensions completely determine Weyl$(F)$.  In fact, the Chevalley-Shephard-Todd theorem \cite{Shephard1954,Chevalley1955} tells us that the set of degrees of the polynomials in the linear masses uniquely determines the discrete complex reflection group symmetry acting linearly on the linear masses which leaves the $M_a$ invariant.  But most complex reflection groups are not Weyl groups of compact reductive Lie algebras.\footnote{Weyl groups are real reflection groups which are lattice isometries.}

The Coleman-Mandula theorem \cite{Coleman:1967ad} (and its conformal version \cite{Maldacena:2011jn}) states that unitary local CFTs can only have compact reductive Lie algebras as internal (flavor) symmetries, and so we expect physical CB deformations should have only Weyl group symmetries.  In fact we find that for all deformations consistent with the safely irrelevant conjecture the reflection symmetry is indeed Weyl.  If one lifts this condition, and allows deformation patterns that include type $II$, $III$, or $IV$ Kodaira singularities, it is easy to construct examples with non-Weyl reflection group symmetry (see \cite{Argyres:2015gha} for some examples).\footnote{Higher-rank examples of SK CB geometries with non-Weyl reflection symmetries have appeared implicitly in \cite{Bordner:1999us,Hurtubise:2001Mh} where the complex integrable systems corresponding to such SK geometries is constructed.}  This is additional evidence for the correctness of the safely irrelevant conjecture.

Thus the Weyl group of the flavor symmetry of a CFT is encoded in the dependence of its CB geometry on the deformation parameters.  This dependence is not directly related in any simple way to the topological data of the deformation (its pattern of splitting into undeformable singularities and the $\SL(2,\Z)$ monodromies of those singularities).  In a companion paper \cite{Argyres:2015gha} we show how to explicitly construct CB geometries realizing this topological data.  From these solutions the Weyl group can then be read off.  

The Weyl group alone does not uniquely determine the flavor symmetry algebra since the Weyl groups of the rank-$n$ simple Lie algebras $B_n$ and $C_n$ are identical.\footnote{$B_n = \SO(2n+1)$ and $C_n=\Sp(2n)$.}  The flavor symmetry enters the CB effective theory in a different way through the spectrum of allowed BPS masses.  As reviewed in appendix \ref{app:details}, the BPS masses get contributions from linear integral combinations of the $m_i$ mass parameters corresponding to the flavor root lattice, $\L_F$.   The flavor-invariant \emph{metric} structure of this lattice, however, does not enter in the CB geometry, and without this data the only information about the flavor symmetry that can be deduced is its rank: $\L_F\simeq\Z^{\text{rank}(F)}$.  But, by combining the Weyl symmetry action on the masses together with this lattice, one can reconstruct (up to an overall scale factor) the metric structure on $\L_F$, and thus use it to distinguish between $B_n$ and $C_n$ flavor symmetries.  This is explained more fully in \cite{Argyres:2015gha} where this procedure is used to deduce the flavor symmetries associated to the CB deformations listed in table \ref{table:theories}.

Additionally, the discrete Weyl group, $\G$, deduced from the CB geometry might be larger than the Weyl group of the SCFT flavor symmetry Lie algebra, $F$.  In fact, it is possible for $\G = \G' \ltimes \text{Weyl}(F)$, so that the actual flavor symmetry Lie algebra is of smaller dimension than ``expected'', and the subgroup $\G' \subset \G$ acts as an additional discrete flavor symmetry of the theory.  These possible reductions of the SCFT flavor symmetry from the maximal symmetry consistent with the CB geometry are classified and analysed further in \cite{Argyres:2016xua}.


\section{Dirac quantization, special K\"ahler, and RG flow conditions\label{sec:udefsing}}

In the previous section we concluded that upon turning on a generic relevant deformation, the scale invariant Kodaira singlarities in table \ref{Table:Kodaira} must split into singularities which cannot be further deformed.  We identified undeformable singularities as being one of the $\{I_n,I^*_n,II^*,III^*,IV^*\}$ type singularities.  We can thus classify the possible safe deformations of the seven scale-invariant Kodaira singularities by their \emph{deformation pattern}, which is the list of undeformable singularities that they split into upon turning on a generic relevant deformation.  As discussed earlier, the sum of the orders of vanishing of the discriminants of the Seiberg-Witten curves associated to each singularity cannot change under deformation.  This then limits the number of possible deformation patterns to 122.

These deformation patterns are not all physically allowed.  There are further constraints that they must satisfy coming from demanding low energy $\cN=2$ supersymmetry, and from demanding consistent low energy field theory interpretations.   The low energy supersymmetry constraint is satisfied if the deformed CB is special K\"ahler.  As reviewed in appendix \ref{app:details} this means that there should be consistent EM duality monodromies around the singularities and that a SW 1-form exists with residues which are integral linear combinations of the linear mass parameters.  A consistent low energy field theory interpretation exists if at singularities on the CB there are approriate massless BPS states in the theory, and if the charges of the BPS states satisfy the Dirac quantization condition.  It turns out that all these requirements are independent in the sense that examples can be constructed where only one of them fails.  Furthermore, these requirements should not only hold for generic mass deformations, but for \emph{all} values of the mass deformation parameters.  This last is equivalent to demanding that the low energy consistency requirements are satisfied for the whole web of theories connected by RG flows.

\subsection{Classification strategy}

An efficient way to impose these constraints is in order of increasing difficulty of computation, so as to eliminate as many inconsistent deformation patterns with as little work as possible.  This means that we should impose the conditions as follows.
\begin{description}
\item[1. Dirac quantization condition:]
The undeformable $I_n$ singularities correspond to very specific IR free theories contributing states with electric and magnetic $\U(1)$ charges on the CB which are (integer multiples of) $Q=\sqrt{n}$.   Dirac quantization implies that all charges must be commensurate.  This immediately eliminates all but 43 deformation patterns as unphysical.
\item[2. Special K\"ahler conditions:]
There are two parts to this condition.  

The first is the necessary condition that the total EM duality monodromy around the undeformed singularity should be realizable as the product of the monodromies around all the singularities in the deformation pattern; see figure \ref{fig:monod}.  There are some simple, though somewhat technical, tests for this condition that further eliminate all but 28 deformation patterns as unphysical.   

The second part is the construction of the SW curve and 1-form satisfying the special K\"ahler conditions reviewed in appendix \ref{app:details}.  This step is quite technical and requires computer help.  However, in all 28 cases we find a unique special K\"ahler solution.  They are listed in table \ref{table:theories}.  Furthermore, all the solutions turn out to be related to restrictions of the maximal deformation solutions in a simple way, as described above in section \ref{SKsplits}.  The explicit solution allows us to deduce the maximal flavor symmetry of the underlying SCFT, as described in \ref{ssec:sings.2}.  The result is that for 27 of the 28 remaining deformation patterns we can deduce a unique maximal flavor symmetry, and the remaining one is ambiguous.  
\item[3. RG flow condition:] 
Upon turning on mass terms the flavor symmetry of the SCFT will be broken in various ways depending on the pattern of masses.  For special patterns of masses, some of the undeformable singularities will merge into other singularities on the CB.  A necessary consistency condition is that for any pattern of masses the resulting singularities have an interpretation in terms of either IR free or conformal theories with the required flavor symmetry to account for the unbroken part of the flavor symmetry.  Testing this condition is also technical, as it requires the explicit mass-dependence of the SW curves.  The end result is that of the 28 special K\"ahler deformation patterns, 3 fail to  meet this requirement with their maximal flavor symmetry assignment.  These are the ones whose entries in the maximal flavor symmetry column are shaded red in table \ref{table:theories}.
\end{description}
In the remainder of this section we will discuss each of these conditions in turn.  The Dirac quantization constraint is simple to implement, so will be fully described here.  The details of the implementation of the special K\"ahler and RG flow conditions will be left to a companion paper \cite{Argyres:2015gha}, where explicit expressions for the SW curves and 1-forms for all the special K\"ahler deformations will be given.  Here we just summarize the results of imposing these constraints and comment on a few interesting features of the solutions.

\subsection{Dirac quantization condition\label{ssec:DQC}}

We have seen in section \ref{froSing} that only the type $I_n$ undeformable singularities for $1\le n \le 9$ and the type $I^*_n$ frozen singularities for $0\le n \le 3$ can appear among deformation patterns of the scale invariant Kodaira singularities.  Also the $III^*$ and $IV^*$ singularities may appear if there exist corresponding frozen SCFTs.  

The undeformable $I_n$ singularities correspond to very specific IR free theories contributing dyonic states with electric and magnetic $\U(1)$ charges on the CB which are integer multiples of $Q=\sqrt{n}$.  If a given dyon has magnetic and electric charges $(p,q)$, then $Q$ is the EM duality invariant charge defined by $Q^2 = \gcd(p^2,q^2)$; see the discussion in appendix \ref{app:details}.  The Dirac quantization condition implies that the invariant charges of all states in the theory must be commensurate.  In particular, the charges of the massless hypermultiplets at each $I_n$ singularity on the CB must be commensurate.\footnote{Charges at different points on the Coulomb branch are measured relative to different $\U(1)$ gauge couplings, so one really needs to compare charges at the same point.  Even moving adiabatically on the Coulomb branch, charged states may decay upon crossing a curve of marginal stability.  However, by charge conservation, and since this process only changes a charged 1-particle state into an integer number of charged 1-particle states, the commensurateness (or not) of the charge basis cannot change.}  

This gives a simple but strong constraint on the allowed deformation patterns.  For instance, the deformation pattern $II^* \to \{ I_1, I_1, I_8\}$ is allowed by the condition that the sum of the orders of vanishing of the discriminants of the singularities on the two sides are equal (see table \ref{Table:Kodaira}).  But the undeformable $I_1$ singularities are $\U(1)$ gauge theories with massless hypermultiplets of invariant charge 1, while the undeformable $I_8$ singularity is a $\U(1)$ gauge theory with a massless hypermultiplet of invariant charge $2\sqrt2$.  Therefore such a deformation pattern is not compatible with the Dirac quantization condition.

\begin{table}[tbp]
\centering
\begin{tabular}{|cll|}
\hline
sing.& ord$_0(D_x)$ & deformation pattern \\ 
\hline
$II^*$ & 10  & $\{{I_1}^{10}\},\ \{{I_1}^6, I_4\},\ 
\{{I_1}^2, I_4^2\},\ \{{I_1}^4,I_0^*\},\ \{{I_1}^3,I_1^*\},\ \{I_3,I_1^*\},\ \{{I_1}^2,I_2^*\}$, \\
&&$\qquad \{I_1,I_3^*\},\ \{{I_1}^2, IV^*\},\ 
\{I_2, IV^*\},\ \{I_1,III^*\},\ \color{black!40} 
\{I_1, I_9\},\ \{I_2^5\}$, \\
&&$\qquad \color{black!40} \{I_2, I_8\},\ \{I_5^2\},\ 
\{I_4,I_0^*\},\ \{{I_2}^2,I_0^*\},\ \{I_2,I_2^*\}$ \\
$III^*$ & 9    & $\{{I_1}^9\},\ \{{I_1}^5, I_4\},\ 
\{{I_1}^3,I^*_0\},\ \{{I_1}^2,I^*_1\},\ \{I_2,I^*_1\},\ 
\{I_1,I^*_2\},\ \{I_1, IV^*\}$, \\ 
&&$\qquad\color{black!40}\{I_1, {I_4}^2\},\ \{{I_3}^3\},\ \{I_3,I^*_0\}$  \\
$IV^*$ & 8    & $\{{I_1}^8\},\ \{{I_1}^4, I_4\},\ \{{I_1}^2,I^*_0\},\ \{I_1,I^*_1\},\ \color{black!40}\{{I_4}^2\},\ 
\{{I_2}^4\},\ \{I_2,I^*_0\}$  \\
$I_0^*$ & 6 & $\{{I_1}^6\},\ \{{I_1}^2, I_4\},\ \{{I_2}^3\},\ 
\color{black!40}\{{I_3}^2\}$ \\
$IV$ & 4 & $\{{I_1}^4\},\ 
\color{black!40}\{{I_2}^2\}$ \\
$III$ & 3 & $\{{I_1}^3\}$ \\
$II$ & 2 & $\{{I_1}^2\}$ \\
\hline
\end{tabular}
\caption{List of deformation patterns of Kodaira singularities allowed by the Dirac quantization condition.  The ones in grey are incompatible with the $\SL(2,\Z)$ monodromies of the Kodaira singularities.\label{t2}}
\end{table}

It is now a simple matter to list all the commensurate deformation patterns satisfying the Dirac quantization constraint.  The resulting list of 31 deformation patterns is shown in table \ref{t2}.  Note that some of these commensurate-charge deformation patterns have non-integral invariant charges.

In table \ref{t2} we have included deformation patterns which use frozen $I_n^*$, $III^*$ and $IV^*$ singularities without making any assumption about their associated units of charge quantization.  Without a priori knowledge of the normalization of the CB charges for these theories, the Dirac quantization condition does not provide much of a constraint.

\subsection{Special K\"ahler conditions}

As explained above, the allowed deformation patterns are also constrained by the condition that the monodromy around a curve enclosing all the singularities resulting from the splitting should be in the same $\SL(2,\Z)$ conjugacy class as that of the monodromy around the initial singularity.  A simple test involving a relation between the $Q^2$'s of the undeformable singularities and the trace of the total monodromy serves to eliminate many of the possible deformation parameters in table \ref{t2}.  Also, many others can be easily shown to satisfy the monodromy constraint by using the explicit monodromies of the maximal deformation patterns described in \cite{Gaberdiel:1997ud, Gaberdiel:1998mv, DeWolfe:1998zf, DeWolfe:1998eu, Hauer:2000xy}.  A few deformation patterns require more work to rule in or out.  These arguments are given in detail in \cite{Argyres:2015gha}.  The result is that of the patterns allowed by Dirac quantization, the ones in grey in table \ref{t2} fail the monodromy condition, leaving 28 viable possibilities.

We then construct a single special K\"ahler geometry for each deformation pattern.  We are not able to prove that these are the only special K\"ahler geometries with these deformation patterns, since our methods of construction rely on making some assumptions.  But in many cases we find the same solution using different methods of construction which rely on quite different assumptions.  The details of the construction and the resulting solutions are presented in \cite{Argyres:2015gha}.   

One immediate result of constructing explicit solutions is that it enables us to determine the maximal flavor symmetry of the SCFT, as explained in section \ref{ssec:sings.2}, and listed in table \ref{table:theories}.  

Another interesting result of constructing the explicit solutions is that we find that the special K\"ahler geometries for the $\{{I_2}^3\}$ and $\{{I_1}^2,I_4\}$ deformations of the $I_0^*$ singularity are very closely realted \cite{Argyres:2015gha}.  Their SW curves are related by a 2-isogeny: a 2-to-1 map which rescales the low energy $\U(1)$ gauge coupling by a factor of $\sqrt2$.  This rescales electric and magnetic charges by opposite factors of $\sqrt2$, which is what is required to change from the $\{{I_2}^3\}$ pattern which has both charges normalized to $Q=\sqrt2$ to the $\{{I_1}^2,I_4\}$ pattern which has some $Q=1$ and some $Q=2$ states.  Both these curves are expected to describe the $\cN=2^*$ $\SU(2)$ gauge theory which has an adjoint hypermutiplet and has $\cN=4$ supersymmetry when the hypermultiplet is massless.\footnote{The $\{{I_2}^3\}$ form of the curve is the one constructed originally in \cite{Seiberg:1994aj}.}  It is interesting to note that, copying the discussion given in section 4.5 of \cite{Tachikawa:2013kta}, the $\{{I_1}^2,I_4\}$ form of the curve has one weak coupling limit where the light electric degrees of freedom have charges normalized to be multiples of 1 and the magnetic monopoles have charges normalized to be multiples of 2, and a second weak coupling limit which is the same but with ``electric" and ``magnetic" interchanged.  These match with the expected \cite{Goddard:1976qe} charge quantizations in the two S-dual weakly coupled descriptions of the $\cN=2^*$ $\SU(2)$ theory, the first with global form of the gauge group the simply connected ${\rm SU}(2)$, and the second with gauge group ${\rm SO}(3)\cong{\rm SU}(2)/\Z_2$.  Indeed, the dependence of the $I^*_0\to\{{I_1}^2,I_4\}$ geometry on the exactly marginal coupling shows that that coupling takes values in a fundamental domain of the modular group $\G^0(2)\subset PSL(2,\Z)$, which is the expected GNO duality group \cite{Goddard:1976qe}.  The $I^*_0\to\{{I_2}^3\}$ geometry, on the other hand, has both electric and magnetic charges quantized in the same units, and its marginal coupling takes values in a fundamental domain of the full $PSL(2,\Z)$ modular group.  The interpretation of this latter theory is less certain, and is discussed in \cite{Argyres:2016yzz}.

\subsection{RG flow condition}

For generic values of the mass deformation parameters, the flavor symmetry, $F$, is broken to rank$(F)$ $\U(1)$ factors.  Also for generic values of the deformation parameters, the singularity is split into the undeformable singularities of the deformation patterns listed in table \ref{table:theories}.  We have seen that the $I_n$ undeformable singularities have a $\U(1)$ flavor symmetry, while the $I_n^*$, $III^*$ and $IV^*$ undeformable singularities have no flavor symmetry.  One checks that all the entries in tables \ref{table:theories} have at least rank$(F)$ separate $\U(1)$ factors for a generic deformation.  (They are allowed to have more than rank$(F)$ low energy $\U(1)$ flavor factors for the reasons outlined in section \ref{ssec:sings.1}.)

This consistent account of the flavor symmetries must also persist for non-generic values of the mass parameters.  Certain patterns of mass parameters do not completely break $F$ to abelian factors but leave some nonabelian factors unbroken.  For these mass patterns, groups of the undeformable singularities must merge to form new singularities on the CB.  Furthermore, these resulting singularities must have an interpretation in terms of either IR free or conformal theories with the required flavor symmetry to account for the unbroken part of the flavor symmetry.

An example where this test fails is the special K\"ahler geometry describing the generic deformation pattern $II^*\to\{{I_1}^2,{I_4}^2\}$ with $\Sp(4)$ maximal flavor symmetry, shown as entry number 3 in table \ref{table:theories}.  One finds from the SW curve for this geometry that upon turning on masses breaking $\Sp(4) \to \SU(2)\oplus \U(1)$ in one way,\footnote{There are two inequivalent ways of doing this for $\Sp(4)$.} the singularity pattern on the CB is $\{I_3^*,I_1\}$.  The $I_1$ singularity carries a $\U(1)$ flavor symmetry.  The $I^*_3$ singularity is an IR free $\SU(2)$ gauge theory with one-loop beta function coefficient satisfying $b_0 = 3/a^2$ by \eqref{I*n-su2} for some charge normalization factor $a$ given in \eqref{a2cases}.  Combining this with \eqref{su2b0}, \eqref{su2dynkin}, and the rules for deducing the flavor symmetry from the hypermultiplet representation content, one finds that the only theory giving an $\SU(2)$ flavor symmetry and an $I_3^*$ singularity is the $\SU(2)$ gauge theory with half-hypermultiplet representation content $2\cdot{\bf 2} + 2\cdot{\bf 3}$ and with charge rescaling factor $a=\sqrt3$.  But this charge normalization is incompatible with Dirac quantization since there is also an $I_1$ singularity which necessarily has charges normalized to multiples of $Q=1$.  Thus this special K\"ahler deformation does not have a consistent interpretation as a low energy field for special values of its mass parameters.  

Note that this argument assumes there was no larger-rank accidental nonabelian flavor symmetry \cite{Argyres:2015gha}.  This is justified by the safely irrelevant conjecture, as discussed in section \ref{SKsplits}.  Thus the RG flow condition can be thought of as testing the consistency of RG flows with the safely irrelevant conjecture.

Some of the geometries listed in table \ref{table:theories} which are consistent under RG flows satisfy this condition in an interesting way.  For example, consider the $II^*\to\{{I_1}^6,I_4\}$ deformation which has maximal flavor symmetry $\Sp(10)$.  When one turns on masses so as to break $\Sp(10)\to\Sp(8)\oplus\U(1)$, the $II^*$ singularity splits into the pattern $\{I_1,I^*_3\}$ (just as in the mass breaking in the last example).  But now there is an IR free $\SU(2)$ gauge theory giving the $I^*_3$ singularity and an $\Sp(8)$ flavor symmetry: it is the one with half-hypermultiplet representation content $8\cdot{\bf 3}$ and with charge rescaling factor $a=1/2$.  Since all the fields are in the $\bf 3$, the CB charge normalization is given, by \eqref{u1su2norm}, as $Q=1$, compatible with Dirac quantization and the existence of the $I_1$ singularity.  

There are further conditions that consistency under RG flow imposes.  There can be special patterns of masses for which some of the undeformable singularities will merge into other singularities on the CB, even though there is no enhanced unbroken flavor symmetry.   In this case consistency requires that the new singularities correspond to SCFTs which only have $\U(1)$ or empty flavor symmetry.  An example of a special K\"ahler geometry for which this test fails is the $III^* \to \{{I_1}^2,I^*_1\}$ deformation pattern with $\SU(2)\oplus\SU(2)$ maximal flavor symmetry, shown as entry 15 in table \ref{table:theories}.  In this case there turns out to be a mass deformation breaking $\SU(2)\oplus\SU(2) \to \U(1)\oplus\U(1)$ for which the CB singularities are $\{II, I^*_1\}$.  But both the $I^*_1$ and the $II$ singularities correspond to theories with no flavor symmetry.

These examples are meant to give an indication of the tightness of the RG flow condition, and of the intricate way that it is satisfied by the geometries listed in table \ref{table:theories}.  A more detailed account is given in \cite{Argyres:2015gha} for the maximal flavor symmetry assignments.  A similar (but substantially more elaborate) analysis of the RG flow condition is given in \cite{Argyres:2016xua} for all the allowed sub-maximal flavor symmetry assignments.


\section{Concluding remarks\label{sec:conclusion}}

We have described how to organize a systematic study of the Coulomb branches of $\cN=2$ SCFTs, at least in the rank 1 case.  We started our analysis from the old result that for one-dimensional Coulomb branches, there are only 7 geometries which can be interpreted as Coulomb branches of $\cN=2$ SCFTs, classified first by Kodaira, and all of them do arise as Coulomb branches of particular $\cN=2$ SCFTs \cite{Seiberg:1994aj,Argyres:1995xn,Minahan:1996fg,Minahan:1996cj}.  But these scale-invariant geometries do not specify a corresponding SCFT uniquely since it is known \cite{Seiberg:1994aj,Argyres:2007cn} that some can arise as Coulomb branches of multiple inequivalent theories.  

In the case of gauge theories, it is easy to characterize inequivalent SCFTs with the same Coulomb branch: they correspond to the different possible choices of matter representations for which the beta-function vanishes.  We approached this problem in the non-lagrangian case by associating to the vicinity of each singularity on the CB a SCFT or an IR free theory, and then identified (or at least constrained) those theories by associating their deformations by local relevant operators to deformations of the CB geometry.  We showed that relevant operators only affect the vicinity of the singularity in question leading to the splitting of the initial singularity into lesser ones. This follows from our analysis of the properties of RG flows between theories that have moduli spaces of vacua, which gives results that have a rather different flavor than those for flows between fixed point theories without moduli spaces. 

We then found --- essentially by computing in examples as reported in \cite{Argyres:2015gha} --- that the deformed geometry which specifies a given SCFT is determined by the topological data encoded in the ``deformation pattern".   The deformation pattern is the list of singularities (i.e., their associated EM duality monodromy conjugacy classes) to which the original singularity splits for generic values of the deformation parameter.  By classifying the possible local $\cN=2$ deformations of $\cN=2$ SCFTs and by a careful analysis of charge normalizations on the CB, we showed how to obtain the flavor symmetries of these SCFTs from the deformations of their CB geometries.  

Furthermore, these deformed special K\"ahler geometries are all found to be linearly related to certain maximally deformed geometries, implying, in particular, that there are no obstructions to turning on different deformations simultaneously.  We argued that this is evidence for the ``safely irrelevant conjecture": $\cN=2$ field theories do not have dangerously irrelevant operators.

Accepting this conjecture, together with an additional technical planarity condition on special K\"ahler geometries in the vicinity of singularities, the absence of interacting rank-0 $\N=2$ SCFTs and assuming --- as we also argued --- that $\cN=2$ supersymmetry is not consistent with gauging discrete nonabelian subgroups of flavor symmetries, it becomes a straightforward task to scan all the possible deformation patterns and eliminate those which do not have a consistent low energy field theory interpretation.  The main low energy consistency conditions are that the Dirac quantization condition be satisfied on the Coulomb branch and the IR flavor symmetry under all patterns of mass deformations be consistent with the safely irrelevant conjecture.  The resulting physically consistent deformed special K\"ahler geometries are those listed in table \ref{table:theories}.  Explicit formulas describing their geometries are given in \cite{Argyres:2015gha}.  A third paper \cite{Argyres:2015ccharges} describes the calculation of some Higgs branches and conformal and flavor central charges of these theories. 

There are many questions raised here that we have not been able to answer:
\begin{itemize}
\item Is there a physical reason to impose the planarity condition on CB geometries?
\item Is there a proof that every special K\"ahler near deformation of a singularity can be extended to a maximal deformation of that singularity?
\item Is there other evidence for or against the safely irrelevant conjecture coming from flows between higher-rank theories or from AdS/CFT arguments?
\item Do any examples of interacting rank-0 $\cN=2$ SCFTs exist?
\item What discrete symmetries can be gauged in $\cN=2$ field theories?
\item Can conformal bootstrap, RG flow, S-duality, or string constructions be used to show the existence of SCFTs with the new CB geometries found here?
\end{itemize}

We believe that the systematic approach presented here both improves our understanding of rank 1 $\cN=2$ SCFTs and provides a coherent framework for extending this understanding.   The two most obvious extensions are: lifting the restriction to planar CB geometries, and lifting the restriction to rank 1 CB geometries.  Both of these extensions lead to dramatically richer CB geometries and are under active investigation \cite{future}.

Finally, it would be very interesting to be able to connect the information which can be extracted with the geometric methods explained here, with the current efforts on studying conformal theories using bootstrap techniques.  The main obstacle in this direction is the lack of satisfactory understanding of the relation between local CFT data and effective actions on moduli spaces though the results in \cite{Hellerman:2017sur} might be illuminating in this respect.


\begin{acknowledgments}
It is a pleasure to thank C. Beem, C. Cordova, J. Distler, T. Dumitrescu, J. Eby, P. Esposito, S. Gukov, A. Hanany, K. Intriligator, C. Long, D. Morrison, L. Rastelli, V. Schomerus, N. Seiberg, A. Shapere, Y. Tachikawa, A. Vainshtein, R. Wijewardhana, J. Wittig, and D. Xie for helpful comments and discussions.  We are especially grateful to K. Intriligator, C. Cordova, and T. Dumitrescu for pointing out errors in table 3 in the first version of this paper.  This work was supported in part by DOE grant DE-SC0011784.  MM was also partially supported by NSF grant PHY-1151392.
\end{acknowledgments}


\begin{appendix}
\section{Rank 1 planar special K\"ahler geometry}\label{app:details} 

We review some technical details \cite{Seiberg:1994rs,Seiberg:1994aj,Donagi:1995cf} of the description of rank 1 (1-dimensional) Coulomb branch (CB) geometries. 

Rank 1 CBs are locally coordinatized by a single complex variable $u$ which can be thought of as (some holomorphic function of) the vev of the complex scalar field in the low energy massless $\U(1)$ vector multiplet.  We are interested in scale-invariant CBs and their deformations which have the complex structure of an $n$-punctured sphere, and we will take $u$ to be a global complex coordinate on this sphere.   We put one puncture at $u=\infty$ which will also be metric infinity; all other punctures will be at finite distances.  Examples with different structures are discussed in \cite{Argyres:2017tmj}.   

Following the discussion in section \ref{sec:localdefs}, a deformation of a scale-invariant CB will depend holomorphically on some number of complex (marginal or relevant) \emph{deformation parameters} $q$, $\m$, and $\bm$.  Thus we will really be interested in a whole family of rank-1 CBs parameterized by some set of deformation parameters.
Here $q$ is a dimensionless exactly marginal coupling, $\m$ is a chiral term coupling with dimension $0<\D(\m)<1$, and $\bm$ is a vector of masses of dimension 1.  We will generally suppress explicit mention of the $q$, $\m$, and $\bm$ dependence of the geometrical objects on the CB:  whenever a quantity depends on the coordinate $u$, it will also be understood to depend on the deformation parameters as well. 

\paragraph{CB formulation of special K\"ahler geometry.}

The massless vector multiplet IR (2-derivative) effective action on the CB is specified by the kinetic terms for its scalar fields and by the coupling of the Maxwell field.   The complex scalar gives a K\"ahler metric on the CB.  The $\U(1)$ gauge coupling, $\t(u,\bm):= \frac{\th}{2\pi}+i\frac{4\pi}{g^2}$, is complex analytic in $u$ and the $\bm$ by standard $\cN=2$ non-renormalization theorems.  $\t$ is allowed to be discontinuous by $\U(1)$ electric-magnetic (EM) duality transformations.  The EM duality group, $\SL(2,\Z)$, is generated by two elements,
\begin{align}\label{STdef}
S=\begin{pmatrix} 0&-1\\ 1&0\end{pmatrix},
\qquad\text{and}\qquad
T=\begin{pmatrix} 1&1\\ 0&1\end{pmatrix},
\end{align}
which satisfy $S^2=(ST)^3=-I$ and acts on $\t$ by fractional linear transformations \eqref{SL2monods}.  The central charge of the supersymmetry algebra is a linear combination of global $\U(1)$ charges on the CB, and is given by
\begin{align}\label{rank1CC}
Z(u) = c\, a_D(u) + d\, a(u) +  \bw(\bm)
\end{align}
where $(c,d)\in\Z^2$ are the integer magnetic and electric charge numbers and $\bw\in\L_F\simeq \Z^{\text{rank}(F)}$ are the vector of $\U(1)^{\text{rank}(F)}$ ``quark number" charges (i.e., the charges under the part of the flavor symmetry, $F$, of the CFT which is generically unbroken by mass terms).  The lattice of these charges, $\L_F$, is thus identified with a weight lattice of $F$.  The normalization of the magnetic, electric, and quark number charges is discussed further below.

The pair $(a_D,a)$ are the ``special coordinates" on the CB and form a holomorphic section of an $\SL(2,\Z)$ vector bundle over the CB.  Equivalently, both the low energy coupling and the special coordinates are holomorphic functions on the CB which may be multivalued up to an $\SL(2,\Z)$ action.  Specifically, upon traversing a closed path $\g$ in the CB, they may change by
\begin{align}\label{SL2monods}
\begin{pmatrix} a_D\\ a \end{pmatrix} \to
\begin{pmatrix} A&B\\ C&D \end{pmatrix}
\begin{pmatrix} a_D\\ a \end{pmatrix},
\qquad
\t \to \frac{A\t+B}{C\t+D},
\qquad\text{for}\qquad
M_\g := \begin{pmatrix} A&B\\ C&D \end{pmatrix}
\in\SL(2,\Z).
\end{align}
$M_\g$ is the EM duality \emph{monodromy} of $\g$.  By continuity, if $\g$ is contractible then $M_\g=I$ since $\SL(2,\Z)$ is discrete.  Thus non-trivial monodromies are associated to paths encircling singularities on the CB.  More generally, the monodromies give a representation of the fundamental group of the CB in $\SL(2,\Z)$.

$\cN=2$ supersymmetry and unitarity imply the \emph{SK conditions}
\begin{align}\label{SKconds}
\frac{\del a_D}{\del a} = \t,\qquad
\text{and}\qquad
\Im\t >0.
\end{align}
(There is a further integrability condition at higher rank, which is trivially satisfied at rank 1.)  Furthermore, the K\"ahler potential on the CB is given by $\cK = \Im(a_D a^*)$, giving the metric $ds^2 = (\Im\t) \, da\, da^*$ in a special coordinate system.

\paragraph{Normalization of the central charge.}

A linear transformation $\left(\begin{smallmatrix}a_D\\a\end{smallmatrix}\right)\to M \left(\begin{smallmatrix}a_D\\a\end{smallmatrix}\right)$ with $M\in\SL(2,\Z)$ takes special coordinates to special coordinates, and is called a \emph{change of EM duality basis}.  The central charge \eqref{rank1CC} is invariant under this change of basis if the magnetic and electric charge numbers are redefined as $(c,d) \to (c,d) M^{-1}$.  Since the choice of EM duality basis is a matter of convention, charges and monodromies are only meaningful up to an overall action or conjugation in $\SL(2,\Z)$.

Denote by $\bz$ the row vector of the physical magnetic and electric charges of a particle,
\begin{align}\label{}
\bz := (p,q).
\end{align}
We want to understand the relation of the magnetic and electric charge numbers $(c,d)$ appearing in the central charge which we have normalized to be integers, with the physical charges $\bz$.  The usual Dirac-Schwinger-Zwanziger quantization condition \cite{Dirac:1931kp, Schwinger:1969ib, Zwanziger:1968rs} is the statement that the charge inner product of two dyons,
\begin{align}\label{dszform}
\vev{\bz_1,\bz_2} := 
p_1 q_2 - q_1 p_2,
\end{align}
is given by a multiple of an integer,\footnote{In cgs units the charge inner product measures the angular momentum in the EM field of a static pair of dyons, and so is normalized to be a half-integer.  Here we are simply redefining what we mean by ``magnetic charge" by a factor of two to make \eqref{dszform} correct.} $\vev{\bz_1,\bz_2} \in \Z$.  This condition does not actually imply that magnetic and electric charges are integral, rather only that $p=c\, m$ and $q=d\, e$ for integers $(c,d)$ (with both $c=1$ and $d=1$ realized in the spectrum) and some fixed positive real $(m,e)$ satisfying $m\,e:=P\in\N$.  When $P=1$ we absorb $e$ and $m$ in the $\U(1)$ gauge coupling, which are themselves absorbed in the normalization of the special coordinates on the Coulomb branch.  But when $P \neq 1$, they cannot be reabsorbed into the gauge coupling, and an additional rescaling of $a(u)$ relative to $a_D(u)$ is needed.  This can be interpreted as changing the relation between the low energy $\U(1)$ coupling, $\t := \del a_D/\del a$, and the $\U(1)$ (electric) coupling from $\Im\t = 4\pi/e^2$ to $\Im\t=P \cdot 4\pi/e^2$.  See, for example, the discussion in \cite{Seiberg:1994aj}.  For general $P$, we choose to rescale the $\U(1)$ coupling so that $m=e=\sqrt P$, and define
\begin{align}\label{zvect}
\bz := (p,q) := \sqrt P \,(c,d),
\qquad\text{with}\qquad c,d\in \Z,
\qquad\text{and}\qquad P \in \N,
\end{align}
to be the magnetic and electric charge vector of a state.

As mentioned above, the freedom to make a change of EM duality basis reflects the freedom to choose different definitions of what we call electric and magnetic charges.  A choice of a pair of charge vectors $\{\bz_1,\bz_2\}$ forming a basis of the charge lattice with invariant product $\vev{\bz_1,\bz_2}=P$ is equivalent to a choice of EM duality basis.  Indeed, the subgroup of $\GL(2,\Z)$ transformations (which take bases of the charge lattice to other bases) which preserve the invariant product is ${\rm Sp}(2,\Z)\simeq \SL(2,\Z)$, independent of the value of $P$.  This is the group of EM duality transformations.  An $\SL(2,\Z)$ invariant of a single dyon charge, $\bz$, is its \emph{EM duality invariant charge}, $Q$, defined by
\begin{align}\label{Qdef}
Q^2 := \gcd(p^2,q^2)
\qquad \text{and} \qquad 
Q>0.
\end{align}
Note that the EM duality invariant charge $Q$ must be an integer multiple of $\sqrt P$.  This possibility that the charge vector may not be integral plays an important role in the classification of rank 1 SCFTs.

The $\bw(\bm)$ term in the central charge \eqref{rank1CC} involves the vector of ``quark number" charges, $\bw$, under the generically unbroken $\U(1)^{\text{rank}(F)}$ symmetry.  These charges span a lattice of rank$(F)$.  Since the linear mass parameters $\bm$ transform in the adjoint of $F$, whenever $\ba(\bm)=0$ for $\ba$ a root of $F$ there should be a degeneracy in the BPS spectrum since on these subspaces the flavor symmetry is not completely abelianized (i.e., it has some unbroken non-abelian factors).  Conversely, if there were some $\bw$ not in the root lattice of $F$, then there will be additional degeneracies in the BPS mass spectrum at every point on the CB for masses satisfying $\bw(\bm)=0$ which are not due to an enhanced symmetry.  Discounting the existence of such accidental degeneracies which persist for all values of $u$, one concludes that the lattice of quark number charges should be the root lattice of $F$;  however, it is a logical possibility that $\L_F$ spans a larger weight lattice of $F$ (i.e., properly containing the root lattice), but that all the states in the theory with quark number charges not in the root lattice are not BPS.

The normalization of the masses $\bm$ relative to the CB vev $u$ is in principle fixed by a choice of normalizations of the local CB and Higgs branch primary operators in the SCFT.  But the relation of the local operator algebra of the SCFT to the effective action on the moduli space is not currently understood, so in practice the relative normalizations of $\bm$ and $u$ in the CB geometry is arbitrary.  Similarly, a metric on the quark number charge lattice, $\L_F$, is in principle determined up to an overall normalization by the Killing form on $F$.  But the Killing form on $F$ does not enter directly in the low energy effective action on the CB, and can only be inferred indirectly by how Weyl$(F)$ acts on it; see the discussion in section \ref{ssec:sings.2}.

\paragraph{Total space formulation of SK geometry.}

The IR effective action on a rank-$r$ Coulomb branch can be encoded in terms of EM duality invariant geometrical data  \cite{Donagi:1994, Seiberg:1994aj, Donagi:1995cf} as the total space of a family of $r$-complex-dimensional abelian varieties (with a choice of polarization) holomorphically fibered over the CB, together with a holomorphic symplectic form on the total space for which the CB is a lagrangian submanifold.  (See \cite{Freed:1997dp} for a discussion of various geometrical formulations of SK geometry and their interrelations.)  A specific algebraic realization of this data in the rank 1 case is \cite{Seiberg:1994rs, Seiberg:1994aj} as a holomorphic family of elliptic curves, $\S(u,\bm)$,  identified with the complex torus fiber of the total space, written in Weierstrass form
\begin{align}\label{rank1curve}
\S(u,\bm): \qquad y^2 &= x^3 + f(u,\bm)\,x + g(u,\bm).
\end{align}
$f$ and $g$ are complex analytic in the relevant parameters $\bm$.  Some simple examples are given in the second column of table \ref{Table:Kodaira}.\footnote{The curves for the $I_n^*$ and $I_n$ singularities in the table are not in Weierstrass form, but can be easily put in that form by an appropriate shift of $x$.}  The holomorphic symplectic form (really, its integral along some path in the CB) is specified by the SW 1-form, $\l(u,\bm)$, a meromorphic 1-form on the fiber satisfying the following \emph{SK conditions} constraining its $u$- and $\bm$-dependence,
\begin{align}\label{SWform}
\frac{\del \l}{\del u} &=  
\Omega + {\rm d}\f, &
\text{Res}(\l) &\in \{\bw(\bm)\ |\ \bw\in\L_F\}&
&\text{and}&
\{\text{Res}(\l)\} \ &\text{spans} \ \L_F.
\end{align}
Here $\Omega$ is a non-vanishing holomorphic one-form on the fiber  $\S(u,\bm)$, and will be discussed in more detail below; $\f$ is an arbitrary meromorphic function on the fiber;  Res$(\l)$ means the residue of $\l$ at any of its poles; and $\L_F$ is the weight lattice of $F$.  (We use a notation in which weights and masses are dual: if $\ff$ is the Cartan subalgebra of $F$, then $\bw\in\ff^*$ and $\bm\in\ff_\C$, and $\bw(\bm) := \w^i m_i$ is the dual pairing.)

The curve and 1-form are related to the low energy data on the CB as follows.  The complex structure of the torus fiber is the low energy $\U(1)$ gauge coupling.  The magnetic and electric $\U(1)$ charge numbers $(c,d)$ and quark numbers $n^i$ of a BPS state, parameterize the homology class of a cycle $[\g(c,d,n^i)] = c\, [\a] + d\, [\b]+n^i\, [\d_i]$ on the fiber, which determines the central charge (and BPS mass) of these states by $Z([\g]) = \oint_\g \l$.  Here $\a$ and $\b$ are a canonical basis of 1-cycles on the torus, and the $\d_i$ are a basis\footnote{There may be more than rank($F$) poles of $\l$, but since the residues are all integral combinations of the rank($F$) mass parameters, $m_i$, there are effectively only rank($F$) independent cycles $\d_i$.} of cycles around the poles of $\l$.   This then reproduces \eqref{rank1CC} with special coordinates $a_D := \oint_\a \l$ and $a := \oint_\b \l$, and flavor charge of the BPS state given by $\bw = n^i \bw_i$ where $\bw_i(\bm)$ is the residue at the $i$-th pole of $\l$.  The differential SK condition on the 1-form \eqref{SWform} then ensures the SK conditions \eqref{SKconds}. 

Since charges are encoded in 1-homology classes of the fiber, the charge inner product \eqref{dszform} is encoded in a non-degenerate integral antisymmetric pairing, $\vev{\cdot,\cdot}$, of 1-cycles on a $2r$-torus (the abelian variety fiber).  This is called a choice of polarization on the $2r$-torus.  For a 2-torus (the rank-1 case) a polarization is thus a positive integral multiple of the intersection form for 1-cycles.  On a canonical basis $\{\a,\b\}$ of 1-cycles, the polarization is given by $\vev{\a,\a} = \vev{\b,\b} = 0$, {$\vev{\b,\a} = -\vev{\a,\b} = P$}, for $P\in\N$.\footnote{{The choice of sign of $\vev{\a,\b}$ is determined by the condition \eqref{SKconds} that $\Im\t>0$.}}  ($P$ is the elementary divisor of the Hodge form associated with the polarization.)  The subgroup of $\GL(2,\Z)$ transformations of $H_1(\S,\Z)$ which preserve the polarization is ${\rm Sp}(2,\Z)\simeq \SL(2,\Z)$, independent of the value of $P$.  This is the group of EM duality transformations.   (For higher-rank cases, ${\rm Sp}(2r,\Z)$ is the EM duality group only for principal polarizations.)  

The location of CB singularities can be readily obtained from \eqref{rank1curve}.  They are given by the zeros of the discriminant of the SW curve,
\begin{align}\label{disc}
D_x(u):=4f^3(u)+27g^2(u).
\end{align}
For scale invariant theories $D_x(u)$ is a homogeneous polynomial in $u$.  Its degree is thus the same as the order of its vanishing at $u=0$, which we denote by ord$_0(D_x)$.  In the case of IR free theories, the value of ord$_0(D_x)$ is closely tied to the number and gauge charges of states becoming massless at $u=0$, as explained in section \ref{sec:udefsing}.

It is important to note that $f$ and $g$ appearing in the Weierstrass form of the curve \eqref{rank1curve}, as well as the 1-form, $\l$, are not uniquely determined by the CB geometry.  In addition to a constant rescaling of $f$ relative to $g$ which can be absorbed in an appropriate rescaling of $x$ and $y$, there can also be discrete nonlinear transformations of $f$ and $g$ which leave the fiber geometry invariant.  The 1-form is even less well-specified since it may be shifted by $u$ integrals of arbitrary total derivatives on the fiber.  These ambiguities play an important role in \cite{Argyres:2015gha} where we construct explicit representatives of $\S(u,\bm)$ and $\l(u,\bm)$ for each entry in table \ref{table:theories}.

The general holomorphic one-form entering in \eqref{SWform} is of the form $\Omega = P(u) dx/y$ for some holomorphic function $P(u)$.  In a neighborhood of any point of the CB we can make an analytic change of variables $u$, $x$, and $y$ which set $P=1$ while preserving the Weierstrass form \eqref{SWform}.  So we fix
\begin{align}\label{Oregular}
\Omega(\bm) = \frac{dx}{y}.
\end{align}
Similarly, using the freedom to rescale $x$ and $y$ in \eqref{rank1curve} by holomorphic functions of $u$ and $\bm$ while preserving the Weierstrass form of the curve and \eqref{Oregular} implies that $f$ and $g$ can be taken to depend polynomially on $u$ and the $\bm$.

\paragraph{Kodaira curves.}

For scale-invariant geometries (i.e., with $\bm=0$) it is then straightforward to classify all the possible CB geometries by taking $f$ and $g$ in \eqref{rank1curve} to be homogeneous polynomials in $u$ and imposing a symmetry under complex rescalings.  This determines the relative scaling weights of $x$, $y$, and $u$, and since $Z$ has mass dimension 1, so must $\l$, and then \eqref{SWform} together with the planarity condition, \eqref{Oregular}, determine the scaling dimension of $u$.  The result is that the only scale-invariant CBs with $\D(u)>1$ (which follows from unitarity of the CFT) and a singularity at the origin are given by the first 7 entries in table \ref{Table:Kodaira} which lists the Kodaira classification  \cite{KodairaI, KodairaII} of singularities of families of elliptic curves varying holomorphically with $u$.   In that table, the freedom to shift and rescale the $x$, $y$, and $u$ variables appropriately has been used to put the curves in Weierstrass form, to put the singularity at $u=0$, and to fix the normalizations of the terms.  

For the first seven entries in the table, the equation \eqref{SWform} for the SW one-form is solved by $\l \sim u\, dx/y$.  The discriminants of these singularities are homogeneous in $u$, $D_x \sim  u^{n}$ for some $n:=\text{ord}_0(\D)$ listed in the table.  In particular, the only singular fiber on the Coulomb branch is at the origin.  A representative, $M_0$, of the $\SL(2,\Z)$ conjugacy class of the monodromy around the singularity is shown.  The overall geometry is that of a cone with positive deficit angle.  The complex $\U(1)$ gauge coupling, $\t_0$, is constant on the CB.  (The $I^*_0$ singularity is actually a family of singularities with arbitrary $\t_0$.)

The last two rows of the table show two infinite series of singularities which depend on an extra dimension-one parameter, $\L$.  In the limit $\L\to0$, they are scale-invariant but have geometries which are degenerate cones (i.e., with deficit angle $2\pi$).  The value of the $\U(1)$ gauge coupling at the tip is $\t_0=i\infty$, corresponding to the weak coupling limit.  These CB geometries thus have the interpretation as the CBs of IR free field theories near $u=0$, with $\L$ playing the role of the strong coupling scale (Landau pole).  They have singular fibers not only at $u=0$ but also at points $u\sim\L^{\D(u)}$, but, given their interpretation as IR free theories, we are only interested in the vicinity of the origin, $|u|\ll \L^{\D(u)}$.  In this limit the SW one-form is again of the form $\l\sim u\, {\rm d}x/y$ near $u=0$, and ord${}_0 (D_x)$ is given in the table.  We discuss the physical interpretation of these singularities in detail in section \ref{froSing}.    

Finally, we comment on how this classification of scale-invariant and IR-free rank-1 CBs changes once the planarity assumption is lifted.  Without the planarity assumption, one cannot assume that the CB coordinate $u$ is necessarily the vev of a scalar field in the CFT:  instead it may be a fractional power of such a vev; see \cite{Argyres:2017tmj} for a discussion.  The $\D(u)>1$ constraint is then lifted and one finds \cite{Argyres:2017tmj} an infinite number of copies of each entry in table \ref{Table:Kodaira} with negative deficit angles (i.e., negative curvature at the tips of their cones) and with $0<\D(u)<1$. We call these \emph{irregular singularities}, for a more detailed discussion see \cite{Argyres:2017tmj}. 


\section{Review of $\cN=2$ superconformal null states and representations}\label{app:repre}

The $\cN=2$ superconformal algebra is generated by $\SO(4)\simeq \SU(2)\times\til\SU(2)$ Lorentz rotations, $\U(1)_R\times\SU(2)_R$ rotations, dilatations, translations $P^{\a\ad}$, special conformal transformations $K^{\a\ad}$, supertranslations $Q^\a_i$ and $\tQ_i^\ad$, and superconformal transformations $S^\a_i$ and $\tS^\ad_i$.  Here $(\a,\ad,i)$ are spinor indices under the $\SU(2)\times\til\SU(2)\times\SU(2)_R$ symmetry.  The nonvanishing $\U(1)_R$ charges, $r$, of the generators are 
\begin{align}\label{U1Rcharges}
r(Q)=r(\tS)=-r(\tQ)=-r(S)=\frac12,
\end{align}
which follows the normalization used in \cite{Dolan:2002zh,Beem:2014zpa}.  

Positive energy representations of a superconformal algebra are  formed by acting repeatedly on a primary field with the $(Q,\tQ)$ operators to form a multiplet of descendant fields.  The primary (annihilated by $(S,\tS)$ at the origin) is characterized by its dimension $\D$, Lorentz spins $(j,\tj)$, $\SU(2)_R$ spin (``R-spin") $R$, and $\U(1)_R$-charge $r$.  If a primary $X$ has dimension $\D$, its descendant is said to be at ``level $\ell$" if it has dimension $\D+(\ell/2)$. 

We use a dot to denote contraction of spinor indices, while a vee on supercharges, $\tQ^\v$ etc., acting on a field means pick out the irrep with highest R-spin.  For example, if $X$ is a field with R-spin $R$, then $Q^\v X$ has R-spin $R+\frac12$.  (Equivalently, the vee means symmetrize on the $\SU(2)_R$ spinor indices.)  Note that when the R-spin of $X$ vanishes, then $Q X = Q^\v X$ since it can only have R-spin $\tfrac12$.  Finally, we use the shorthand ``$G_1G_2\cdots G_n X$" for the nested (anti)commutators of charges acting on a field $[G_1,[G_2,\cdots [G_n,X\}\cdots\}\}$.

Positivity of the norm of the following descendants of superconformal primary $X^R_{j,\tj}$ of $\U(1)_R$ charge $r$ and dimension $\D$ leads to the inequalities \cite{Dobrev:1985qv, Dolan:2002zh}
\begin{subequations}
\begin{align}
&\, \tj =0: \qquad\ ||\tQ_\ad^\v X|| \ge0 & 
&\Rightarrow &
\D &\ge 2R-r \, , \label{ss1} \\
\Biggl\{ &
\begin{array}{rr}
\tj=0: & ||(\tQ^\v)^2 X|| \ge0 \\[1mm]
\tj>0: & \qquad ||\tQ^\v\cdot X|| \ge0 \\
\end{array}
\Biggr\} &
&\Rightarrow &
\D &\ge 2R+2-r + 2\tj \, , \label{ss3}
\end{align}
\end{subequations}
and
\begin{subequations}
\begin{align}
&\, j =0: \qquad\ ||Q_\a^\v X|| \ge0 & 
&\Rightarrow &
\D &\ge 2R+r \, , \label{ss2} \\
\Biggl\{ &
\begin{array}{rr}
j=0: & ||(Q^\v)^2 X|| \ge0 \\[1mm]
j>0: & \qquad ||Q^\v\cdot X|| \ge0 \\
\end{array}
\Biggr\} &
&\Rightarrow &
\D &\ge 2R+2+r + 2j \, . \label{ss4}
\end{align}
\end{subequations}
When any of these inequalities is saturated there is a corresponding null state, and these are all the independent null states that can occur, i.e., any other null state is a descendant of one of these.  These unitarity bounds lead to the list of unitary $\cN=2$ representations shown in table \ref{N2table}.  

\begin{table}[ht]
\centering\small
$\begin{array}{|cl|crlrl|c|}\hline
\multicolumn{8}{|c|}{\mbox{\bf $\cN=2$ superconformal unitary, positive-energy representations}}\\ \hline\hline
\multicolumn{2}{|c|}{\mbox{Name \cite{Dolan:2002zh}}}&
\multicolumn{5}{c|}{\mbox{Primary field quantum numbers}}
&\mbox{Null states}\\ \hline
\cA^\D_{R,r(j,\tj)} & 
&&\D&
\multicolumn{3}{l|}{>2R+2+|r+j-\tj|+j+\tj\ \ }
&\text{none}\\
\bar\cC_{R,r(j,\tj)} & 
&&\D&=2R+2-r+2\tj,\ \ &r&< \tj-j 
&\text{\eqref{ss3}}\\
\cC_{R,r(j,\tj)} &
&&\D&=2R+2+r+2j,\ \ &r&> \tj-j
&\text{\eqref{ss4}}\\
\hat\cC_{R\,(j,\tj)} & 
&&\D&=2R+2+j+\tj,\ \ &r&= \tj-j
&\text{\eqref{ss3} \& \eqref{ss4}}\\
\bar\cB_{R,r(j,0)} & (\equiv 
\bar\cE_{r(j,0)}\ \text{if}\ R{=}0)
&\tj=0&\D&=2R-r,\ \ &-r&>j+1
&\text{\eqref{ss1}}\\
\cB_{R,r(0,\tj)} & (\equiv
\cE_{r(0,\tj)}\ \text{if}\ R{=}0)
&j=0  &\D&=2R+r,&+r&>\tj+1
&\text{\eqref{ss2}}\\
\bar\cD_{R\,(j,0)} & 
&\ \tj=0\ &\quad \D&=2R-r,&
-r&=j+1
&\text{\eqref{ss1} \& \eqref{ss4}}\\
\cD_{R\,(0,\tj)} & 
&\ j=0\ &\quad \D&=2R+r,&
+r&=\tj+1
&\text{\eqref{ss3} \& \eqref{ss2}}\\
\cBh_{R} & 
&\ j=\tj=0\ &\quad \D&=2R,&\qquad r&=0\qquad
&\text{\eqref{ss1} \& \eqref{ss2}}\\
\hline
\end{array}$
\caption{The names of the representations are given along with the constraints satisfied by the Lorentz spins $(j,\tj)$, R-spin $R$, $\U(1)_R$ charge $r$, and dimension $\D$ of their primaries.  The independent null states of each type of representation are also shown.\label{N2table}}
\end{table}

The most important representations for our purposes are the ``semi-chiral" $\cB$ multiplets (and their conjugates) and their ``bi-chiral" $\cBh$ shortening.  Vevs of the Lorentz-scalar primaries of these multiplets can parameterize the moduli space of vacua of the SCFT.

The $\cB$ multiplet primaries have $j=0$ and satisfy the right-chiral condition \eqref{ss2}, so a product of $\cB$ primaries coupled to the maximal R-spin is also a $\cB$ primary.  This implies that Lorentz scalar $\cB$ multiplet primaries with maximal R-spin ($R_3=R$) form a chiral ring.  A $\cBh$ multiplet primary is a Lorentz scalar satisfying both left and right chiral conditions \eqref{ss1} and \eqref{ss2}, so the product of two $\cBh$ primaries coupled to the maximal R-spin is also a $\cBh$ primary.  So their complex scalar primaries with highest R-spin ($R_3=R$) also form a chiral ring.  Also, it is easy to see that the product of a $\cBh$ maximal R-spin primary with a scalar $\cB$ multiplet maximal R-spin primary is another scalar $\cB$ primary.  Thus these operators together satisfy the chiral ring relations
\begin{align}\label{chiralring}
\cB_{R,r\,(0,0)}\vee \cB_{S,s\,(0,0)} 
&= \cB_{R{+}S\,,\,r{+}s\,(0,0)},&
\cBh_R \vee \cBh_S 
&= \cBh_{R{+}S},
\notag\\
\cBh_R \vee \cB_{S,s\,(0,0)} 
&= \cB_{R{+}S\,,\,s\,(0,0)},
\end{align}
where the vee means couple to the maximal R-spin.

The $\cBh$ multiplet with $R=0$ ($\D=0$) is the identity, and one with $R=\tfrac12$ ($\D=1$) is a free massless hypermultiplet. Those with $R=1$ ($\D=2$) have a conserved current at the second level, so transform in the adjoint of the $\cN=2$ flavor symmetry algebra.  Since the non-trivial $\cBh$ primaries carry no $\U(1)_R$ charge and have non-zero $\SU(2)_R$ spins, their vevs are identified with the Higgs branch complex coordinate ring.  The $R=0$ $\cB$ multiplets are anti-chiral, and are also called ``$\cE$"  multiplets.  The $\cB_{0,r(0,0)} \equiv \cE_{r(0,0)}$ scalar primaries may be identified with Coulomb branch chiral ring operators, as they have zero R-spin but non-zero $\U(1)_R$ charge.  The $\cB_{R,r(0,0)}$ scalar primaries with $R>0$ may be identified with mixed branch chiral ring operators.  Finally, the Lorentz scalar $\cD_{R\,(0,0)}$ maximal R-spin primaries are just specializations of the $\cB_{R,r\,(0,0)}$ scalar primaries to $r=1$, and satisfy the same chiral ring relations \eqref{chiralring}.  In general, the R-spin singlet $\cD$ multiplets are free fields, and in particular the $\cD_{0\,(0,0)}$ are free (anti-chiral) vector multiplets.

We should note that the identification of all these chiral ring operators with Higgs, Coulomb, and mixed branch operators is conjectural \cite{Tachikawa:2013kta,Beem:2014zpa}: it is possible that some or all of them may occur in the SCFT operator algebra but do not correspond to flat directions.


\section{Local $\cN=2$ supersymmetric deformations of SCFTs}\label{app:defo}

The possible candidate local operators which can deform the ``lagrangian" of an $\cN=2$ SCFT are given by all combinations of supercharges, $(\tQ^n Q^m)^R_{(j,\tj)}$, acting on primary operators.  Lorentz invariance implies the spins must be contracted to form scalars, so the spins of the primaries must be the same as those of the supercharge combinations acting on them.  We make no assumption about the R-spin of the primaries, nor do we require that the total R-spin of the deformation terms vanish.  ($\cN=2$ supersymmetry does not require $\SU(2)_R$ invariance.)   

$\cN=2$ supersymmetry invariance requires that each of these possible terms be transformed into total derivatives by both $Q^i_\a$ and $\tQ^i_\ad$ for all $i,\a,\ad$ after using the $\cN=2$ supertranslation algebra.  The only place derivatives come from in the supertranslation algebra is from the anticommutator of a $Q$ with a $\tQ$.  After dropping total derivatives, the $\cN=2$ supertranslation algebra simply says that all $Q$'s and $\tQ$'s anticommute.  Thus $\cN=2$ supersymmetry invariance requires that each term is annihilated upon antisymmetrizing with the $Q$ and $\tQ$.  The only extra conditions that primaries can satisfy are the vanishing of the null states \eqref{ss1}--\eqref{ss4}, so these supersymmetry annihilation conditions must follow as descendants of one or more of these null states. 

Start by considering just left-chiral spins (``l-spins"); we will later combine them with right-spins.  So we consider ``$Q$-primaries" $Y^R_j$ of R-spin $R$ and l-spin $j$ satisfying one or none of the left-chiral shortening conditions \eqref{ss2} or \eqref{ss4}.  These $Q$-primaries may therefore be of the form $Y = \tQ^n X$ for $X$ a superconformal primary.  

The first thing to notice is that the possible l-spins ($j$) and R-spins ($R$) that can be realized by combinations of the left supercharges are
\begin{align}\label{QnjI}
(Q^n)^R_j \ =\  \Bigl\{\ \
(Q&)^{\frac12}_{\frac12}&
(Q^2&)^{1}_{0}&
(Q^2&)^{0}_{1}&
(Q^3&)^{\frac12}_{\frac12}&
(Q^4&)^{0}_{0} \ \ \Bigr\}
\end{align}
All other l- and R-spin combinations vanish by the supertranslation algebra.  Acting on a $Q$-primary $Y$ of l-spin $j$ and R-spin $R$, it follows from \eqref{QnjI} that the possible l- and R-spins of $Y$ $Q$-descendants  are (no correlations among the signs)
\begin{align}\label{jdesc}
(Q Y)^{R\pm\frac12}_{j\pm\frac12} &&
(Q^2 Y)^{R\pm1}_j &&
{(Q^2 Y)'}^R_j && {(Q^2 X)''}^R_j &&
(Q^2 Y)^R_{j\pm1} &&
(Q^3 Y)^{R\pm\frac12}_{j\pm\frac12} &&
(Q^4 Y)^R_j .
\end{align}
Here ${(Q^2 Y)'}^R_j := ((Q^2)^1_0 Y)^R_j$ and ${(Q^2 Y)''}^R_j := ((Q^2)^0_1 Y)^R_j$ are the two different ``channels" for making total R-spin $R$ and l-spin $j$ combinations.  In all other cases there is only a single channel.  
If $j<1$ or $R<1$ then the entries in this list with negative total $j$ or $R$ should be discarded.  Note also that if $R=0$ then the ${(Q^2 Y)'}^R_j$ channel vanishes, while if $j=0$, ${(Q^2 Y)''}^R_j \equiv 0$.

Now compute the l- and R-spins of the $Q$-descendants of the various null states in \eqref{ss2} and \eqref{ss4}.  Acting on a $Q$-primary $Y_0$ of l-spin $j=0$ and R-spin $R$, the \eqref{ss2} null state gives the null descendants
\begin{align}\label{Anull}
(QY_0)^{R+\frac12}_{\frac12}=0\ \Rightarrow \ 
\Bigl\{
(Q^2 Y_0)^{R}_{\frac12\pm\frac12} ,\ 
(Q^2 Y_0)^{R+1}_{0} ,\ 
(Q^3 Y_0)^{R\pm\frac12}_{\frac12} ,\ 
(Q^4 Y_0)^R_0 \ \Bigr\}=0.
\end{align}
Since $j=0$, the $(Q^2 Y_0)^R_0$ descendant is in the ${(Q^2 Y)'}^R_0$ channel, since the ${(Q^2 Y)''}^R_0$ channel vanishes identically.  Comparing \eqref{Anull} to \eqref{jdesc} with $j=0$ we see that the existence of the \eqref{ss2} null state puts no restrictions on the  $Y_0$ descendants
\begin{align}\label{Anotnull}
0 \neq \Bigl\{
(Q Y_0)^{R-\frac12}_{\frac12} ,\ 
(Q^2 Y_0)^{R-1}_0 
\Bigr\} .
\end{align}
Note, however, that the first vanishes if $R<\frac12$ and the second vanishes if $R<1$.
Similarly, the null descendants of the first null state in \eqref{ss4} are
\begin{align}\label{Bnull}
(Q^2 Y_0)^{R+1}_{0}=0\ \Rightarrow \ 
\Bigl\{
(Q^3 Y_0)^{R+1\pm\frac12}_{\frac12} , \ 
(Q^4 Y_0)^R_0 \ \Bigr\}=0,
\end{align}
and comparing to \eqref{jdesc} with $j=0$, we see that the existence of this null state puts no restrictions on the following $Y_0$ descendants
\begin{align}\label{Bnotnull}
0 \neq \Bigl\{
(Q Y_0)^{R\pm\frac12}_{\frac12} ,\ 
(Q^2 Y_0)^{R-1}_0 ,\ 
(Q^2 Y_0)^{R}_{\frac12\pm\frac12} ,\ 
(Q^3 Y_0)^{R-\frac12}_{\frac12} 
\Bigr\} .
\end{align}
Again, since $j=0$, the $(Q^2 Y_0)^R_0$ descendant in this list is in the ${(Q^2 Y)'}^R_0$ channel, whiles the other channel vanishes identically.
Finally, consider the null state in \eqref{ss4} on a $Q$-primary $Y$ with l-spin $j>0$.  Its null descendants are
\begin{align}\label{Cnull}
(QY)^{R+\frac12}_{j-\frac12} =0\qquad &\Rightarrow \\  
\Bigl\{
(Q^2 Y)^{R}_{j-1} ,\ 
&{(Q^2 Y)'}^R_j{+}{(Q^2 Y)''}^R_j ,\ 
(Q^2 Y)^{R+1}_{j} ,\ 
(Q^3 Y)^{R\pm\frac12}_{j\pm\frac12} ,\ 
(Q^4 Y)^R_j \ \Bigr\}=0.
\notag
\end{align}
Comparing to \eqref{jdesc}, we see that the existence of this null places no restrictions on the $Y$ descendants
\begin{align}\label{Cnotnull}
0 \neq \Bigl\{
(Q Y)^{R-\frac12}_{j-\frac12} ,\ 
(Q Y)^{R\pm\frac12}_{j+\frac12} ,\ 
(Q^2 Y)^{R-1}_j ,\ 
{(Q^2 Y)'}^R_j{-}{(Q^2 Y)''}^R_j ,\ 
(Q^2 Y)^{R}_{j+1} 
\Bigr\} .
\end{align}
Note, however, that if $R=0$, since ${(Q^2 Y)'}^0_j \equiv 0$, the $(Q^2Y)^0_j$ descendant is in fact null.

Now consider potential deformations of the form $Q^m Y$.  We first demand that it be annihilated by $Q$, which we'll call the ``l-susy condition".  Then for each value of $0\le m\le 4$ it is easy to deduce the necessary conditions for the deformation to be l-susy and non-vanishing:  
\begin{itemize}
\item[$m{=}0$:]
The l-spin of $Y$ is $j=0$, so the possible null states are \eqref{Anull} or \eqref{Bnull}.  Looking at the $QY$ descendants in \eqref{Anotnull} and \eqref{Bnotnull}, we deduce that a necessary condition for it to be an l-susy deformation is that $Y$ is an \eqref{Anull} null state with $R=0$.
\item[$m{=}1$:]
The l-spin of $Y$ is $j=\frac12$, so the only possible null states are \eqref{Cnull}.  Of the two possible l-spin singlet $QY$ combinations, $(QY)^{R+(1/2)}_0$ and $(QY)^{R-(1/2)}_0$, the first vanishes by \eqref{Cnull}, so does not deform the action.  The $Q^2Y$ descendants of the second combination are $(Q^2Y)^{R-1}_{1/2}$ and ${(Q^2Y)'}^R_{1/2}-{(Q^2Y)''}^R_{1/2}$.  The first of these  vanishes only for $R<1$.  The second vanishes, by the comment after \eqref{Cnotnull}, only when $R=0$.  But $(QY)^{R-(1/2)}_0\equiv0$ when $R=0$, so there are no l-susy $QY$ deformations.
\item[$m{=}2$:]
There are two l- and R-spin $Q^2$ combinations to examine:
\begin{itemize}
\item[$(Q^2)^1_0$:]
The l-spin of $Y$ is $j=0$, so the possible null states are \eqref{Anull} or \eqref{Bnull}.  From the $Q^3 Y$ descendants in \eqref{Anotnull} and \eqref{Bnotnull}, we deduce that the deformation either: (a) has $Y$ satisfying the \eqref{Anull} condition, or (b) is the $R+1$ R-spin combination with $Y$ satisfying the \eqref{Bnull} condition.  In case (a) the deformations with total R-spin $R$ and $R+1$ vanish by \eqref{Anull}, and in case (b) the total R-spin $R+1$ component vanishes by \eqref{Bnull}.  So these R-spin components do not deform the action.  The only remaining potential deformation is the $[(Q^2)^1_0 Y^R_0]^{R-1}_0$ component with $Y$ satisfying \eqref{Anull}.
\item[$(Q^2)^0_1$:]
The l-spin of $Y$ is $j=1$, so the possible null states are \eqref{Cnull}.  Since there are no $Q^3 Y$ descendants in \eqref{Cnotnull}, it follows that the deformation is annihilated by $Q$ if $Y$ satisfies \eqref{Cnull}.  But this potential deformation vanishes by \eqref{Cnull} so does not actually deform the action.
\end{itemize}
\item[$m{=}3$:]
The l-spin of $Y$ is $j=\frac12$, so the possible null states are \eqref{Cnull}.  Since the $Q^4 Y$ descendant is not in \eqref{Cnotnull}, the potential deformation is l-susy if $Y$ satisfies \eqref{Cnull}.  But the potential deformation vanishes by \eqref{Cnull}, so does not deform the action.
\item[$m{=}4$:]
This deformation is automatically annihilated by $Q$, so is l-susy for $Y$ a general multiplet (i.e., having no null descendants), and vanishes identically if $Y$ satisfies any null state condition.
\end{itemize}
To summarize, there are only 3 combinations of $Q$'s acting on $Q$-primaries, $Y^R_j= [\tQ^n X_j]^R$, which can give non-vanishing deformations annihilated by $Q^i_\a$.  They all have $j=0$, so we only show their R-spin superscripts:
\begin{align}
\text{(a)}&: &
&[\tQ^n X]^0  &
&\text{and $X$ satisfies the \eqref{ss2} null condition,}
\notag\\
\text{(b)}&: &
&[(Q^2)^1_0 (\tQ^n X)^R]^{R-1}  &
&\text{and $X$ satisfies the \eqref{ss2} null condition and $R\ge1$,}
\notag\\
\text{(c)}&: &
&Q^4 [\tQ^n X]^R  &
&\text{and $X$ satisfies no null condition.}
\label{abccase}
\end{align}

Similar conditions hold as necessary conditions for a deformation to be annihilated by $\tQ$.  Of the 9 possible ways of combining these conditions between left and right, the only one with non-trivial R-spins  is $(\tQ^2)^1_0\, (Q^2)^1_0\, X^R_{0,0}$, which satisfies both the left- and right-chiral version of case (b) in \eqref{abccase} only for the $[(\tQ^2)^1_0\, (Q^2)^1_0\, X^R_{0,0}]^{R-2}$ R-spin combination.  Thus we get the 9 types of possible $\cN=2$ deformations of a SCFT shown in table \ref{N2deftable}, where we have used the R-charge assignments of $Q$ and $\tQ$ \eqref{U1Rcharges} and the restrictions on dimensions and R-charges in table \ref{N2table} to get the listed R-charge and dimension conditions.


\end{appendix}

\bibliographystyle{JHEP}
\providecommand{\href}[2]{#2}\begingroup\raggedright\endgroup

\end{document}